\documentclass[sigplan,screen]{acmart}

\AtBeginDocument{%
  \providecommand\BibTeX{{%
    \normalfont B\kern-0.5em{\scshape i\kern-0.25em b}\kern-0.8em\TeX}}}

\hypersetup{bookmarks=false}

\newcommand{\macrospath}{.}

\usepackage{ifthen}
\usepackage{xspace}

\newboolean{talk}
\setboolean{talk}{false}
\newboolean{paper}
\setboolean{paper}{false}

\newboolean{IEEEstyle}
\setboolean{IEEEstyle}{false}
\newboolean{lipicsstyle}
\setboolean{lipicsstyle}{false}
\newboolean{eptcsstyle}
\setboolean{eptcsstyle}{false}
\newboolean{entcsstyle}
\setboolean{entcsstyle}{false}
\newboolean{sigplanstyle}
\setboolean{sigplanstyle}{false}
\newboolean{easychairstyle}
\setboolean{easychairstyle}{false}
\newboolean{scrartclstyle}
\setboolean{scrartclstyle}{false}
\newboolean{lncsstyle}
\setboolean{lncsstyle}{false}

\newboolean{acmmart}
\setboolean{acmmart}{false}

\newboolean{needstheorems}
\setboolean{needstheorems}{false}
\newboolean{withimages}
\setboolean{withimages}{false}
\newboolean{withproofs}
\setboolean{withproofs}{true}

\newboolean{french}
\setboolean{french}{false}

\setboolean{acmmart}{true}
\usepackage{booktabs}   
\usepackage{subcaption} 
\usepackage[all]{xy}

\usepackage{ebproof}

\usepackage{booktabs}   
\usepackage{subcaption} 
\usepackage{graphicx}
\usepackage{cleveref}
\usepackage[normalem]{ulem}

\usepackage[utf8]{inputenc}

\usepackage{amsmath, amsfonts, amssymb, amsthm, mathtools}
\usepackage{stmaryrd}
\usepackage{mathtools}
\usepackage{enumerate}
\usepackage{bbold}
\usepackage{listings}

\usepackage{fancybox}
\usepackage{hhline}
\usepackage{marginnote}
\usepackage{soul}
\usepackage{xcolor}

\newcommand{\ignore}[1]{}

\newcommand{\colspace}{@{\hspace{.5cm}}}
\newcommand{\myinput}[1]{\ifthenelse{\boolean{withimages}}{\input{#1}}{}}

\newcommand{\reflemma}[1]{Lemma~\ref{l:#1}}
\newcommand{\reflemmas}[2]{Lemmas~\ref{l:#1} and \ref{l:#2}} 
\newcommand{\reflemmap}[2]{Lemma~\ref{l:#1}.\ref{p:#1-#2}}

\newcommand{\reflemmasps}[3]{Lemmas~\ref{l:#1}.\ref{p:#1-#2}-\ref{p:#1-#3}} 

\newcommand{\refpoint}[1]{Point~\ref{p:#1}}

\newcommand{\refcorollaryp}[2]{Corollary~\ref{c:#1}.\ref{p:#1-#2}}

\newcommand{\refthm}[1]{Thm.~\ref{thm:#1}}

\newcommand{\refprop}[1]{Prop.~\ref{prop:#1}}
\newcommand{\refpropp}[2]{Prop.~\ref{prop:#1}.\ref{p:#1-#2}} 
\newcommand{\refsect}[1]{Sect.~\ref{sect:#1}}

\newcommand{\refsubsect}[1]{Subsect.~\ref{subsect:#1}}

\newcommand{\refapp}[1]{Appendix~\ref{app:#1} (p.~\pageref{app:#1})}
\newcommand{\reftab}[1]{Table~\ref{tab:#1}}
\newcommand{\reffig}[1]{Fig.~\ref{fig:#1}}
\newcommand{\refcoro}[1]{Cor.~\ref{coro:#1}}

\newcommand{\refrmk}[1]{Remark~\ref{rmk:#1}} 
\newcommand{\refrmkp}[2]{Remark~\ref{rmk:#1}.\ref{p:#1-#2}} 
\newcommand{\refex}[1]{Ex.~\ref{ex:#1}}

\newcommand{\ie}{\textit{i.e.}\xspace}
\newcommand{\eg}{\textit{e.g.}\xspace}
\newcommand{\ih}{\textit{i.h.}\xspace}

\newcommand{\lat}{\mbox{term}\xspace}
\newcommand{\Lat}{\mbox{Term}\xspace}
\newcommand{\lav}{value\xspace}
\newcommand{\Lav}{Value\xspace}

\newcommand{\defeq}{\coloneqq} 
\newcommand{\eqdef}{\eqqcolon} 
\newcommand{\grameq}{\Coloneqq} 
\newcommand{\set}[1]{\{#1\}}

\newcommand{\size}[1]{|#1|}

\newcommand{\firesym}{{\! f}}

\renewcommand{\l}{\lambda}
\newcommand{\isub}[2]{\{#1/#2\}}
\newcommand{\replace}[2]{#1{\shortleftarrow}#2}
\renewcommand{\isub}[2]{\{\replace{#1}{#2}\}}
\newcommand{\esub}[2]{[\replace{#1}{#2}]}
\renewcommand{\esub}[2]{[#1{\shortleftarrow}#2]}
\newcommand{\subs}[4]{\{\replace{#1}{#2}, \ldots, \replace{#3}{#4}\}}
\newcommand{\sub}{\sigma}
\newcommand{\subunf}[1]{\sub(#1)}
\newcommand{\fv}[1]{{\sf fv}(#1)}

\newcommand{\rootRew}[1]{\mapsto_{#1}}
\newcommand{\Rew}[1]{\rightarrow_{#1}}

\newcommand{\rtom}{\rootRew{\msym}}

\newcommand{\expo}{\mathsf{sub}} 
\newcommand{\expovar}{\expo_{\varsym}} 

\newcommand{\rtobv}{\rootRew{\betav}} 

\newcommand{\iftsym}{\textsf{ift}}
\newcommand{\iffsym}{\textsf{iff}}
\newcommand{\ifesym}{\textsf{ife}}
\newcommand{\apesym}{\textsf{@e}}
\newcommand{\rtoift}{\rootRew{\iftsym}} 
\newcommand{\rtoiff}{\rootRew{\iffsym}} 
\newcommand{\rtoife}{\rootRew{\ifesym}} 
\newcommand{\rtoape}{\rootRew{\apesym}} 
\newcommand{\toift}{\Rew{\iftsym}} 
\newcommand{\toiff}{\Rew{\iffsym}} 
\newcommand{\toife}{\Rew{\ifesym}} 
\newcommand{\toape}{\Rew{\apesym}} 

\newcommand{\topif}{\Rew{\mathsf{pif}}} 

\newcommand{\rtoin}{\rootRew{\inert}}

\newcommand{\tob}{\Rew{\beta}}

\newcommand{\betav}{{\beta_v}} 
\newcommand{\betavsym}{\betav}
 
\newcommand{\varsym}{y}

\newcommand{\tobv}{\Rew{\betav}} 

\newcommand{\betain}{\beta_{\isym}} 
\newcommand{\inert}{\betain} 
\newcommand{\toin}{\Rew{\inert}}

\newcommand{\esym}{{\mathsf e}}
\newcommand{\isym}{i}

\newcommand{\msym}{\mathsf{m}}
\newcommand{\fsym}{f}

\newcommand{\psym}{{\mathsf p}}

\newcommand{\shufeqext}{\shufeqext} 

 \newcommand{\tom}{\Rew{\msym}}

\newcommand{\evsym}{\expo_\mathit{var}}
\newcommand{\elsym}{\expo_{l}}
\newcommand{\ersym}{\expo_{r}}
\newcommand{\eitesym}{\expo_{\mathsf{if}}}
\newcommand{\rtoev}{\rootRew{\evsym}}
\newcommand{\rtoel}{\rootRew{\elsym}}
\newcommand{\rtoeite}{\rootRew{\eitesym}}
\newcommand{\toeite}{\Rew{\eitesym}}

\newcommand{\toev}{\Rew{\evsym}}
\newcommand{\toel}{\Rew{\elsym}}
\newcommand{\toer}{\Rew{\ersym}}

\newcommand{\toevar}{\Rew{\expovar}}

\newcommand{\alphaequiv}{=_\alpha}

\newcommand{\tm}{t}
\newcommand{\tmtwo}{u}
\newcommand{\tmthree}{s}
\newcommand{\tmfour}{r}
\newcommand{\tmfive}{q}
\newcommand{\tmsix}{p}

\newcommand{\var}{x}
\newcommand{\vartwo}{y}
\newcommand{\varthree}{z}
\newcommand{\varfour}{w}

\newcommand{\val}{v}
\newcommand{\valtwo}{\val'}
\newcommand{\valthree}{\val''}

\newcommand{\mol}{b}
\newcommand{\moltwo}{b'}
\newcommand{\molthree}{b''}

\newcommand{\molv}{v}
\newcommand{\molvtwo}{w}

\newcommand{\ctxholep}[1]{\langle #1\rangle}
\newcommand{\ctxhole}{\ctxholep{\cdot}}

\newcommand{\ctx}{C}

\newcommand{\ctxp}[1]{\ctx\ctxholep{#1}}

\newcommand{\arbctxp}[1]{\arbctxp{#1}}
\newcommand{\arbctxtwop}[1]{\arbctxtwop{#1}}

\newcommand{\revctx}{R}
\newcommand{\revctxtwo}{\revctx'}
\newcommand{\revctxthree}{\revctx''}
\newcommand{\revctxp}[1]{\revctx\ctxholep{#1}}

\newcommand{\revctxthreep}[1]{\revctxthree\ctxholep{#1}}

\newcommand{\tomachhole}[1]{\leadsto_{#1}}

\newcommand{\tomacho}{\tomachhole{\osym}}

\newcommand{\tomachp}{\tomachhole{\psym}}

\newcommand{\env}{e}
\newcommand{\envtwo}{\env'}
\newcommand{\envthree}{\env''}

\newcommand{\emptyenv}{\epsilon}
\newcommand{\genv}{E}

\newcommand{\genvp}[1]{\genv\ctxholep{#1}}

\ifthenelse{\boolean{acmmart}\or \boolean{IEEEstyle}}{
  \renewcommand{\state}{s}
  }{
  \newcommand{\state}{s}
  }
\newcommand{\statetwo}{s'}

\newcommand{\rename}[1]{\renamenop{#1}}
\newcommand{\renamenop}[1]{#1^\alpha}

\newcommand{\exec}{\rho}

\newcommand{\deriv}{d}

\newcommand{\sizehole}[2]{|#2|_{#1}}

\newcommand{\sizepr}[1]{\sizehole{\psym}{#1}} 

\newcommand{\mach}{{\sf M}}

\newcommand{\consttwo}{b}

\newcommand{\itm}{i}

 \newcommand{\gconst}{i}

\newcommand{\fire}{f}

\newcommand{\betaf}{\beta_{\!\fsym}} 
\newcommand{\betafsym}{\beta_{\!\fsym}} 

\newcommand{\rtof}{\rootRew{\betaf}}
\newcommand{\tof}{\Rew{\betaf}}
\newcommand{\cbetafsym}{\mathsf c\betaf}
\newcommand{\tocf}{\Rew{\cbetafsym}}

\newcommand{\csym}{\mathsf{sea}}

\newcommand{\osym}{{\mathsf o}}

\newcommand{\la}[1]{\lambda #1.}

\newcommand{\tmseven}{m}

\newcommand{\myproof}[1]{
\ifthenelse{\boolean{omitproofs}}{\begin{IEEEproof} Proof available but omitted for readability. \end{IEEEproof}}{#1}}

\newcommand{\Open}{\text{Open}\xspace}
\newcommand{\aglam}{\Crumble \glam}
\newcommand{\apglam}{Pointed \aglam}
\newcommand{\oaglam}{\Open \aglam}
\newcommand{\oapglam}{\Open \apglam}
\newcommand{\glam}{G\lam}
\newcommand{\lam}{LAM\xspace} 

\newcommand{\crumble}{crumble\xspace}
\newcommand{\crumblep}{crumble\xspace}

\newcommand{\crumbled}{crumbled\xspace}
\newcommand{\crumbledt}{bite\xspace}
\newcommand{\Crumbledt}{Bite\xspace}
\newcommand{\crumbling}{crumbling\xspace}
\newcommand{\Crumble}{Crumble\xspace}
\newcommand{\Crumblep}{Crumble\xspace}
\newcommand{\Crumbled}{Crumbled\xspace}
\newcommand{\Crumbling}{Crumbling\xspace}

\newcommand{\gregoire}{Gr{\'{e}}goire\xspace}

\newcommand{\withproofs}[1]{\ifthenelse{\boolean{withproofs}}{#1}{}}

\newcommand{\withoutproofs}[1]{\ifthenelse{\boolean{withproofs}}{}{#1}}

\newcommand{\firecalc}{\lambda_\mathsf{fire}}
\newcommand{\cfirecalc}{\lambda_\mathsf{fire}^{\ifsym}}

\newcommand{\plotcalc}{\lambda_\mathsf{Plot}}
\newcommand{\plotcalcif}{\lambda_\mathsf{Plot}^{\mathsf{if}}}

\newcommand{\doubt}[1]{}

\newcommand{\letexp}{\mathsf{let}}
\newcommand{\letin}[3]{{\sf let}\ #1=#2\ {\sf in}\ #3}

\ifthenelse{\boolean{acmmart} \or \boolean{IEEEstyle}}{
  
  }{
  
  }

\newcommand{\append}[2]{#1 \mathrel{@} #2}
\newcommand{\envctx}{E}
\newcommand{\envctxp}[1]{\envctx\ctxholep{#1}}

\newcounter{numberone}
\newcounter{numberoneroman}
\newcounter{numberonealph}

\newcommand{\cbv}{CbV\xspace}
\newcommand{\ocbv}{Open \cbv}
\newcommand{\ccbv}{Closed \cbv}

\newcommand{\tostrat}{\rightarrow}
\newcommand{\tomachine}{\tomachhole\mach}

\newcommand{\domain}[1]{\mathsf{dom}(#1)}

\newcommand{\cell}{c}
\newcommand{\celltwo}{d}
\newcommand{\cellthree}{e}

\newcommand{\cctx}{C}
\newcommand{\cctxtwo}{\cctx'}
\newcommand{\cctxthree}{\cctx''}
\newcommand{\cctxfour}{\cctx'''}
\newcommand{\cctxp}[1]{\cctx\ctxholep{#1}}
\newcommand{\cctxtwop}[1]{\cctxtwo\ctxholep{#1}}
\newcommand{\cctxthreep}[1]{\cctxthree\ctxholep{#1}}

\newcommand{\ectx}{E}

\newcommand{\ectxp}[1]{\ectx\ctxholep{#1}}

\newcommand{\mytr}[1]{\underline{#1}}
\newcommand{\auxtr}[1]{\overline{#1}}

\newcommand{\rb}[1]{#1_\downarrow}
\newcommand{\rbp}[1]{(#1)_\downarrow}

\newcommand{\aenv}{\env_v}
\newcommand{\aenvtwo}{\envtwo_v}
\newcommand{\aenvthree}{\envthree_v}
\newcommand{\acell}{\cell_v}
\newcommand{\fenv}{\env_\firesym}
\newcommand{\fenvtwo}{\envtwo_\firesym}
\newcommand{\fenvthree}{\envthree_\firesym}
\newcommand{\fcell}{\cell_\firesym}

\newcommand{\len}[1]{|#1|}
\newcommand{\lenv}[1]{\len{#1}_\textup{var}}
\newcommand{\no}[2]{\##1(#2)}
\renewcommand{\no}[2]{\len{#2}_{#1}}
	
\newtheorem*{proofof}{Proof of}

\makeatletter
\newcommand\Copy[2]{
        \marginpar{\scriptsize \ \ \hyperlink{hl-appendix-#1}{Proof p.\,{\pageref*{appendix-#1}}}}
	\immediate\write\@auxout{\unexpanded{\global\long\@namedef{mytext@#1}{#2}
  }}%
	#2%
}

\newcommand\Paste[1]{%
        \hypertarget{hl-appendix-#1}{}\label{appendix-#1}
	\renewcommand{\inappendix}[1]{}
	\ifcsname mytext@#1\endcsname
	\@nameuse{mytext@#1}%
	\else
	``??''
	\fi
	\renewcommand{\inappendix}[1]{#1}
}
\makeatother

\newcommand{\mylabel}[1]{\inappendix{\label{#1}}\inappendix{}}
\newcommand{\inappendix}[1]{#1}

\newcommand{\disj}{\mathrel{\#}}

\newcommand{\sep}{\mathrel{\!\textnormal{\texttt{¦}}\!}}
\newcommand{\benv}{\env_{{\,\sep}}}
\newcommand{\benvtwo}{\envtwo_{{\,\sep}}}
\newcommand{\myiota}[1]{\iota(#1)}
\newcommand{\atoi}[1]{(#1)_{\Downarrow}}
\newcommand{\toc}{\Rew{\csym}}

\newcommand{\length}[1]{\len{#1}_\textup{len}}
\newcommand{\lcontext}{context\xspace}
\newcommand{\lContext}{Context\xspace}
\newcommand{\valuectx}{v-context\xspace}
\newcommand{\firectx}{f-context\xspace}

\newcommand{\tolet}{\Rew{\text{let}}}

\newcommand{\ifsym}{\mathsf{if}}
\newcommand{\xite}[3]{\ifsym\,#1\,\mathsf{then}\,#2\,\mathsf{else}\,#3}
\newcommand{\xitecompact}{\mathsf{if}\mbox{-}\mathsf{then}\mbox{-}\mathsf{else}}
\newcommand{\err}{\mathsf{err}}
\newcommand{\true}{\mathsf{true}}
\newcommand{\false}{\mathsf{false}}

\newcommand{\pif}{Pif\xspace}

\newcommand{\lcell}{$v$-\crumblep}

\newcommand{\lenvironment}{$v$-environment\xspace}
\newcommand{\firecell}{$f$-\crumblep{}\xspace}
\newcommand{\fireenvironment}{$f$-environment\xspace}
\newcommand{\pcell}{pointed \crumble{}\xspace}
\newcommand{\Pcell}{Pointed \crumble{}\xspace}
\newcommand{\penvironment}{pointed environment\xspace}
\newcommand{\Penvironment}{Pointed environment\xspace}

\newcommand{\pval}{\val_{\neg \var}}
\newcommand{\pvalue}{practical value\xspace}

\newcommand{\Pvalues}{Practical values\xspace}

\newcommand{\crumblesym}{\mathsf{Cr}}
\newcommand{\tocrumble}{\Rew{\crumblesym}}
\newcommand{\topcrumble}{\Rew{\mathsf{pCr}}}
\newcommand{\ocrumblesym}{\mathsf{oCr}}
\newcommand{\toocrumble}{\Rew{\ocrumblesym}}
\newcommand{\topocrumble}{\Rew{\mathsf{poCr}}}

\theoremstyle{acmplain}
{Fact}[section]
\theoremstyle{acmdefinition}
\newtheorem{remark}
{Remark}[section]

\renewcommand{\rtom}{\rootRew{\betav}} 

\renewcommand{\tom}{\Rew{\betav}}

\renewcommand{\msym}{\betav}

\begin{document}

\title[Crumbling Abstract Machines]{\texorpdfstring{Crum\raisebox{-.02in}{bl\raisebox{-.02in}{in\raisebox{-.02in}{\rotatebox{10}{g}}}}}{Crumbling} Abstract Machines}  

\author{Beniamino Accattoli}
\affiliation{
  \department{LIX}              
  \institution{Inria \& \'Ecole Polytechnique}            
  \country{France}                    
}
\email{beniamino.accattoli@inria.fr}          

\author{Andrea Condoluci}
\affiliation{
  \department{Department of Computer Science and Engineering}             
  \institution{University of Bologna}           
  \country{Italy}                   
}
\email{andrea.condoluci@unibo.it}         

\author{Giulio Guerrieri}
\affiliation{
  \department{Department of Computer Science}             
  \institution{University of Bath}           
  \country{United Kingdom}                   
}
\email{g.guerrieri@bath.ac.uk}

\author{Claudio Sacerdoti Coen}
\affiliation{
  \department{Department of Computer Science and Engineering}             
  \institution{University of Bologna}           
    \country{Italy}                   
}
\email{claudio.sacerdoticoen@unibo.it}         

\renewcommand{\shortauthors}{Accattoli, Condoluci, Guerrieri, and Sacerdoti Coen}

\begin{abstract}
Extending the $\l$-calculus with a construct for sharing, such as $\letexp$ expressions, enables a special representation of terms: iterated applications are decomposed by introducing sharing points in between any two of them, reducing to the case where applications have only values as immediate subterms. 

This work studies how such a crumbled representation of terms impacts on the design and the efficiency of abstract machines for call-by-value evaluation. About the design, it removes the need for data structures encoding the evaluation context, such as the applicative stack and the dump, that get encoded in the environment. 
About efficiency, we show that there is no slowdown, clarifying in particular a point raised by Kennedy, about the potential inefficiency of such a representation.

Moreover, we prove that everything smoothly scales up to the delicate case of open terms, needed to implement proof assistants. 
Along the way, we also point out that continuation-passing style transformations---that may be alternatives to our representation---do not scale up to the open case.

\end{abstract}


\maketitle

\section{Introduction}
This paper is about the extension of $\l$-calculus with explicit constructors for \emph{sharing}. 
The simplest such construct is a $\letin \var \tmtwo \tm$ expression, standing for \emph{$\tm$ where $\var$ will be substituted by $\tmtwo$}, that we also write more concisely as $\tm\esub\var\tmtwo$ and call ES (for \emph{explicit sharing}, or \emph{explicit subsitution}\footnote{$\letexp$ expressions and explicit substitutions usually come with different operational semantics: $\letexp$ expressions substitute in just one step, while explicit substitutions substitute in many \emph{micro} steps, percolating through the term structure. They follow however the same typing principles. Moreover, explicit substitutions have many different formulations. In this paper we see $\letexp$ expressions as yet another form of explicit substitutions, and thus conflate the two terminologies.}).  
Thanks to ES, $\beta$-reduction can be decomposed into more atomic steps.
The simplest decomposition splits $\beta$-reduction as
$(\la\var\tm)\tmtwo \Rew{\beta_\textsf{ES}} \tm\esub\var\tmtwo \Rew{\textsf{ES}} \tm\isub\var\tmtwo$.

It is well-known that ES are somewhat redundant, as they can always be removed, by simply coding them as $\beta$-redexes. They are however more than syntactic sugar, as they provide a simple and yet remarkably effective tool to understand, implement, and program with $\l$-calculi and functional programming languages. 

From a logical point of view, ES are the proof terms corresponding to the extension of natural deduction with a cut rule, and the cut rule is \emph{the} rule representing computation, according to Curry-Howard. 
From an operational semantics point of view, they allow elegant formulations of subtle strategies such as call-by-need evaluation---various presentations of call-by-need use ES \cite{Wad:SemPra:71,DBLP:conf/popl/Launchbury93,DBLP:journals/jfp/MaraistOW98,DBLP:conf/popl/AriolaFMOW95,DBLP:journals/jfp/Sestoft97,DBLP:conf/icfp/KutznerS98} and a particularly simple one is by Accattoli, Barenbaum, and Mazza in \cite{DBLP:conf/icfp/AccattoliBM14}.
From a programming point of view, they are part of most functional languages we are aware of. 
From a rewriting point of view, they enable proof techniques not available within the $\l$-calculus (\eg reducing a global rewriting properties such as standardization to a local form, see \citet{DBLP:conf/rta/Accattoli12}). 
Finally, sharing is used in all implementations of tools based on the $\l$-calculus to circumvent \emph{size explosion}, the degenerate behaviour for which the size of $\l$-terms may grow exponentially with the number of $\beta$-steps.

\paragraph{\Crumbled Forms} Once sharing is added to the $\l$-calculus, it enables a representation of terms 
where a sharing point is associated with every constructor of the term.  
Such a special form, roughly, is obtained by (recursively) decomposing iterated applications by introducing an ES in between any two of them. 
For instance, the representation of the term $(((\la\var \var (\var \var)) \vartwo) ((\la\varthree\varthree) \vartwo)) \vartwo$ is
$$(\varfour'' \vartwo) \esub {\varfour''}{\varfour' \varfour} \esub{\varfour'}{(\la\var(\var\var')\esub{\var'}{\var\var}) \vartwo} \esub\varfour{(\la\varthree\varthree) \vartwo}$$
Note that the transformation involves also function bodies (\ie $\la\var\var (\var \var)$ turns into $\la\var(\var\var')\esub{\var'}{\var\var}$), that ES are grouped together unless forbidden by abstractions, and that ES are flattened out, i.e. they are not nested unless nesting is forced by abstractions.

This work studies such a representation, 
called \emph{\crumbled} as it crumbles a \lat by means of ES.
Our \emph{\crumbling transformation} closely resembles---while not being exactly the same---the transformation
into \emph{a(dministrative) normal form} (shortened to ANF), introduced by  \citet{DBLP:conf/pldi/FlanaganSDF93a}, building on work by \citet{DBLP:journals/lisp/SabryF93}, itself a variant of the \emph{continuation-passing style} (CPS) transformation. 

A delicate point is to preserve \crumbled forms during evaluation. 
ES often come together with commutation rules to move them around the term structure. These rules are often used to unveil redexes during evaluation or to preserve specific syntactic forms. 
They may introduce significant overhead that, if not handled carefully, can even lead to asymptotic slowdowns as shown by \citet{DBLP:conf/icfp/Kennedy07}. One of the contributions of this work is to show that \crumbled forms can be evaluated and preserved with no need of commutation rules, therefore avoiding Kennedy's potential slowdown.


\paragraph{This Paper} The focus of our work is on the impact of \crumbled forms on the design and 
asymptotic overhead of abstract machines with \emph{weak} evaluation (\ie out of abstractions) on closed terms, and the scalability to (possibly) open terms. 
Bounding the overhead of abstract machines is a new trend, according to which the machine overhead has to be proved polynomial or even linear in the number of $\beta$-steps \cite{DBLP:conf/icfp/AccattoliBM14,DBLP:conf/aplas/AccattoliBM15,fireballs,DBLP:conf/wollic/Accattoli16,AccattoliGuerrieri17b,DBLP:conf/ppdp/AccattoliB17}. 
Open terms---that are not needed to implement functional languages---are used to implement the more general and subtle case of proof assistants. The two topics actually motivate each other: the naive handling of open terms with the techniques for functional languages gives abstract machines with exponential overhead \cite{fireballs,AccattoliGuerrieri17b} which then pushes to develop more efficient machines. 

We anticipate here the main results of the paper: \crumbled forms induce abstract machines for weak evaluation with less data structures and the transformation does not introduce any asymptotic overhead. Moreover, these facts smoothly scale up to open terms. 

\paragraph{Why study \crumbled forms.} 
Our interest in studying \crumbled forms comes precisely from the fact that they remove some data structures from the design of abstract machines. The relevance of this fact becomes evident when one tries to design abstract machines for strong evaluation (that is, evaluating under abstraction). 
The study of such machines is extremely technical (see also section \refsect{extensions}) because they have more data structures and more transitions than in the closed and open cases. The many additional transitions are in particular due to the handling of the various data structures. 
In call-by-name, the situation is still manageable \cite{DBLP:conf/lfp/Cregut90,DBLP:conf/ppdp/Garcia-PerezNM13,DBLP:conf/aplas/AccattoliBM15,DBLP:conf/wollic/Accattoli16}, but in call-by-value/need the situation becomes quickly desperate---it is not by chance that there is not a single strong abstract machine for call-by-value/need in the literature.

This work is then preliminary to a detailed study of strong abstract machines for call-by-value and call-by-need. 
The aim is to explore the subtleties in frameworks that are well understood, such as the closed and open call-by-value cases, and show that there are no slow downs in turning to a \crumbled representation.\medskip

The next sub-sections continue the introduction with a lengthy overview of the role of environments, the content of the paper, the relationship with the ANF, the asymptotic study of abstract machines, and related work.

\subsection{Environments} 
ES are often grouped together instead of being scattered all over the term, in finite sequences called \emph{environments}. 
Abstract machines typically rely on environments. 
\Crumbled forms also rely on packing ES together, as pointed out before, but  depart from the ordinary case as environments may appear also under abstractions. 

\paragraph{\Crumbled Environments.} 
The notion of environment induced by \crumbled forms, named here \emph{\crumbled environments}, is peculiar. 
\Crumbled environments indeed play a double role: they both store delayed substitutions, as also do ordinary environments, and \emph{encode evaluation contexts}.
In ordinary abstract machines, the evaluation context is usually stored in data structures such as the \emph{applicative stack} or the \emph{dump}. 
Roughly, they implement the search for the redex in the ordinary applicative structure of \lat{s}. 
For \crumbled forms, the evaluation context is encoded in the \crumbled environment, and so the other structures disappear.

\paragraph{Operations on \Crumbled Environments.} 
There are two subtle implementative aspects of \crumbled environments, that set them apart from ordinary ones. 
Ordinary environments are presented with a sequential structure but they are only accessed randomly (that is, not sequentially)---in other words, their sequential structure does not play a role. 
\Crumbled environments, as the ordinary ones, are accessed randomly, to retrieve delayed substitutions, but they are also explored sequentially---since they encode evaluation contexts---in order to search for redexes. 
Therefore, their implementation has to reflect the sequential structure. 

The second subtlety is that \crumbled machines also have to concatenate environments, that is an operation never performed by ordinary machines, and that has to be concretely implemented as efficiently as possible, \ie in constant time. 
That this point is subtle is proved by the fact that Kennedy's slowdown  \cite{DBLP:conf/icfp/Kennedy07} amounts to a quadratic overhead in  evaluating terms in ANF due to the concatenation of environments.

To address these points, we provide a prototype OCaml implementation of \crumbled environments in \refapp{ocamlimpl}, to be compared with the one of global environments in \citet{DBLP:conf/ppdp/AccattoliB17}, that does not concretely implement the sequential structure. 
In particular, our implementation concatenates environments in constant time and does not suffer from Kennedy's slowdown. 
Essentially, Kennedy's slowdown amounts to the fact that his implementation concatenates ANF environments in linear rather than constant time (see \Cref{sect:kennedy-danvy}).


\subsection{Content of the Paper}
\paragraph{The Closed Case.} 
First, 
we define \crumbled forms and an abstract machine evaluating them, the \aglam, and show that it implements Plotkin's \emph{closed} small-step call-by-value (\cbv for short) $\l$-calculus (extended with conditionals, see below). 
Moreover, we study the overhead of the machine, and show that it is linear in the number of $\beta$-steps and in the size of the initial term, exactly as the best machines for \cbv executing ordinary \lat{}s. 
Therefore, the \crumbling transformation does not introduce any asymptotic overhead. The study is detailed and based on a careful and delicate spelling of the invariants of the machine. In particular, our approach does not suffer from Kennedy's potential slowdown.

\paragraph{Open Terms.} The second ingredient of the new trend of abstract machines \cite{DBLP:conf/icfp/AccattoliBM14,DBLP:conf/aplas/AccattoliBM15,fireballs,DBLP:conf/wollic/Accattoli16,AccattoliGuerrieri17b,DBLP:conf/ppdp/AccattoliB17}---the first being complexity analyses---is studying evaluation in  presence of (possibly) open terms or even strong evaluation (\ie under abstraction), which is required in order to implement proof assistants. 
Apart from few exceptions---\citet{DBLP:conf/lfp/Cregut90}, \citet{DBLP:conf/icfp/GregoireL02}, and \citet{DBLP:conf/ppdp/Garcia-PerezNM13}---the literature before the new wave mostly neglected these subtle cases, and none of those three papers addressed complexity. 

The open case, in which evaluation is weak but terms are possibly open is strictly harder than the closed one, and close in spirit to the strong case, but easier to study---it is for instance the one studied by \citet{DBLP:conf/icfp/GregoireL02} when modelling (an old version of) the abstract machine of the kernel of Coq.

\paragraph{Open Call-by-Value.} Open evaluation for \cbv---shortened \ocbv---is particularly subtle because, as it is well-known, Plotkin's operational semantics is not adequate when dealing with open terms---see  \citet{DBLP:conf/aplas/AccattoliG16,aplas18}. 
\ocbv has been studied deeply by 
\citet{fireballs,DBLP:conf/aplas/AccattoliG16,AccattoliGuerrieri17b,aplas18}, exploring different presentations, their rewriting, cost models, abstract machines, and denotational semantics. 
One of the motivations of this work is to add a new piece to the puzzle, by lifting the crumbling technique to the open case. 

Our second contribution is to show that the \crumbling technique smoothly scales up to \ocbv. 
We provide an abstract machine, the \oaglam, and we show that it implements \emph{the fireball calculus}---the simplest presentation of \ocbv---and that, as in the closed case, it only has a linear overhead. Two aspects of this study are worth pointing out.
First, the technical development follows almost identically the one for the closed case, once the subtler invariants of the new machine have been found. 
Second, \emph{the substitution of abstractions on demand}, a technical optimisations typical of open/strong cases (introduced in \citet{DBLP:journals/corr/AccattoliL16} and further studied in \citet{fireballs,AccattoliGuerrieri17b}), becomes superfluous as it is subsumed by the \crumbling transformation.

\subsection{The Relationship with ANF} 
As long as one sticks to the untyped $\l$-calculus, \crumbled forms coincide with ANF. The ANF, we said, is a variant of the CPS transformation. 
Roughly, the difference is that the ANF does not change the type, when terms are typed (here we work without types).


\citet{DBLP:conf/icfp/Kennedy07} pointed out two problems with the ANF. One is the already discussed quadratic overhead, that does not affect our approach. The second one is the fact that the ANF does not smoothly scale up when the $\l$-calculus is extended to further constructs such as conditionals or pattern matching. Essentially, the ANF requires conditionals and pattern matching to be out of ES, that is, to never have an expression such as $\tmthree\esub \var {(\xite \val \tm \tmtwo)}$. Unfortunately, these configurations can be created during evaluation. To preserve the ANF,  one is led to add so-called \emph{commuting conversions} such as:
\begin{equation}
  \tag{CC}
\small \tmthree\esub \var {(\xite \val \tm \tmtwo)}  
\to
 \xite \val {(\tmthree\esub \var \tm )} {(\tmthree \esub \var \tmtwo )}
\end{equation}
Clearly, there is an efficiency issue: the commutation causes the duplication of the subterm $\tmthree$. A way out is to use a continuation-like technique, which makes Kennedy conclude that then there is no point in preferring ANF to CPS.

This is where our \crumble representation departs from the ANF, as we do not require conditionals and pattern matching to be out of ES. 
Kennedy only studies the closed case. Our interest in open and strong evaluation is to explore the theory of implementation needed for proof assistants. 
In these settings, commutations of conditionals and pattern matching such as those hinted at by Kennedy are not valid, as they are not validated by dependent type systems like those of Coq or Agda.
For example, adding the CC rule above when the conditional is dependently typed breaks the property of \emph{subject reduction}, as typed terms reduce to ill-typed terms. Consider the term:
\begin{center}
\small $ (\var+1) \esub {\var\colon (\xite {true} {nat} {bool})} {\xite {true} 0 {\textit{false}}} \colon nat$
\end{center}
that has type $nat$ because the type of $\var$ is convertible to $nat$. 
By applying rule CC, we obtain:
$$
\xite {true} {(
(\var+1) \esub \var 0 
)} {(\underline{(\var+1) \esub \var {\textit{false}} })}
$$
which is clearly ill-typed.

The problem in the open case is actually more general, as not even the CPS would work: its properties do not scale up to open terms. 
In \Cref{sect:kennedy-danvy}, indeed, we provide a counter-example to the simulation property in the open case.\footnotemark
\footnotetext{\citet{DBLP:journals/mscs/DanvyF92} claim that the CPS transformation scales up to open terms (their Theorem 2). However, as we discuss in \Cref{sect:kennedy-danvy}, they consider only Plotkin's operational semantics, which is not adequate for open terms.}

To sum up, neither commuting conversions nor the CPS transformation can be used in our framework. Therefore, we accept that conditionals and pattern matching may appear in ES (in contrast to Kennedy) and so depart from the ANF.

In the paper we treat the cases of the closed and open \cbv calculi extended with conditionals. The essence of the study is the crumbling of $\beta$-reduction, not the conditionals. Conditionals are included only to stress the difference with respect to the ANF (pattern matching can be handled analogously), but they do not require a special treatment.

\subsection{The Complexity of Abstract Machines}

\paragraph{Asymptotic Bounds $vs$ Benchmarking} The study of asymptotic bounds for abstract machines is meant to complement the use of benchmarking, by covering \emph{all} possible cases, that certainly cannot be covered via benchmarking.

The relevance of such a study is particularly evident when one considers open terms or strong evaluation. For strong evaluation, for instance, for more than 25 years in the literature there has been only Cregut's abstract machines \cite{DBLP:conf/lfp/Cregut90}, which on size exploding families of terms actually has exponential overhead (in the number of $\beta$-steps and the size of the initial term). A polynomial machine, developed via a careful asymptotic study, is in \citet{DBLP:conf/wollic/Accattoli16}. 
Similarly, the abstract machine for open terms described in \citet{DBLP:conf/icfp/GregoireL02} suffers of exponential overhead on size exploding families (even if the authors then in practice implement a slightly different machine with polynomial overhead). The asymptotic study of this case is in \citet{fireballs,AccattoliGuerrieri17b}.


\paragraph{Abstract machines \textit{vs} compilation}

Abstract machines and compilation to machine language are two distinct techniques to execute a program. 
Compilation is typically more efficient, but it only handles the case where terms are closed and evaluation is weak, that is, the one of functional languages. 
Strong evaluation is sometimes employed during compilation to optimise the compiled code, but typically only on linear redexes where size and time explosions are not an issue. 
Abstract machines are the only execution technique implemented in interactive theorem provers based on dependent types, that need strong evaluation.

\citet{DBLP:conf/icfp/Kennedy07} argues that CPS-based translations are superior to ANF also because the CPS makes join points explicit as continuations, so that invocation of the continuation can be compiled efficiently using jumps. The argument is only valid for compilation and it does not affect abstract machines.

\paragraph{Garbage collection} We study \emph{abstract} machines, that on purpose ignore many details of concrete implementations, such as garbage collection, that is an orthogonal topic. 
In particular, garbage collection is always at most polynomial, if not linear, and so its omission does not hide harmful blowups. As far as we know, no abstract machine implemented in interactive provers performs garbage collection.

\subsection{Related Work} 
\paragraph{Environments.} In a recent work, \citet{DBLP:conf/ppdp/AccattoliB17} compare various kinds of environments, namely, \emph{global}, \emph{local}, and \emph{split}, from implementative and complexity points of view. The \crumbling transformation can be studied with respect to every style of environment. 
Here we focus on \crumbled \emph{global} environments because they are simpler and because we also consider the open case, where all kinds of environment induce the same complexity.

\paragraph{Administrative Normal Forms.} The literature on ANF is scarce. Beyond the already cited original papers, Danvy has also studied them and their relationship to CPS, but usually calling them \emph{monadic normal forms} \cite{DBLP:journals/scp/Danvy94,DBLP:conf/popl/HatcliffD94,DBLP:conf/cc/Danvy03} because of their relationship with Moggi's monadic $\l$-calculus \cite{DBLP:journals/iandc/Moggi91}. 
That terminology however sometimes describes a more liberal notion of terms, for instance in \citet{DBLP:conf/icfp/Kennedy07}. 
Kennedy's paper is also another relevant piece in the literature on ANF.

\section{The \pif Calculus}\label{sect:plotkin}
The grammars and the small-step operational semantics of the \emph{\pif calculus} $\plotcalcif$, that is, Plotkin's calculus $\plotcalc$ \cite{DBLP:journals/tcs/Plotkin75} for \ccbv evaluation extended with booleans and an $\xitecompact$ construct, plus error handling for clashing constructs, are in \reffig{plotkin-calculus}.

A \lat is either an application of two \lat{s}, an $\xitecompact$, or a \emph{\lav}, which is in turn either a variable, a ($\l$-)abstraction, $\true$, $\false$, or an error $\err$. We distinguish values that are not variables, noted $\pval$ and called \emph{practical values}, following \citet{DBLP:journals/iandc/AccattoliC17}. The \emph{body} of an abstraction $\la\var\tm$ is $\tm$ and the \emph{bodies} of a conditional $\xite\tm\tmtwo\tmthree$ are its two branches $\tmtwo$ and $\tmthree$.
\Lat{s} are always identified up to $\alpha$-equivalence and the set of free variables of a \lat $\tm$ is denoted by $\fv{\tm}$;
$\tm$ is \emph{closed} if $\fv{\tm} = \emptyset$, \emph{open} otherwise.
We use $\tm\isub\var\tmtwo$ for the term obtained by the capture-avoiding substitution of $\tmtwo$ for each free occurrence of $\var$ in $\tm$.

\paragraph{\lContext{s}} In general, \emph{contexts} 
are denoted by $\ctx$ and are terms with exactly one occurrence of a special constant $\ctxhole$ called \emph{the hole}, that is a placeholder for a removed subterm.
In the paper we  use various notions of contexts in different calculi---for $\plotcalcif$ the relevant notion is \emph{right (evaluation) \valuectx} $\revctx$ (see \reffig{plotkin-calculus}).
The basic operation on (whatever notion of) contexts is the \emph{plugging} $\ctxp\tm$ of a \lat $\tm$ for the hole $\ctxhole$ in $\ctx$: simply the hole is removed and replaced by $\tm$, possibly capturing variables. 

\paragraph{Evaluation} According to the definition of right \valuectx, \cbv evaluation $\topif$ in $\plotcalcif$ is \emph{weak}, \ie it does not reduce under $\lambda$-abstractions and in the branches of an $\xitecompact$.
\cbv evaluation is defined for \emph{any} (possibly open) \lat.
But it is well-known that this operational semantics is \emph{adequate} only for \emph{closed} \lat{s}, as first noticed by Paolini and Ronchi Della Rocca \cite{parametricBook}, see also Accattoli and Guerrieri \cite{DBLP:conf/aplas/AccattoliG16,aplas18}.
When restricted to closed \lat{s}, $\plotcalcif$ is called \emph{Closed (Conditional) \cbv}:
in this setting,
evaluation can fire a $\beta$-redex $(\la{\var}\tm)\tmtwo$ only if the argument $\tmtwo$ is a closed value, \ie a closed $\lambda$-abstraction, a boolean, or $\err$; and in the production $\revctx\val$ for the definition of right \valuectx{s}, $\val$ is always a closed value. Note that we work with right-to-left evaluation---this is forced by the production $\revctx\val$ in the definition of right evaluation \valuectx{s}. In the closed case one could as well work with left-to-right evaluation, the choice is inessential. 

The error constant $\err$ is generated during evaluation by the two cases of construct clashes: when the condition for an $\xitecompact$ is an abstraction and when a boolean is applied to a term. Both cases would be excluded by typing, but in our untyped setting they are possible, and handled via errors. Similarly, errors are also propagated when they appear as conditions for $\xitecompact$ and as left terms of an application. These cases are handled by rules $\toife$ and $\toape$. Note that errors do not propagate when they occur as arguments of applications: if the left sub-term of the application becomes an abstraction that erases the error then the error is \emph{handled} and it is not observable.

A key property of Plotkin's \ccbv is \emph{harmony}: a closed \lat is $\betav$-normal if and only if it is a (closed) \lav \ie a (closed) $\lambda$-abstraction.
Therefore, every closed \lat either diverges or it evaluates to a (closed) $\lambda$-abstraction. Harmony extends to $\plotcalcif$.

\begin{proposition}[\pif harmony]
\label{prop:pif-harmony}\Copy{prop:pif-harmony}{
  Let $\tm$ be a closed term. $\tm$ is $\topif$-normal if and only if $\tm$ is a value.
}\end{proposition}

\begin{figure}[t]
  \centering

  \scalebox{0.9}{
\begin{tabular}{c}
    $\begin{array}{r@{\hspace{.3cm}}rlllllllll}
\textsc{Terms} & \tm,\tmtwo, \tmthree &\grameq& \val \mid \tm\tmtwo \mid \xite \tm \tmtwo \tmthree\\
\textsc{Values} & \val & \grameq & \var \mid \pval\\
\textsc{\Pvalues} & \pval & \grameq &  \la\var\tm \mid \true \mid \false \mid \err\\
\textsc{Right \valuectx{}} & \revctx &\grameq & \ctxhole\mid \tm\revctx\mid \revctx\val \mid \xite\revctx \tmtwo \tmthree
	          \end{array}$
	          \\\\
    $\begin{array}{rclllllllll}
	\multicolumn{3}{c}{\textsc{Reduction Rules at Top Level}}
\\
	    (\la\var\tm)\val
	    & \rtobv &
	    \tm\isub\var{\val} &
 \\

		    \xite\true\tm\tmtwo
		    & \rtoift &
		     \tm\\

		    \xite\false\tm\tmtwo
		    & \rtoiff &
		     \tmtwo
		     \\[0.5em]
		    \xite\tm \tmtwo \tmthree
		    & \rtoife &
		    \err \ \ \ \ \  \mbox{if $\tm=\la\var\tmtwo$ or $\tm = \err$}
		    \\
		    \tm \tmtwo
		    & \rtoape &
		    \err \ \ \ \ \ \mbox{if $\tm\in \set{\true,\false,\err}$}
		    \end{array}$
\\\\
$\begin{array}{rclllllllll}
	    \multicolumn{3}{c}{\textsc{Contextual closure} }\\
	    \revctxp \tm \Rew{a} \revctxp \tmtwo & \textrm{~~~if } & \tm \rootRew{a} \tmtwo \ \ \ \ \mbox{for $a\in\set{\betav,\iftsym,\iffsym,\ifesym,\apesym}$}
	    \\[2pt]
	    \topif & \defeq & \tobv \cup \toift \cup \toiff \cup \toife \cup \toape

    \end{array}$
  \end{tabular}}
  \Description{Pif calculus $\plotcalcif$.}
  \caption{Pif calculus $\plotcalcif$.}
  \label{fig:plotkin-calculus}
\end{figure}

\section{\Crumbled Evaluation, Informally}
\label{sect:intro-abn-env}

\paragraph{Decomposing applications.} The idea is to forbid the nesting of non-value constructs such as applications  and $\xitecompact$ without losing expressive power. To ease the explanation, we focus on nested applications and forget about $\xitecompact$---they do not pose any difficulty. Terms such as $(\tm \tmtwo) \tmthree$ or $\tm (\tmtwo \tmthree)$ are then represented as $(\la \var(\var\tmthree)) (\tm \tmtwo)$ and $(\la\var(\tm\var))(\tmtwo \tmthree)$ where $\var$ is a fresh variable. It is usually preferred to use $\letexp$ expressions rather than introducing $\beta$-redexes, so that one would rather write $\letin \var {\tm\tmtwo} {(\var\tmthree)}$ and $\letin \var {\tmtwo\tmthree} {(\tm\var)}$, or, with ES
(aka environment entries), rather write
\begin{center}
$(\var\tmthree)\esub\var{\tm\tmtwo}$ \ \ \ and \ \ \ $(\tm\var)\esub\var {\tmtwo\tmthree}$
\end{center}
If the \crumbling transformation $\mytr{\,\cdot\,}$ is applied to the whole \lat---recursively on $\tm$, $\tmtwo$ and $\tmthree$ in our examples---all applications have the form $\val \val'$, \ie they only involve values.
If moreover \cbv evaluation is adopted, then such a \crumbled form is stable by evaluation (reduction steps are naturally defined so that a \crumbled form reduces to a \crumbled form), as variables can only be replaced by values.

\paragraph{Simulation and no evaluation contexts.} Let us now have a look at a slightly bigger example and discuss the recursive part of the \crumbling transformation. Let $I=\la\var\var$ be the identity and consider the term $\tm \defeq ((\la\vartwo\vartwo\vartwo) I) ((I I) I)$ whose right-to-left evaluation is
\[\begin{array}{rcllllll}
\tm & \tobv & ((\la\vartwo\vartwo\vartwo) I) (I I) & \tobv & ((\la\vartwo\vartwo\vartwo) I) I 
\\
& \tobv & (II) I & \tobv & II & \tobv & I   
  \end{array}\]

The \crumbling transformation decomposes all applications, taking special care of grouping all the environment entries together, flattening them out (that is, avoiding having them nested one into the other), and reflecting the evaluation order in the arrangement of the environment.
For instance, the \crumbled representation $\mytr\tm$ of the \lat $\tm$ 
above is
\begin{align*}
\mytr{\tm} &= (\varfour\varthree)\esub\varfour{(\la\vartwo\vartwo\vartwo) I} \esub\varthree{\var I} \esub\var{I I}
\end{align*}
and evaluation takes always place at the end of the environment, as follows:
\[\begin{array}{llllllllll}
 \mytr\tm
 & \tobv & \multicolumn{3}{l}{
 (\varfour\varthree)\esub\varfour{(\la\vartwo\vartwo\vartwo) I} \esub\varthree{\var I} \esub\var I
 }\\
 & \Rew{[\ ]} & \multicolumn{3}{l}{
 (\varfour\varthree)\esub\varfour{(\la\vartwo\vartwo\vartwo) I} \esub\varthree{I I}
 }\\
 & \tobv & \multicolumn{3}{l}{
 (\varfour\varthree)\esub\varfour{(\la\vartwo\vartwo\vartwo) I} \esub\varthree I 
 }\\
 & \Rew{[\ ]} & 
 
 (\varfour I)\esub\varfour{(\la\vartwo\vartwo\vartwo) I}
 
 & \tobv & (\varfour I)\esub\varfour{I I} 
 \\
 & \tobv & (\varfour I)\esub\varfour I
 & \Rew{[\ ]} & I I  \ \ \ \  \tobv  \ \ \ \ I
\end{array}\]
where the $\tobv$ steps correspond exactly to steps in the ordinary evaluation of $\tm$ and $\Rew{[\ ]}$ steps simply eliminate the explicit substitution when its content is a value. Note how the transformation makes the redex always appear at the end of the environment, so that the need for searching for it---together with the notion of evaluation context---disappears.

Let us also introduce some terminology. Values and applications of values are 
\emph{\crumbledt{}s}.
The transformation, called \emph{\crumbling translation}, turns a \lat into a 
\crumbledt with an environment---such a pair is called a \emph{\crumblep}.

\paragraph{Turning to micro-step evaluation.} The previous example covers what happens when the \crumbling transformation is paired with small-step evaluation.
Abstract machines, however, employ a finer mechanism that we like to call \emph{micro-step} evaluation, where the substitutions due to $\beta$-redexes are delayed and represented as new environment entries, and moreover substitution is decomposed as to act on one variable occurrence at a time.
In particular, such a more parsimonious evaluation never removes environment entries because they might be useful later on---garbage collection is assumed to be an orthogonal and independent process.
To give an idea of how micro steps work, let's focus on the evaluation of the subterm $(\varfour\varthree)\esub\varfour{(\la\vartwo\vartwo\vartwo) I}$ of our example (because micro-step evaluations are long and tedious), that proceeds as follows:
\[\small\begin{array}{llll}
  (\varfour\varthree)\esub\varfour{(\la\vartwo\vartwo\vartwo) I}
 & \tobv &
 (\varfour\varthree)\esub\varfour{\vartwo\vartwo} \esub\vartwo I
 &
 \Rew{[\ ]} 
 \\
 
 (\varfour\varthree)\esub\varfour{\vartwo I} \esub\vartwo I
 & \Rew{[\ ]} &
 (\varfour\varthree)\esub\varfour{I I} \esub\vartwo I
 &
 \tobv 
 \\
 (\varfour\varthree)\esub\varfour{\var } \esub\var I \esub\vartwo I
 & \Rew{[\ ]} &
 (\varfour\varthree)\esub\varfour I \esub\var I \esub\vartwo I
 &
 \Rew{[\ ]} 
 \\
 (I \varthree)\esub\varfour I \esub\var I \esub\vartwo I
\end{array}\]
where $\tobv$ steps now introduce new environment entries.
Now the redex is not always at the end of the environment, but it is always followed on the right by an environment whose entries are all abstractions, so that the search for the next redex becomes a straightforward visit from right to left of the environment---the evaluation context has been coded inside the sequential structure of the environment.

\paragraph{Abstraction bodies and the concatenation of environments.}
There is a last point to explain. We adopt weak evaluation---that only evaluates out of abstractions---but the \crumbling transformation also transforms the bodies of abstractions and the branches of $\xitecompact$ into \crumblep{s}.
Let us see another example.
The \crumbled representation of $\tmtwo \defeq (\la\var((\var\var)(\var\var))) (I I)$ then is
$$\mytr\tmtwo = ((\la\var((\vartwo\varthree) \esub\vartwo {\var\var} \esub\varthree {\var\var})) \varfour) \esub \varfour {I I}$$
Micro-step evaluation goes as follows:
\begin{center}
$\begin{array}{lllllllllllll}
\mytr\tmtwo
 & \tobv &
((\la\var((\vartwo\varthree) \esub\vartwo {\var\var} \esub\varthree {\var\var})) \varfour) \esub \varfour {\varfour'} \esub {\varfour'} I
\\
 & \Rew{[\ ]} &
((\la\var((\vartwo\varthree) \esub\vartwo {\var\var} \esub\varthree {\var\var})) \varfour) \esub \varfour I \esub {\varfour'} I
\\
& \Rew{[\ ]} &
((\la\var((\vartwo\varthree) \esub\vartwo {\var\var} \esub\varthree {\var\var})) I) \esub \varfour I \esub {\varfour'} I.
\end{array}$
\end{center}
At this point, the reduction of the $\beta$-redex (involving $\l\var$) has to combine the \crumblep of the redex itself with the one of the body of the abstraction, by concatenating the environment of the former (here $\esub \varfour I \esub {\varfour'} I$) at the end of the environment of the latter ($\esub\vartwo {\var\var} \esub\varthree {\var\var}$), interposing the entry created by the redex itself ($\esub\var I$), thus producing the new \crumblep:
\[(\vartwo\varthree) \esub\vartwo {\var\var} \esub\varthree {\var\var} \esub\var I \esub \varfour I \esub {\varfour'} I.\]
The key conclusion is that evaluation needs to \emph{concatenate} \crumbled environments, which is an operation that ordinary abstract machines instead never perform.

Note that transforming abstraction bodies may produce nested ES, if the abstraction occurs in an ES. This is the only kind of nesting of ES that is allowed.

\section{The \Crumbling Transformation}
\label{sect:preliminaries}
In this section we formally define the language of \crumbled forms and the \crumbling transformation.

\paragraph*{\Crumbled forms.} \Lat{s} are replaced by \emph{\crumblep{}s}, which are formed by a
\crumbledt and an environment, where in turn 
\begin{itemize}
	\item a 	\emph{\crumbledt} is either a \emph{\crumbled value} (\ie a variable, a boolean, an error, or an abstraction over a \crumblep), an application of \crumbled values, or a $\xitecompact$ on a crumbled value whose alternatives are \crumblep{s}, and 
	\item an \emph{environment} is a finite sequence of explicit substitutions of 
	\crumbledt{s} for variables.
\end{itemize}
Formally, the definition is by mutual induction:
\[\begin{array}{rrcl}
 \textsc{\Crumbledt{}s} & \mol,\moltwo  &\grameq &\molv \mid \molv\molvtwo \mid \xite \molv \cell \celltwo
 \\
  \textsc{\Crumbled values } & \molv, \molvtwo &\grameq &\var \mid \la\var\cell \mid \true \mid \false \mid \err
\\
\textsc{Environments } & \env, \envtwo  &\grameq &\emptyenv \mid \env   \esub\var\mol 
\\
 \textsc{\Crumblep{s} } & \cell,\celltwo &\grameq &(\mol, \env)
\end{array}\]
\begin{itemize}

  \item \emph{Bodies}: the bodies of abstractions and $\xitecompact$ are themselves \crumblep{}s---the forthcoming \crumbling transformation is indeed \emph{strong}, as it also transforms bodies.
  \item \emph{\Crumblep{s} are not closures}: the definition of \crumblep{s} may remind one of \emph{closures} in abstract machines with local environments, but the two concepts are different. 
  The environment $\env$ of a \crumblep $(\mol,\env)$, indeed, does not in general bind all the free variables of the 
  \crumbledt~$\mol$.
\end{itemize}
We freely consider environments as lists extendable on both ends, and whose concatenation is obtained by simple juxtaposition.
Given a \crumblep $(\mol, \env)$ and an environment $\envtwo$ the \emph{appending} of $\envtwo$ to $(\mol, \env)$ is $\append{(\mol,\env)}{\envtwo}  \defeq (\mol, \env \envtwo)$.

\paragraph{Free variables, $\alpha$-renaming, and all that.}
All syntactic expressions are \emph{not} considered up to $\alpha$-equivalence.
Free variables are defined as expected for \crumbledt{}s. For environments and \crumblep{}s they are defined as follows (via the auxiliary notion of \emph{domain} of environments; 
this is 
because global environments are used here):
\[\begin{array}{rcl \colspace rcl}
\domain{\env\esub{\var}{\mol}} &\defeq & \domain{\env} \cup \{\var\}
&
\domain{\emptyenv} &\defeq & \emptyset
\\
\domain{(\mol,\env)} &\defeq & \domain{\env}
&
\fv{\emptyenv} &\defeq & \emptyset
\\
\fv{\env\esub{\var}{\mol}} &\defeq & \multicolumn{4}{l}{(\fv{\env} \smallsetminus \{\var\}) \cup \fv{\mol}}
\\
\fv{(\mol,\env)} &\defeq & \multicolumn{4}{l}{(\fv{\mol} \smallsetminus \domain{\env}) \cup \fv{\env}.}
\end{array}\]

\noindent Let $\env = \esub{\var_1}{\mol_1} \dots \esub{\var_k}{\mol_k}$ be an environment: we denote the lookup of $\var_i$ in $\env$ by $\env(\var_i) \defeq \mol_i$.
We say that a \crumblep $\cell$ or an environment $\env$ are \emph{well-named} if all the variables occurring on the lhs of ES outside abstractions in $\cell$ or $\env$ are pairwise distinct.

\paragraph*{The \crumbling translation.}
A \lat is turned into a \crumblep via the following \emph{\crumbling translation} 
$\mytr{\,\cdot\,}\,$, which uses an auxiliary translation 
$\auxtr{\,\cdot\,}$ from \lav{}s into \crumbled values.
\[\begin{array}{c\colspace\colspace c\colspace\colspace c}
  \auxtr \var  \defeq \var 
&
    \auxtr \true \defeq \true
    &
     \auxtr \false \defeq \false
     \\
      \auxtr \err  \defeq \err
    &
      \auxtr {\la\var\tm}  \defeq \la\var{\mytr\tm}
    \end{array}\]
\[\begin{array}{rcl\colspace\colspace c\colspace\colspace c\colspace\colspace c \colspace\colspace cccccccccc}
 \mytr \val  & \defeq &  (\auxtr\val, \emptyenv) \\
 \mytr {\val\valtwo}  &\defeq& (\auxtr\val \auxtr\valtwo, \emptyenv) \\
  \mytr {\tm\val}  & \defeq  & (\var \auxtr\val, \esub\var\mol\env)  
  & (*) \\
 \mytr {\tmtwo\tm} & \defeq & \append{\mytr{\tmtwo\var}}{ (\esub\var\mol\env)}  
  & (*) \\
  \mytr {\xite \val \tmtwo \tmthree} &\defeq& (\xite {\auxtr\val} {\mytr\tmtwo} {\mytr\tmthree}, \emptyenv) \\
  \mytr {\xite\tm\tmtwo\tmthree}  & \defeq & (\xite\var{\mytr\tmtwo}{\mytr\tmthree}, \esub\var\mol\env)  
  & (*) 
    \end{array}\]

$(*)$ if $\tm$ is not a \lav and $\mytr\tm = (\mol,\env)$, and $\var$ is fresh.

According to the definition, if $\tmtwo$ and $\tm$ are not values,
$\mytr{\tmtwo\tm} = (\vartwo\var, \esub{\vartwo}{\moltwo}\envtwo\esub{\var}{\env})$ with $\mytr{\tm} = (\mol,\env)$, $\mytr{\tmtwo} = (\moltwo, \envtwo)$ and $\var, \vartwo$ fresh.
\begin{example}\label{ex:delta}
	Let $\delta \defeq \la{\var}{\var\var}$ and $I \defeq \la{\var}\var$:
	$\auxtr{I} = \la{\var}\mytr{\var} = \la{\var}{(\var, \emptyenv)}$, $\auxtr{\delta} = \la{\var}{\mytr{\var\var}} = \la{\var}{(\var\var, \emptyenv)}$ (as $\mytr{\var\var} = (\var\var, \emptyenv)$), and $\mytr{\delta\delta} = (\bar{\delta}\bar{\delta},\emptyenv)$. So,
	\[\begin{array}{rclccc}
	\mytr{\delta\delta I} &= &(\varthree \auxtr{I}, \esub{\varthree}{\bar{\delta}\bar{\delta}}) 
	\\
	&=& (\varthree \la{\var}{(\var, \emptyenv)}, \esub{\varthree}{(\la{\var}{(\var\var, \emptyenv)})\la{\var}{(\var\var, \emptyenv)}})
	\\[.2cm]
	\mytr{\delta\delta (\var\var)} &= &(\varthree \varfour, \esub{\varthree}{\bar{\delta}\bar{\delta}} \esub{\varfour}{\var\var}) \\
	&=& (\varthree \varfour, \esub{\varthree}{(\la{\var}{(\var\var, \emptyenv)})\la{\var}{(\var\var, \emptyenv)}}\esub{\varfour}{\var\var}).
	\end{array}\]
\end{example}


The \crumbling translation $\mytr{\,\cdot\,}$ is not surjective: the \crumblep $\cell \defeq (\var\var,\esub{\var}{\vartwo})$ is such that $\mytr{\tm} \neq \cell$ for any \lat $\tm$.

\paragraph*{Read back.} There is a left inverse for the \crumbling translation, called \emph{read-back} and defined by:
\[\begin{array}{rcl\colspace \colspace rcl}
	\rb{\var} &\defeq&  \var 
	&
	\rbp{\la{\var}{\cell}}  &\defeq& \la{\var}{\rb{\cell}}
	\\
	\rb{\true} &\defeq&  \true
	&
	\rb{\false} &\defeq&  \false
	\\
	\rb{\err} &\defeq&  \err
	&
	\rbp{\molv\molvtwo}  &\defeq&  \rb{\molv}\rb{\molvtwo}
	\\
	\rbp{\xite\molv\cell\celltwo}  & \defeq  &\multicolumn{4}{l}{\xite{\rb{\molv}}{\rb{\cell}}{\rb{\celltwo}}}
	\\
	\rb{(\mol,\emptyenv)} & \defeq &
	\rb\mol
	\\
	\rb{(\mol,\env{\esub\var\moltwo})}  & \defeq &
	\multicolumn{4}{l}{\rb{(\mol,\env)} \isub{\var}{\rb{\moltwo}} }
\end{array}\]

\begin{proposition}[Read-back and the \crumbling translation]
\label{prop:read-back-inverse-transl}
\Copy{prop:read-back-inverse-transl}{
For every \lat $\tm$ and every \lav $\val$, one has $\rb{\mytr{\tm}} = \tm$ and $\rb{\auxtr{\val}} = \val$.}
\end{proposition}

\begin{remark}[\Crumbling translation, free variables]
	\label{rmk:preservation-fvs}
\qquad
	\begin{enumerate}
		\item For any \lat $\tm$ and any value $\val$, one has $\fv{\tm} = \fv{\mytr{\tm}}$ and $\fv{\val} = \fv{\auxtr{\val}}$; in particular, $\tm$ is closed if and only if $\mytr{\tm}$ is so.

		\item For any 
		\crumbledt $\mol$ and any \crumblep $\cell$,  $\fv{\rb{\mol}} = \fv{\mol}$ and $\fv{\rb{\cell}} = \fv{\cell}$.
		\item \label{p:preservation-fvs-commutation-fvs}
  The \crumbling translation commutes with the renaming of free variables.

      \item \label{p:preservation-fvs-val-to-val}
  The \crumbling translation and the read-back map values to values.
	\end{enumerate}
\end{remark}

\paragraph*{\Crumbled contexts.}
For \crumbled forms, we need \emph{contexts} both for environments~and \crumblep{s}:
\begin{align*}
 \textsc{Environment contexts} \ \ \ \ & \ectx \defeq \env   \esub\var\ctxhole \\
 \textsc{\Crumble contexts}\ \ \ \ & \cctx \defeq \ctxhole \mid (\mol, \ectx)
\end{align*}
\Crumblep{s} can be plugged into both notions of contexts.
Let us point out that the following definition of plugging is slightly unusual as it does a little bit more than just replacing the hole, because simply replacing would not provide a well-formed syntactic object: plugging indeed extracts the environment from the plugged \crumblep and concatenates it with the environment of the context.
Such an unusual operation---that may seem ad-hoc---is actually one of the key technical points in order to obtain a clean proof of the implementation theorem.

    \begin{definition}[Plugging in \crumbled contexts]
  Let $\ectx=\env\esub\var\ctxhole$ be an environment context, $\cctx$ be a \crumble context, and $\cell= (\moltwo,\envtwo)$ be a \crumblep.
  The \emph{plugging} $\ectxp{\cell}$ of $\cell$ in $\ectx$ and the \emph{plugging} $\cctxp{\cell}$ of $\cell$ in $\cctx$ are defined by
  \begin{align*}
      (\env\esub\var\ctxhole) \ctxholep {(\moltwo,\envtwo)} \defeq  \env
      \esub\var\moltwo  \envtwo
      \qquad      \\
      \ctxhole \ctxholep {\cell} \defeq \cell \quad \quad
      (\mol,\ectx) \ctxholep {\cell} \defeq (\mol,\ectxp{\cell})
  \end{align*}
\end{definition}

\begin{example}
	\label{ex:plugging}
	In \refex{delta} we have seen that \sloppy $\mytr{\delta\delta I} = (\varthree \la{\var}{\var_\emptyenv}, \esub{\varthree}{(\la{\var}{(\var\var)_ \emptyenv})\la{\var}{(\var\var)_\emptyenv}})$, where we set $\mol_\emptyenv \defeq (\mol, \emptyenv)$ for any 
	\crumbledt $\mol$.
	We have that $\mytr{\delta\delta I} =  \cctxp{\cell}$ with $\cctx \defeq (\varthree \la{\var}{\var_\emptyenv}, \esub{\varthree}{\ctxhole})$ and  $\cell \defeq ((\la{\var}{(\var\var)_ \emptyenv})\la{\var}{(\var\var)_\emptyenv}, \emptyenv)$
\end{example}

\noindent The notions of well-named, $\fv{\cdot}$, and $\domain{\cdot}$ 
can be naturally extended to \crumble contexts.
The definition of read back is extended to \crumble contexts by setting $\rb{\ctxhole} \defeq \ctxhole$ and $\rb{(\mol,\env{\esub\var\ctxhole})}  \defeq
	\rb{(\mol,\env)} \isub{\var}{\ctxhole}$.
Note however that the unfolding of a \crumble context is not necessarily a \lcontext, because the hole can be duplicated or erased by the unfolding. For instance, let $\cctx \defeq (\var\,\var, \esub\var\ctxhole) $.
  Then $\rb\cctx = \ctxhole\ctxhole$ is not a \lcontext.

\reflemma{basecases-both} provides the properties of the translation 
needed to prove the invariants of machines in the next sections.

\begin{lemma}[Properties of \crumbling]
\label{l:basecases-both} 
 	\Copy{l:basecases-both}{For every \lat{} $\tm$:
  \begin{enumerate}
  	\item\mylabel{p:basecases-both-freshness}
     \emph{Freshness:} $\mytr\tm$ is well-named.
    \item\mylabel{p:basecases-both-closure}
     \emph{Closure:} if $\tm$ is closed, then $\fv{\mytr\tm}=\emptyset$.
    \item\mylabel{p:basecases-both-disjointness}
     \emph{Disjointedness}: $\domain\cctx \cap \fv\mol = \emptyset$ if $\mytr{\tm} = \cctxp{(\mol,\env)}$.
    \item\mylabel{p:basecases-both-abstractions}
     \emph{Bodies:} every body in $\mytr\tm$ is the translation of a \lat{}.
     \item\mylabel{p:basecases-both-unfold-right}
     \emph{Contextual decoding:} if $\mytr\tm = \cctxp\cell$, then $\rb\cctx$ is a right \valuectx.
  \end{enumerate}
	}
\end{lemma}

\section{The Closed Case}
\label{sect:closed}


Here we show how to evaluate \crumbled forms with a micro-step operational semantics. 
We builds over the work of Accattoli and co-authors, who employ the following terminology:
\begin{itemize}
  \item \emph{Calculus}: for a small-step semantics where both substitution and search for the redex are meta-level operations;
  \item \emph{Linear calculus}: for a micro-step semantics where substitution is decomposed---the calculus has ES and possibly a notion of environment if the ES are grouped together---but the search for the redex is still meta-level and expressed via evaluation contexts;
  \item \emph{Abstract machine}: for a micro-step semantics where both substitution and search for the redex are decomposed. The search for redexes is handled via one or more stacks called \emph{applicative stack}, \emph{dump}, \emph{frame}, and so on; the management of names is also explicit, i.e. \emph{not} up-to $\alpha$-equivalence.
\end{itemize}
The \crumbling transformation blurs the distinction between a linear calculus and an abstract machine because it allows using the sequential structure of the environment as the only stack needed to search for redexes. 

The operational semantics for \crumbled forms that we present next is in the style of a linear calculus, because spelling out the straightforward search for redexes is not really informative. 
Nonetheless, we do call it an \emph{abstract machine}, because of the blurred distinction in the \crumble case and because we manage names explicitly. 
In \refapp{towards-impl-closed-pre} we spell out the actual abstract machine.

\subsection{The \aglam}
\label{subsect:micro-step-closed} 

\paragraph*{Transitions.}
To introduce the \aglam (\glam stands for Global Leroy Abstract Machine) we need some definitions. 
First, environments and \crumblep{s} made out of \pvalue{}s only are defined and noted as follows:
\[\begin{array}{lrcll}
  \textsc{\lenvironment{}s} & \aenv & \grameq &\epsilon \mid \aenv  \esub\var{\pval} \qquad 
  \\
    \textsc{\lcell{}s}  & \acell & \grameq & (\pval, \aenv) 
\end{array}\]
Essentially, a \lenvironment{} stands for the already evaluated coda of the environment described in the paragraph about micro-steps in \refsect{intro-abn-env}, while \lcell{}s are fully evaluated \crumblep{s} (\ie final states of the machine), as we show below. 

Second, given a \crumblep $\cell$ we use $\rename{\cell}$ for a \crumblep obtained by $\alpha$-renaming the names in the domain of $\cell$ with fresh ones so that $\rename{\cell}$ is well-named.

The transitions act on \crumblep{s} whose environments are \lenvironment{}s. The top level transitions are:
\begin{align*}
  ((\la\var\cell)\,\molv, \aenv)
  &  \rtom 
  \append{(\append{\cell}{\esub\var\molv})^\alpha}{\aenv}
  && 
  \\%
  \\[-.7\baselineskip]
  (\xite\true\cell\celltwo, \aenv)
		    & \rtoift 
		     \append\cell\aenv\\
		    (\xite\false\cell\celltwo, \aenv)
		    & \rtoiff 
		     \append\celltwo\aenv
		     \\
		    (\xite\molv \cell \celltwo, \aenv)
		    & \rtoife 
		    (\err, \aenv) & (1)
		    \\
		    (\molv \molvtwo, \aenv)
		    & \rtoape 
		    (\err, \aenv)& (2)
		    \\
		      \\[-.7\baselineskip]
  (\var,\aenv)
  & \rtoev 
  (\aenv(\var),\aenv)
  & (3)
  \\
  (\var \molv,\aenv)
  & \rtoel 
  (\aenv(\var) \, \molv, \aenv)
  & (3)
  \\
    (\xite\var\cell\celltwo,\aenv)
  & \rtoeite 
  (\xite{\aenv(\var)}\cell\celltwo, \aenv)
  & (3)
\end{align*}
\begin{enumerate}
  \item if $\molv=\la\var\cellthree$ or $\molv = \err$
  \item if $\molv\in \set{\true,\false,\err}$
  \item if $\var \in \domain{\aenv}$
\end{enumerate}
Transitions are then closed by \crumble contexts: for every $a\in \set{\msym, \iftsym, \iffsym,\ifesym,\apesym,\evsym, \elsym, \eitesym}$ define $\cctxp\cell \Rew{a} \cctxp\celltwo$ if $\cell \rootRew{a}\celltwo$. The transition relation $\tocrumble$ of the \aglam is defined as the union of all these rules. 
Let us explain each transition:
\begin{itemize}
  \item \emph{$\tom$}: (forget about the $\alpha$-renaming for the moment---see the next paragraph) the rule removes a $\beta$-redex and introduces an ES $\esub\var\molv$ instead of performing the meta-level substitution. Moreover, the environment of the body $\cell$ of the abstraction and the external environment $\aenv$ are concatenated (via the appending operation $\append{}$) interposing $\esub\var\molv$.
  
  \item \emph{Conditional and error transitions $\toift,\toiff, \toife, \toape$}: these transitions simply mimics the analogous rules on the Pif calculus, with no surprises.

  \item \emph{Substitution transitions $\toel, \toevar,\toeite$}: 
  the variable $\var$ is 
  substituted by the corresponding \crumbled value in the environment $\aenv$, if any. 
  In the closed case, a forthcoming invariant guarantees that $\aenv(\var)$ is always defined so that side-condition (3) is actually always satisfied. There are no rules to substitute on the right of an application, we explain this below.
\end{itemize}

Note that, according to the definitions of plugging and top level transitions, the transition relation follows \emph{right-to-left} evaluation, since the environment on the right of a redex is a \lenvironment, \ie it is made of practical values only, which means that it has already been evaluated (see the harmony property for \aglam in \refprop{harmony-closed} below).
Adopting right-to-left evaluation implies that the \aglam does not need a rule $\toer$ symmetrical to $\toel$, whose top level shape would be $(\molv\var,\aenv) \rootRew{\esym_r} 
(\molv\,\aenv(\var), \aenv)$ with $\var \in \domain{\aenv}$: indeed, if $\molv$ is a variable then $\toel$ applies to the same redex $(\molv\var,\aenv)$, otherwise $\molv$ is an abstraction and then $\tom$ applies to $(\molv\var,\aenv)$.

\paragraph{The cost and the place of $\alpha$-renaming.} Abstract machines with global environments have to $\alpha$-rename at some point, this is standard\footnote{Local environments do allow to avoid renamings, but the simplification is an illusion, as the price is payed elsewhere---see \citet{DBLP:conf/ppdp/AccattoliB17}---there is no real way out.}. 
In our implementation, renaming is implemented as a copy function. And the cost of renaming is under control because of  forthcoming invariants of the machine. 
This is all standard \cite{DBLP:conf/ppdp/AccattoliB17}. 
Often the burden of renaming/copying is put on the substitution rules. It is less standard to put it on the $\betav$-transition, as we do here, but nothing changes. 
Last, a technical remark: in rule $\tom$ the $\alpha$-renaming at top level has to pick names that are fresh also with respect to the \crumble context enclosing it. 
This point may seem odd but it is necessary to avoid name clashes, and it is trivially obtained in our concrete implementation, where variable names are memory locations and picking a fresh name amounts to allocating a new location, that is of course new globally.

\begin{definition}[Reachable \crumblep]
A \crumblep is \emph{reachable} (by the \aglam) if it is obtained by a sequence of transitions starting from the translation $\mytr{\tm}$ of a closed \lat $\tm$. 
\end{definition}

{\paragraph*{Unchaining abstractions.} 
The substitution performed by the rule $\toev$ may seem an unneeded optimization; quite the opposite, it fixes an issue causing quadratic overhead in the machine. 
The culprits are malicious \emph{chains of renamings}, \ie{} environments of the form $\esub{\var_1}{\var_2}\esub{\var_2}{\var_3}\cdots\esub{\var_n}{\la\vartwo\cell}$ substituting variables
for variables and finally leading to an abstraction. 
\citet{fireballs} showed that the key to linear overhead is to perform substitution steps while going through the chain from right to left.

\begin{example}
	\label{ex:evaluation-closed}
	Consider the \crumblep{} $\mytr{\delta\delta} = (\auxtr\delta\,\auxtr\delta,\emptyenv)$, where $\auxtr\delta=\la\var{(\var\var,\emptyenv)}$; then:
  \[\begin{array}{rcllllll}
    \mytr{\delta\delta} & \tom & (\var\var, \esub{\var}{\auxtr\delta}) & \toel & (\auxtr\delta\,\var, \esub{\var}{\auxtr\delta}) 
  	\\
  	& \tom & (\vartwo\vartwo, \esub{\vartwo}{\var}\esub{\var}{\auxtr\delta}) \\
    & \toev & (\vartwo\vartwo, \esub{\vartwo}{\auxtr\delta}\esub{\var}{\auxtr\delta}) & \toel \ldots
  \end{array} \]
		
	In \refex{delta} we introduced the \crumblep $\mytr{\delta\delta I} = (\varthree \,\auxtr I, \esub{\varthree}{\auxtr\delta\,\auxtr\delta})$ where $\auxtr I = (\la\var{(\var,\emptyenv)})$; 
	in accordance with the \crumblep decomposition shown in \refex{plugging}, we have:
	\[\begin{array} {clcccc}
		\mytr{\delta\delta I} & \tom (\varthree \, \auxtr I, \esub{\varthree}{\var\var} \esub{\var}{\auxtr\delta})
		\toel (\varthree \, \auxtr I, \esub{\varthree}{\auxtr\delta\,\var} \esub{\var}{\auxtr\delta}) \\
    & \tom (\varthree\,\auxtr I, \esub\varthree{\vartwo\vartwo} \esub{\vartwo}{\var}\esub{\var}{\auxtr\delta}) 
    \\
    &\toel (\varthree\,\auxtr I, \esub\varthree{\vartwo\vartwo} \esub{\vartwo}{\auxtr\delta}\esub{\var}{\auxtr\delta}) \toel \ldots
  \end{array}\]
  Consider now the \emph{open} \crumblep 
  $$\cell\defeq\mytr{\delta\delta (\var\var)} = (\varthree\varfour, \esub{\varthree}{\auxtr\delta\,\auxtr\delta}\esub\varfour{\var\var}).$$
  The \crumblep $\cell$ is normal because its only possible decomposition of the form $\cctxp{(\mol,\aenv)}$ is for $\aenv=\emptyenv$ (as $\var\var$ is not a \pvalue), and no transitions apply to the rightmost entry $\esub\varfour{\var\var}$ since $\var$ is free.
\end{example}

 The \aglam satisfies a harmony property.
}

\begin{proposition}[Harmony for the \aglam]
  \label{prop:harmony-closed}
  \Copy{prop:harmony-closed}%
  {A closed \crumblep $\cell$ is normal if and only if it is a \lcell.}
\end{proposition}

\subsection{The Implementation Theorem}
\label{subsect:the-implementation-theorem}
To show that the \aglam correctly implements the Pif calculus, we apply an abstract approach introduced by \citet{AccattoliGuerrieri17b}, which we reuse as well in the following sections for other \crumble abstract machines and other evaluation strategies of the $\lambda$-calculus. 

\paragraph{The implementation theorem, abstractly.} 
In  \citet{AccattoliGuerrieri17b} it is proven that, given 
\begin{itemize}
 \item a generic abstract machine $\mach$, which is a \emph{transitions relation} $\tomachine$ over a set of \emph{states} 
 that splits into 
 \begin{itemize}
 \item \emph{principal transitions} $\tomachp$, that corresponds to the evaluation steps on the calculus, and 
 \item \emph{overhead transitions} $\tomacho$, that are specific of the machine,
 \end{itemize}
 \item an evaluation strategy $\to$ in the $\lambda$-calculus, and 
 \item a decoding $\rbp{\cdot}$ of states of $\mach$ into \lat{s},
\end{itemize}
 $\mach$ correctly implements $\tostrat$ via $\rbp{\cdot}$ whenever $(\mach, \tostrat, \rbp{\cdot})$ forms an \emph{implementation system}, \ie whenever the following conditions are fulfilled (where $\state$ and $\statetwo$ stand for generic states of $\mach$):
\begin{enumerate}
	\item \emph{Initialization:} there is an encoding $\mytr{\cdot}$ of \lat{s} such that $\rb{\mytr{\tm}} = \tm$;
	\item\label{p:def-beta-projection} \emph{Principal projection}: $\state \tomachp \statetwo$ implies $\rb\state \tostrat \rb{\statetwo}$;
	\item\label{p:def-overhead-transparency} \emph{Overhead transparency}: $\state \tomacho \statetwo$ implies $\rb{\state} = \rb{\statetwo}$;	
	\item \emph{Determinism}: 
	$\tomachine$ is deterministic;
	\item \emph{Halt}: $\mach$ final states (to which no transition applies) decode to $\tostrat$-normal \lat{s};
		\item 	\emph{Overhead Termination}:  $\tomacho$ terminates.
\end{enumerate}
 
Our notion of implementation, tuned towards complexity analyses, requires a perfect match between the number of steps of the strategy and the number of principal transitions of the execution.

\begin{theorem}[Machine Implementation, \cite{AccattoliGuerrieri17b}]
	\label{thm:implem}
	If a machine $\mach$, a strategy $\tostrat$ on $\lambda$-terms and a decoding $\rb\cdot$ form an implementation system then:
	\begin{enumerate}
		\item\label{p:implem-exec-to-deriv} \emph{Executions to derivations}: for any $\mach$-execution $\exec \colon \mytr\tm \tomachine^* \state$ there is a $\tostrat$-derivation $\deriv \colon \tm \tostrat^* \rb\state$.
		
		\item\label{p:implem-deriv-to-exec} \emph{Derivations to executions}: for every $\tostrat$-derivation $\deriv \colon \tm \tostrat^* \tmtwo$ there is an $\mach$-execution $\exec \colon \mytr\tm \tomachine^* \state$ such that $\rb\state = \tmtwo$.
		
		\item\label{p:implem-beta-matching} \emph{Principal matching}: in both previous points the number $\sizepr\exec$ of principal transitions in $\exec$ is exactly the length $\size\deriv$ of the derivation $\deriv$, \ie $\size\deriv = \sizepr\exec$.
	\end{enumerate}
\end{theorem}


\paragraph{The crumbling implementation system.} The states of the \aglam are \crumblep{}s. Its principal transitions are those labeled with $\set{\betav,\iftsym,\iffsym,\ifesym,\apesym}$, while the overhead transitions are those labeled with $\set{\evsym, \elsym, \eitesym}$.
We can now show that the \aglam, Pif evaluation $\topif$ and the read-back $\rbp{\cdot}$ form an implementation system, that is, that the \aglam implements the Pif calculus.

We are going to provide five of the six sufficient conditions required by the implementation theorem (\refthm{implem});
the sixth one, the termination of overhead transitions, is subsumed by the finer complexity analysis in \refsubsect{complexity-closed}.

The sufficient conditions, as usual, are proved by means of a few invariants of the machine, given by \reflemma{invariants-closed} below. These invariants are essentially the properties of the translation in \reflemma{basecases-both} extended to all reachable \crumblep{s}.  
One of them---namely \emph{contextual decoding}---however, is weaker because reachable \crumblep{s} do not necessarily have the same nice structure as the initial \crumblep{s} obtained by translation of a \lat{}, as the next remark explains.

\begin{remark}
	Even though not all \crumble contexts unfold to \lcontext{s},
	\crumble contexts obtained by decomposing 
	\crumblep{s} given by the translation of \lat{s} do (\reflemmap{basecases-both}{unfold-right})---this is the contextual decoding property. 
	Unfortunately, it is not preserved by evaluation. 
	Consider the \crumblep $\cell \defeq \mytr{(\la\var \var(\var\var))\,I} = ((\la\var (\var\vartwo, \esub\vartwo{\var\var})) \auxtr{I},\emptyenv)$ with $\auxtr{I} = \la{\varthree}{(\varthree,\emptyenv)}$.
	Clearly, $\cell = \ctxholep{\mytr{(\la\var \var(\var\var))\,I}}$ where $\rb{\ctxhole} = \ctxhole$ is a \lcontext.
	After one $\msym$ step, the \crumblep $\cell$ reaches \sloppy $(\var\vartwo, \esub\vartwo{\var\var} \esub\var {\auxtr{I}}) = \cctxp{(\auxtr{I}, \emptyenv)}$ for $\cctx \defeq (\var\vartwo, \esub\vartwo{\var\var} \esub\var\ctxhole)$.
	But $\cctx$ unfolds to $\rb{\cctx} = \ctxhole(\ctxhole\ctxhole)$, which is not a $\lambda$-context.
\end{remark}


\begin{lemma}[Invariants for the \aglam]
	\label{l:invariants-closed} %
	\Copy{l:invariants-closed}{
		For every reachable \crumblep $\cell$ in the \aglam:
		\begin{enumerate}
			\item\mylabel{p:invariants-closed-fresh}
			\emph{Freshness:} $\cell$ is well-named.
			\item\mylabel{p:invariants-closed-fvs}
			\emph{Closure:} $\fv\cell=\emptyset$.
			\item\mylabel{p:invariants-closed-abstractions}
			\emph{Bodies:} every body occurring in $\cell$ is a subterm (up to renaming) of the initial \crumblep.
			\item\mylabel{p:invariants-closed-ctx-decoding}
			\emph{Weak contextual decoding:}
			for every decomposition $\cctxp{(\mol, \aenv)}$ where $\mol$ is not a \crumbled value, if $\cctxthree$ is a prefix of $\cctx$ then $\rb{\cctxthree}$ is a right \valuectx.
			
		\end{enumerate}
	}
\end{lemma}

Freshness and closure are invariants needed to ensure the basic functioning of the machine. 
The bodies invariant corresponds to what is sometimes called \emph{subterm invariant}. 
It is the key invariant for complexity analyses, as it allows to bound the size of duplicated subterms (that are always abstractions) using the size of the initial term. 
Usually, it is only needed for complexity analyses, while here it is needed for the implementation theorem as well (namely, only for the proof of the weak contextual decoding invariant). 
The weak contextual decoding invariant, finally, is essential to show that principal transitions of the \aglam project on evaluation steps in $\plotcalcif$.

\begin{theorem}[Implementation]
  \label{thm:implementation-closed}
\Copy{thm:implementation-closed}{
  Let $\cell$ be a \crumblep reachable by the \aglam.
  \begin{enumerate}
    \item \emph{Initialization: }
     $\rb{\mytr{\tm}} = \tm$ for every \lat{} $\tm$.
    \item\mylabel{p:implementation-closed-projection} \emph{Principal projection: }
      if $\cell \Rew{a} \celltwo $ then $\rb\cell \Rew{a} \rb\celltwo$, for any rule $a\in\set{\betav,\iftsym,\iffsym,\ifesym,\apesym}$.
    \item \emph{Overhead transparency: }
      if $ \cell \Rew{a} \celltwo $ then $\rb\cell = \rb\celltwo$
			for any rule $a \in \set{\evsym,\elsym, \eitesym}$.
    \item \emph{Determinism: } the transition 
    $\tocrumble$ is deterministic.
    \item \emph{Halt: } if $\cell$ is $\crumblesym$-normal then $\rb\cell$ is $\pif$-normal.
    \item \emph{Overhead termination: } $\Rew{a}$ terminates, for any rule $a \in \{\evsym, \elsym, \eitesym\}$.
  \end{enumerate}
	
	Therefore, the \aglam, Pif evaluation $\topif$, and the read-back $\rbp{\cdot}$ form an implementation system.
}
\end{theorem}

\subsection{Complexity for the closed case}
\label{subsect:complexity-closed}
To estimate the cost of the \aglam, we provide first an upper bound on the number of overhead transitions---namely the substitution ones $\evsym$,  $\elsym$, and $\eitesym$---in an execution $\exec$ as a function of the number $\sizepr\exec$ of principal transitions. 
Then we discuss the cost of implementing single transitions. Last, by composing the two analyses we obtain the total cost, that is linear in the number of principal transitions and in the size of the initial term/crumble, that is, the machine is bilinear.

\paragraph{Number of transitions: non-renaming substitutions}
Let $\exec:\cell_0 \tocrumble^* \cell$ be an execution (\ie a sequence of transitions) in the \aglam and let $\sizepr\exec, \no\evsym\exec, \no\elsym\exec, \no\eitesym\exec$ be the number of principal, $\evsym$, $\elsym$, and $\eitesym$ transitions in $\exec$, respectively. 
Clearly, a $\elsym$ transition can only be immediately followed by a $\msym$ or a $\apesym$ transition (since $\tocrumble$ is deterministic), and so $\no\elsym\exec \leq \no\msym\exec + \no\apesym\exec + 1$. Similarly, a $\eitesym$ transition is immediately followed by a $\iftsym$, a $\iffsym$ or a $\ifesym$ transition. 
Therefore, $\no\elsym\exec + \no\eitesym\exec \leq \sizepr\exec + 1$.

\paragraph{Number of transitions: renaming steps} The analysis of $\no\evsym\exec$ steps is subtler. A \emph{variable \crumblep} is a \crumblep of the form $(\var,\env)$. The number of $\evsym$ transitions is bounded by the number of variable crumbles out of bodies appearing in evaluation position along an execution $\exec \colon \cell_0 \to^* \cell$. These can be due to the following reasons:
\begin{enumerate}
	\item \emph{Static}: variable crumbles out of bodies in the initial state $\cell_0$;
	\item \emph{Dynamic}: variable crumbles obtained dynamically. In turn, these are divided into (see also the discussion after \refprop{mytr_constants_new}):
	\begin{enumerate}
		\item \emph{Copy}: variable crumbles occurring in the bodies of abstractions and $\xitecompact$ (and thus frozen) that become active because the construct is evaluated and the body exposed;
		\item \emph{Creation}: variable crumbles that cannot be traced back to variable crumbles appearing in prefixes of the execution.
	\end{enumerate}
\end{enumerate}

We now show that the crumbling translation does not produce any variable crumbles out of bodies, but one, if the original term is itself a variable. Therefore, the contribution of point 1 is at most 1. 
We need a measure, counting variable crumbles out of bodies. Note that a variable crumble $(\var,\env)$ appearing in a crumble context $\cctx$ rather takes the form $\esub\vartwo\var\env$, which is why the following measure counts the substitutions containing only a variable.
\[\begin{array}{rcl\colspace \colspace rclcccc}
\multicolumn{6}{c}{\lenv{\mol} \defeq 0 \ \ \ \ \mbox{if $\mol \neq \var$}} \\
\lenv\var & \defeq &1 &
\lenv{(\mol,\env)} & \defeq &\lenv\mol + \lenv\env &\\
\lenv\emptyenv & \defeq &0 &
\lenv{\env\esub\var\mol} & \defeq &\lenv\env + \lenv\mol \\
\end{array}\]

\begin{proposition}\label{prop:mytr_constants_new}
 \Copy{prop:mytr_constants_new}{
  Let $\tm$ be a \lat and $\val$ a \lav. Then:
  \begin{enumerate}
  	\item $\lenv{\mytr\tm}\leq 1$; and $\lenv{\mytr\tm}=1$ if and only if $\tm$ is a variable;
	  	\item $\lenv{\auxtr\val}\leq 1$; and $\lenv{\auxtr\val}=1$ if and only if $\val$ is a variable.
  \end{enumerate}
  }
\end{proposition}

Let us now discuss the variable crumbles of point 2.a (dynamic copy). By the bodies invariant (\reflemmap{invariants-closed}{abstractions}), these pairs appear in a body of the initial crumble. By the bodies property of the crumbling translation (\reflemmap{basecases-both}{abstractions}), all these bodies are the translation of a \lat, and---by using \refprop{mytr_constants_new} again---we obtain that each such body contributes at most with one variable crumble. Since each body is exposed by one $\tom$ or $\toift$ or $\toiff$ transition, we have that the variable crumbles of point 2.a are bounded by $\sizepr\exec$.

Last, we bound the number of variable crumbles at point 2.b (dynamic creation). There is only one rule that can create a new variable crumble (and exactly one), namely $\tom$ when the argument of the $\beta$-redex is a variable. For instance,
$$((\la\var(\var\var,\emptyenv)) \vartwo, \esub\vartwo{\la\varthree\varthree}) \tom (\var\var,\esub\var\vartwo\esub\vartwo{\la\varthree\varthree})$$
where the created variable crumble is $(\vartwo, \esub\vartwo{\la\varthree\varthree})$.
Then the number of variable crumbles at point 2.b is bounded by the number of $\tom$ transitions, itself bounded by $\sizepr\exec$.

The following lemma sums up the previous discusssions.

\begin{lemma}\label{l:linear-bounds}
Let $\exec:\cell_0 \tocrumble^* \cell$ be a \aglam execution. 
\begin{enumerate}
	\item \emph{Linear number of non-renamings substitutions}: $\no\elsym\exec + \no\eitesym\exec \leq \sizepr\exec + 1$.
	\item \emph{Linear number of renamings}: $\no\evsym\exec \leq 2\sizepr\exec +1$.
	\item \emph{Linear number of substitutions}: $\no\elsym\exec + \no\evsym\exec + \no\eitesym\exec \leq 3\sizepr\exec + 2$.
	\end{enumerate}
\end{lemma}

\paragraph{Cost of single transitions} Performing a single transition $\to$ in the \aglam
consists of four operations:
\begin{enumerate}
\item \emph{Search}: locating the next redex;
\item \emph{Unplugging}: splitting the \crumblep to be reduced into a \crumble context $\cctx$ and the \crumblep $\cell$ that is the redex at top level;
\item \emph{Rewriting}: applying a rewriting rule to the \crumblep $\cell$, obtaining a new \crumblep $\celltwo$;
\item \emph{Plugging}: putting the new \crumblep back into the \crumble context obtaining $\cctxp\celltwo$.
\end{enumerate}

The search for redexes is embedded into the definition of the rules, via the contextual closure. The technical definition of plugging and unplugging of \crumblep{s} into a \crumble context is quite involved and, if implemented literally, is not constant time. 

To ease the reasoning, in this section we assume that search and (un)plugging have negligible cost and show that the total cost of rewriting is bilinear. 
In \refapp{towards-impl-closed-pre} we introduce a slight variant of the \aglam, the \apglam, that adds a transition for searching redexes and removes the need for plugging and unplugging. 
A further analysis of the \apglam shows that the total cost of search and (un)plugging is bilinear and thus negligible, justifying the results of this section.

\paragraph{Cost of single transitions: $\betav$ transitions}
We denote by $\len\tm$, $\len\cell$, $\len{\env}$ and $\len{\mol}$  the size of \lat{s}, \crumblep{s}, environments and \crumbledt{}s, respectively, defined as follows:
\begin{center}
$\begin{array}{rcl\colspace rcl\colspace rcl}
	\size{\var} = \size{\true} = \size{\false} = \size{\err} &\defeq & 1 
	\\
	\size{\la{\var}{\tm}} & \defeq & \size{\tm}+1 
	\\
	\size{\tm\tmtwo} & \defeq & \size{\tm} + \size{\tmtwo} + 1  
	\\
	\size{\xite{\tm}{\tmtwo}{\tmthree}} & \defeq & \multicolumn{5}{l}{\size{\tm} + \size{\tmtwo} + \size{\tmthree} + 1}
	\\
	\size{\emptyenv} & \defeq  & 0
	\\
	\size{\env\esub{\var}{\mol}} &\defeq & \size{\env} + \size{\mol}  	
	\\
	\size{(\mol,\env)} & \defeq & \size{\mol} + \size{\env}
\end{array}$
\end{center}
The cost of each $\msym$ transition (that needs to perform a copy of the \crumblep in the abstraction in order $\alpha$-rename it) is bound by the size of the copied \crumblep. 
By the bodies invariant (\reflemmap{invariants-closed}{abstractions}) the abstraction is the $\alpha$-renaming of one the abstractions already present in the initial \crumblep. Therefore the cost of a $\msym$ transition is bound by the size of the initial \crumblep. The next lemma shows that the size of the initial crumble is linear in the size of the initial term translating to the crumble. Therefore, the cost of a $\betav$ transition is linear by the size of the initial term.

\begin{lemma}[Size of translated terms]\label{l:size-mytr}
 \Copy{l:size-mytr}{
  Let $\tm$ be a \lat and $\val$ a \lav. Then $\size{\mytr\tm}\leq 5\size\tm$ and $\size{\auxtr\val}\leq 5\size\val$.
  }
\end{lemma}

\paragraph{Cost of single transitions: substitutions}
The cost of $\elsym$, $\evsym$, and $\eitesym$ transitions depends on the choice of data structures for implementing the machine. Following the literature on global environment machines~\cite{DBLP:conf/ppdp/AccattoliB17}, we assume the global environment to be implemented as a store and variable occurrences to be implemented as pointers into the store, so that lookup in the environment can be performed in constant time on a Random Access Machine (RAM). 
As for the cost of actually performing the replacement of $\var$ with $\aenv(\var)$ in the $\evsym,\elsym$ and $\eitesym$ rules, it can be done in constant time by copying the pointer to $\aenv(\var)$. 
This is possible because the actual copy, corresponding to $\alpha$-renaming, is done in the $\msym$ step. 
Thus, single substitution transitions have constant cost.

\paragraph{Cost of single transitions: conditionals and errors} It is immediate that---if one excludes plugging and unplugging---these transitions have constant cost.

\paragraph{Cost of executions.} Summing up all the analyses in this section we obtain the following theorem.
   
\begin{theorem}[The \aglam is bilinear up to search and (un)plugging]\label{thm:compl-closed}
 \Copy{thm:compl-closed}{
 For any closed \lat{} $\tm$ and any \aglam execution $\exec\colon\mytr\tm\tocrumble^*\cell$, the cost of implementing $\exec$ on a RAM is $O((\sizepr\exec + 1) \cdot \len{\tm})$ plus the cost of plugging and unplugging.}
\end{theorem}

\paragraph*{OCaml implementation.} In \Cref{app:towards-impl-closed-pre} we introduce the \apglam, a refinement of the \aglam making explicit the search for redexes and removing the need for (un)plugging, and having the \emph{same complexity}: the cost for searching redexes and (un)plugging is negligible.
Moreover, an implementation in OCaml of the \apglam can be found in \refapp{ocamlimpl}, together with the code that implements the \crumbling translation. 
The Appendix also discusses in detail a parsimonious choice of data structures for the implementation of \penvironment{}s.

\section{The Open Case}
\label{sect:open}

\subsection{The Fireball Calculus}
\label{subsect:fireball}

\begin{figure}[t]
  \centering
  \scalebox{0.85}{
\begin{tabular}{c} 
   $
    \begin{array}{r@{\hspace{.5cm}}rll}
    \textsc{Terms} & \tm,\tmtwo &\grameq& \ldots \textup{ \ (as in } \plotcalcif \textup{, see \Cref{fig:plotkin-calculus})} \\
    
    \textsc{{\Lav}s} & \val & \grameq & \ldots \textup{ \ (as in } \plotcalcif \textup{, see \Cref{fig:plotkin-calculus})}\\
    
    \textsc{Fireballs} & \fire & \grameq & \val \mid \gconst\\
	    
	   \textsc{Inert terms} & \gconst& \grameq &  \var\,\fire \mid \gconst\,\fire \mid \xite\var\tm\tmtwo \\&&&\mid \xite\gconst\tm\tmtwo\\
    %
      \textsc{Right \firectx{}} & \revctx  & \grameq & \ctxhole\mid \tm\revctx \mid \revctx\fire \mid \xite\revctx\tmtwo\tmthree\\[0.5em]	
\end{array}$
\\
%
%
	$\begin{array}{rcl}
	\multicolumn{3}{c}{\textsc{Reduction Rules at Top Level}} \\
	(\la\var\tm)\gconst \rtoin \tm\isub\var\gconst & & \rtobv, \rtoift, \rtoiff, \rtoife, \rtoape \textup{ as in } \plotcalcif \\[0.5em]
	\multicolumn{3}{c}{\textsc{Contextual closure} }\\
	\revctxp \tm \Rew{a} \revctxp \tmtwo & \textrm{~~~if } & \tm \rootRew{a} \tmtwo \ \ \ \ \mbox{for $a\in\set{\betav,\betain,\iftsym,\iffsym,\ifesym,\apesym}$} \\[0.5em]
	\tof \ \defeq \ \ \tobv \cup \toin &  &\tocf \ \defeq \ \ \bigcup_{a\in\set{\betaf,\iftsym,\iffsym,\ifesym,\apesym}} \Rew{a}
	\end{array}$
    \end{tabular}
 }
	\Description{The conditional fireball calculus $\cfirecalc$.}
  \caption{The conditional fireball calculus $\cfirecalc$.}
  \label{fig:fireball-calculus}
\end{figure}

In this section we recall the fireball calculus $\firecalc$, the simplest presentation of \ocbv, and extend it with conditionals. The extension is completely modular. For the issues of Plotkin's setting with respect to open terms and for alternative presentations of \ocbv, we refer the reader to 
\citet{DBLP:conf/aplas/AccattoliG16,aplas18}.

The fireball calculus  was introduced without a name and studied first by \citet{DBLP:journals/ita/PaoliniR99,parametricBook}.
It has then been rediscovered by \citet{DBLP:conf/icfp/GregoireL02} to improve the implementation of Coq, and later by \citet{fireballs} to study cost models, where it was also named. 
We present it following \citet{fireballs}, changing only  inessential, cosmetic details.

\paragraph{The fireball calculus.} The conditional fireball calculus $\cfirecalc$ is defined in \reffig{fireball-calculus}. The conditional part is exactly as in the closed case.
The idea is that the \lav{}s of the Pif calculus are generalised to \emph{fireballs}, by adding \emph{inert \lat{s}}. 
Fireballs (noted $\fire$) and inert \lat{s} (noted $\gconst$) are defined by mutual induction (in \reffig{fireball-calculus}). 
For instance, $\var$ and $\la\var\vartwo$ are fireballs as \lav{s}, while $\vartwo(\la\var\var)$, $\var\vartwo$, and $(\varthree(\la\var\var))(\varthree\varthree) (\la\vartwo(\varthree\vartwo))$ are fireballs as inert \lat{s}. 

The main feature of inert \lat{s} is that they are open, normal, and that when plugged in a \lcontext they cannot create a redex, hence the name ``inert''. Essentially, they are the \emph{neutral terms} of \ocbv. 
In \gregoire and Leroy's presentation \cite{DBLP:conf/icfp/GregoireL02}, inert \lat{}s are called \emph{accumulators} and fireballs are simply called values.
Variables are, morally, both \lav{s} and inert \lat{}s. 
In \citet{fireballs} they were considered as inert \lat{}s, while here, for minor technical reasons we prefer to consider them as \lav{}s and not as inert \lat{s}---the change is inessential.

\paragraph{Evaluation rules.} First, \cbv $\beta$-reduction is replaced by  \emph{call-by-fireball} $\beta$-reduction $\tof$: the $\beta$-rule can fire, \emph{lighting} the argument, 
 only if the argument  
 is a fireball (\emph{fireball} is a catchier version of \emph{fire-able term}).
 We actually distinguish two sub-rules: the usual one that \emph{lights} \lav{s}, noted $\tobv$,
 and a new one that \emph{lights} inert \lat{s}, noted $\toin$ (see \reffig{fireball-calculus}). Second, we include all the rules about conditionals and errors, exactly as before, obtaining the evaluation relation $\tocf$.
 Note that evaluation is \emph{weak}: it does not reduce in abstraction nor $\xitecompact$ bodies.
 
 We endow the calculus with the 
 (deterministic) right-to-left evaluation strategy, defined via right \firectx{s} $\revctx$---note the production $\revctx \fire$, forcing the right-to-left order. A more general calculus (without conditionals) is defined in \citet{DBLP:conf/aplas/AccattoliG16}, for which 
 the right-to-left strategy is shown to be complete. 
 We omit details about the rewriting theory of the fireball calculus because our focus here is on implementations. 

\begin{example}
\label{ex:torf}  
  Let $\tm \defeq (\la{\varthree}{\varthree(\vartwo\varthree)})(\la{\var}{\var})$.
  Then, \sloppy $\tm \tof (\la{\var}{\var})(\vartwo \, (\la{\var}{\var})) \tof \vartwo \, (\la{\var}{\var})$, where the final term $\vartwo \, (\la{\var}{\var})$ is a fireball (and $\betaf$-normal).
\end{example}

\paragraph{Properties.} As discussed in \refsect{closed}, \ccbv enjoys harmony (\refprop{pif-harmony}). 
The fireball calculus $\firecalc$ satisfies an analogous property in the  \emph{open} setting by replacing abstractions with fireballs; we here further extend it to conditionals (\refpropp{distinctive-fireball-new}{open-harmony} below).
The key property of inert \lat{s} is summarised by \refpropp{distinctive-fireball-new}{inerts-and-creations}:
substitution of inert \lat{s} does not create or erase $\cbetafsym$-redexes, and hence can always be avoided.
It plays a role in the design of the open abstract machine of the next section.
\begin{proposition}[Properties of $\cfirecalc$]
  \label{prop:distinctive-fireball-new}
  \Copy{prop:distinctive-fireball-new}{Let $\tm, \tmtwo$ be \lat{}s.
  \begin{enumerate}
    \item\label{p:distinctive-fireball-new-open-harmony}
      \emph{Open harmony:}
      $\tm$ is $\cbetafsym$-normal if and only if $\tm$ is a fireball.
    \item\label{p:distinctive-fireball-new-inerts-and-creations}
      \emph{Inert substitutions and evaluation commute:} 
      Let $\gconst$ be an inert \lat. 
      Then $\tm \tocf \tmtwo$ if and only if $\tm \isub\var\gconst \tocf \tmtwo \isub\var\gconst$.
  \end{enumerate}
  }
\end{proposition}

\subsection{The \oaglam}
\label{subsect:micro-step-open}

Here we extend the \aglam defined in \refsect{closed} to the case of open terms, implementing Open (Conditional) CbV, \ie{} the conditional fireball calculus $\cfirecalc$: in this way we obtain the \emph{\oaglam}. The extension impacts on the core $\l$-calculus, while conditionals are essentially orthogonal to the issues of open terms.

\paragraph*{Evaluated environments.} First, we need to discuss the environments under which evaluation takes place. In the open case, \lcell{}s and \lenvironment{}s generalize to \firecell{}s and \fireenvironment{}s, and are denoted as follows:
\begin{align*}
  \textsc{\firecell{}s:} \ & \fcell \qquad &
  \textsc{\fireenvironment{}s:} \ & \fenv 
\end{align*}

Recall that in the \aglam the already evaluated coda of the environment is  made out only of practical values.
Unfortunately, a syntactic characterisation of \fireenvironment{}s (and \firecell{}s) is more involved than the simple definition of \lenvironment{}s.

In the \aglam, to check whether a \crumbledt $\mol$ is in normal form with respect to a \lenvironment{} $\aenv$, it suffices to check whether $\mol$ is a practical value. 
In the open case, looking at the syntactic structure of the term is not enough: some applications are now normal, for example the \crumbledt $\vartwo\,\var$ is normal with respect to the environment $\env\defeq\esub\var I$, but not all of them are normal, for instance $(\var\,\vartwo, \esub\var I) \Rew\elsym (I\,\vartwo, \esub\var I)$ as in the closed case (exact definitions are given below). 
Because of this additional complication, we are going to define \fireenvironment{}s directly in terms of their 'semantics', \ie~of their read-back to \lat{s}. 
Intuitively, fully evaluated \fireenvironment{}s should correspond to substitutions of fully evaluated \lat{s} in $\cfirecalc$. 
And since by 
harmony normal forms in $\cfirecalc$ are simply fireballs, it suffices to request that the read-back of every entry in a~\fireenvironment~is~a~fireball.

Let us now define \fireenvironment{}s formally:
$\fenv$ is a \emph{\fireenvironment{}} (resp.~$\fcell$ is a \emph{\firecell}) if for any environment context $\genv$ (resp.~any \crumble context $\cctx$) and any \crumblep $\cell$ such that $\fenv = \genv\ctxholep\cell$ (resp. $\fcell = \cctxp\cell$) the following two conditions hold:
\begin{enumerate}
  \item \emph{Read-back to fireballs:} $\rb\cell$ is a fireball, and
  \item \emph{Unchaining practical values:} if $\rb\cell$ is a practical value, then $\cell=(\molv,\env)$ for some practical value $\molv$ and some $\env$.
\end{enumerate}

The second requirement forbids $\molv$ to be a variable and is crucial for capturing the correct behaviour of the substitution rule $\toev$, which removes the malicious chains of substitutions (of variables for variables) discussed in \refsect{closed}.

\paragraph*{Transitions.} The transitions of the \oaglam:
\begin{align*}
  ((\la\var\cell)\,\molv, \fenv)
  &  \rtof 
  \append{(\append{\cell}{\esub\var\molv})^\alpha}{\fenv}
  && 
  \\%
  \\[-.7\baselineskip]
  (\xite\true\cell\celltwo, \fenv)
		    & \rtoift 
		     \append\cell\fenv\\
		    (\xite\false\cell\celltwo, \fenv)
		    & \rtoiff 
		     \append\celltwo\fenv
		     \\
		    (\xite\molv \cell \celltwo, \fenv)
		    & \rtoife 
		    (\err, \fenv) & (1)
		    \\
		    (\molv \molvtwo, \fenv)
		    & \rtoape 
		    (\err, \fenv)& (2)
		     \\%
  \\[-.7\baselineskip]
  (\var,\fenv)
  & \rtoev 
  (\fenv(\var),\fenv)
  & (3)
  \\
  (\var \molv,\fenv)
  & \rtoel 
  (\fenv(\var) \, \molv, \fenv)
  & (3)
  \\
      (\xite\var\cell\celltwo,\fenv)
  & \rtoeite 
  (\xite{\fenv(\var)}\cell\celltwo, \fenv)
  & (3)
\end{align*}
\begin{enumerate}
  \item if $\molv=\la\var\cellthree$ or $\molv = \err$
  \item if $\molv\in \set{\true,\false,\err}$
  \item if $\var \in \domain{\fenv}$
\end{enumerate}
\sloppy%
Top level transitions are then closed by \crumble contexts by setting $\cctxp\cell \Rew{a} \cctxp\celltwo$ if $\cell \rootRew{a}\celltwo$ for $a\in \set{\msym, \evsym, \elsym, \eitesym, \iftsym, \iffsym,\ifesym,\apesym}$. 
The transition relation  $\toocrumble$ of the \oaglam is defined as the union of all these rules.
A \emph{principal transition} of the \oaglam is a transition $\Rew{a}$ for any rule $a\in\set{\betaf,\iftsym,\iffsym,\ifesym,\apesym}$.

%

There are only two differences with the transitions of the \aglam. First, $\tom$ is now noted $\rtof$ and yet it is identical to the one in the closed case (the comments about $\alpha$-renaming given in \refsect{closed} still hold). This is because there is a subtle difference: the argument of the $\beta$-redex may be a variable (which is a value) substituted by a inert term in the environment, thus becoming a $\toin$ step (and not a $\tobv$ step) when read-back in $\cfirecalc$. Second, there is a slightly different side condition for the substitution transitions: it requires not only that a variable is defined in $\fenv$ (like in the closed case), but also that the corresponding term in the environment is a practical value (and not an inert term nor a variable). 

{
Note that the substitution transitions substitute values only. The environment $\fenv$ may contain also \crumbledt{}s that are variables or applications, but these bites are not substituted: this choice is justified by the property of $\cfirecalc$ stated in \refpropp{distinctive-fireball-new}{inerts-and-creations}. Besides, avoiding the substitution of inert terms is a prerequisite for efficiency of the machine, that would otherwise be subjected to an exponential overhead due to \emph{size explosion}, see for example \citet{fireballs,AccattoliGuerrieri17b}.
}

The harmony between evaluation rules and the syntactic definition of normal forms is witnessed by the following property.

\begin{proposition}[Harmony for the \oaglam]
  \label{prop:harmony-open}
  \Copy{prop:harmony-open}%
  {A \crumblep $\cell$ is $\ocrumblesym$-normal if and only if it is a \firecell.}
\end{proposition}

\begin{example}
	\label{ex:evaluation-open}
	
  Recall that $\auxtr\delta=(\lambda\var.\var\var,\emptyenv)$.
  In \refex{evaluation-closed} we noted that the (open) \crumblep $\mytr{\delta\delta (\var\var)}$ was stuck in the \aglam.
	Now instead it correctly reduces, never reaching a normal form:
	\[\begin{array}{rllll}
		\mytr{\delta\delta (\var\var)}  &=&
     (\varthree\varfour, \esub\varthree{\auxtr\delta\,\auxtr\delta}\esub\varfour{\var\var})
    \\
    & \tom & (\varthree \varfour, \esub{\varthree}{\vartwo\vartwo} \esub{\vartwo}{\auxtr\delta}\esub{\varfour}{\var\var}) 
    \\
		&\toel &(\varthree \varfour, \esub{\varthree}{\auxtr\delta\,\vartwo} \esub\vartwo{\auxtr\delta}\esub{\varfour}{\var\var}) \to \cdots
	\end{array}\]
\end{example}

\paragraph{Implementation Theorem}
\label{subsect:implement-open}


The proof of the implementation theorem for the \oaglam follows the same structure as for the \aglam in \refsubsect{the-implementation-theorem}, relying on similar but subtler invariants that can be found in the appendix, \reflemma{invariants-open}.

\begin{theorem}[Implementation]
  \label{thm:implementation-open}
\Copy{thm:implementation-open}{
  Let $\cell$ be a \crumblep that is reachable by the \oaglam.
  \begin{enumerate}
    \item \emph{Initialization: }
     $\rb{\mytr{\tm}} = \tm$ 
    \item \emph{Principal projection: }
      if $\cell \Rew{a} \celltwo $ then $\rb\cell \Rew{a} \rb\celltwo$
      for $a \in \set{\betafsym,\iftsym,\iffsym,\ifesym,\apesym}$.
    \item \emph{Overhead transparency: }
      if $ \cell \Rew{a} \celltwo $ then $\rb\cell = \rb\celltwo$
			for any rule $a \in \set{\evsym,\elsym, \eitesym}$.

    \item \emph{Determinism: } the transition 
    $\toocrumble$ is deterministic.
    \item \emph{Halt: } if $\cell$ is $\ocrumblesym$-normal then $\rb\cell$ is $\cbetafsym$-normal.
    \item \emph{Overhead termination: } $\Rew{a}$ terminates, for any rule $a \in \set{\evsym,\elsym, \eitesym}$.
  \end{enumerate}
	Therefore, the \oaglam, the right-to-left conditional fireball evaluation $\tocf$ and the read-back $\rbp{\cdot}$ form an implementation system.
}
\end{theorem}


\paragraph{Complexity}
\label{subsect:complexity-open}

The complexity analysis is identical to the one in \refsubsect{complexity-closed}. 
Indeed, once the search for the next redex and (un)plugging are neglected, the two machines only differ by the additional $O(1)$ side condition for the substitution transitions.

\begin{theorem}[The \oaglam is bilinear up to search and (un)plugging]
\label{thm:compl-open}
		For any \lat{} $\tm$ and any \oaglam execution $\deriv\colon\mytr\tm\toocrumble^*\cell$, the cost of implementing $\exec$ on a RAM is $O((\sizepr\exec + 1) \cdot \len{\tm})$ plus the cost of plugging and unplugging.
\end{theorem}

\paragraph{OCaml implementation} Following the same pattern of the closed case, in \refapp{towards-impl-open-pre} 
we introduce a machine making explicit the search for redexes and removing the need of (un)plugging, so as to show that their cost 
is negligible.
The OCaml code implementing this further machine is in \refapp{ocamlimpl}, together with a detailed discussion of the adopted data structures. 
The code for the open and closed machines is identical but for five lines: three implement the additional check for \pvalue{s} in the substitution transitions, the others consider also inert terms in the search transition.

\section{Extensions}
\label{sect:extensions}
\paragraph{Left-to-right \cbv} The (right-to-left) \aglam can also implement a left-to-right strategy for the Pif calculus. The only change concerns the crumbling transformation, that on applications has to put the environment coming from the (transformation of the) left subterm on the right of the one coming from the right subterm.

\paragraph{Call-by-need} The crumbling technique applies also to call-by-need machines. There are however a few differences. First, the machine does no longer explore sequentially the environment from right-to-left, it rather starts on the left and then jumps back and forth, by need. Then the definition of evaluation contexts is trickier, especially in the open case.

\paragraph{Strong \cbv} Simply designing an abstract machine for strong reduction is relatively easy. However the easy machines are not \emph{bilinear}, and not even polynomial. 

The needed optimisations to make them reasonable (i.e. polynomial or bilinear) are clear, they are the same at work in the open case (or in the call-by-name case): 
\begin{enumerate}
\item substitute only abstraction and not inert terms, and
\item do not substitute abstractions on variable occurrences that are not applied.
\end{enumerate}
These principles however have different consequences in different settings. In particular, 2 implies that some abstraction are kept shared forever, and a \emph{strong} \cbv approach has to evaluate them (while the open setting does not) and only once, thus it has to evaluate them while they are shared, adding a call-by-need flavor. 

There are two difficulties. First, the specification of the search for redexes, that becomes involved and requires many machine transitions---the crumbling technique is meant to help here. Second, the proof of correctness of the machine. 

All proofs of correctness in the literature (including those in this paper) are simulations up to sharing based on a \emph{bijection} of $\beta$-redexes (or principal steps) between the abstract machine and the $\l$-calculus strategy (one half of the bijection is the principal projection property of implementation systems in \refsubsect{the-implementation-theorem}, the other half is implied by the other properties).

The evaluation under \emph{shared} abstraction required by CbV strong evaluation breaks the usual bijection of $\beta$-redexes (as one $\beta$-transition of the machine is mapped to \emph{many} $\beta$-steps on the calculus, and not necessarily those of a standard strategy), thus forbidding to employ the standard technique for proving correctness.

The new proof technique for correctness for reasonable strong CbV and the intricacies of the search for redexes in the strong case, do deserve to be studied carefully, and are thus left to future work.

 
\section{Conclusions}
\label{sect:conclusions}
This paper studies abstract machines working on \crumbled forms with respect to design, efficiency, scalability, and implementations, putting emphasis on the role played by environments and providing a detailed technical development. 
In particular, we study the \crumble setting on top of global environments---in future work we would like to explore the more technical case of local environments.

At the level of design, 
switching to \crumbled forms removes the need for machine data structures such as the applicative stack or the dump, as they are encoded in \crumbled environments.

At the level of efficiency, the evaluation of \crumbled forms does not require any overhead: \crumble abstract machines are linear in the number of steps of the calculus and in the size of the initial term, exactly as ordinary abstract machines with global environments.

At the level of scalability, everything---including the complexity---smoothly scales up from the closed case, relevant for programming languages, to the more delicate case of open terms, needed to implement proof assistants. 
As shown in \Cref{sect:kennedy-danvy}, CPS translations do not smoothly scale up to the open case (contrary to what claimed by \citet{DBLP:journals/mscs/DanvyF92}), so that our work 
shows an advantage of the \crumbling transformation in this setting.

At the level of implementations, we stress the different operations on \crumbled environments (sequential access and concatenation) and provide a concrete implementation, which 
does not suffer from the potential slowdown of \crumbled forms pointed out by \citet{DBLP:conf/icfp/Kennedy07} (see \Cref{sect:kennedy-danvy}).

In future work we plan to apply our results to the design of abstract machines for strong call-by-value and call-by-need evaluation. Preliminary results suggest that the simplification to the code noticed in the open case is preserved and even amplified in the harder case of strong evaluation.

\paragraph{Acknowledgments}
This work has been partially funded by the
ANR JCJC grant COCA HOLA (ANR-16-CE40-004-01) and  by the EPSRC grant EP/R029121/1
``Typed Lambda-Calculi with Sharing and Unsharing''.

\bibliographystyle{ACM-Reference-Format}
\bibliography{\macrospath/biblio}

\newpage
\appendix
\section*{Technical Appendix}
\addcontentsline{toc}{section}{Technical Appendix}


\section{Comments on Related Works}
\label{sect:kennedy-danvy}
Here we discuss in detail Kennedy's potential slowdown and provide a counter-example to the scalability of the CPS transformation to open \lat{s}.

\paragraph{Kennedy}
\citet{DBLP:conf/icfp/Kennedy07} compares three different calculi: a monadic calculus, which has ES, a calculus of administrative normal forms (ANFs) and the image of a CPS transformation. 
In the monadic calculus $\betav$-redexes can be hidden by ES which need to be commuted to reveal the $\betav$-redex. 
Kennedy shows an example (see \reffig{kennedy}) where the number of commutations is not bounded linearly by the number of $\betav$-steps and blames the inefficiency of his compiler on that. 
In his example, the number of commutations 
is \emph{quadratic} in the number of $\betav$-steps, since the $i^\textup{th}$ $\betav$-step is immediately followed by $i$ commutation steps. 

\begin{figure*}[!ht]
\begin{align*}
\tm \defeq & \ (\varthree_1 \var_0) \esub{\varthree_1}{\la{\var_1} \consttwo\vartwo_1 \esub{\vartwo_1}{\varthree_2\var_1}} \esub{\varthree_2}{\la{\var_2} \consttwo\vartwo_2 \esub{\vartwo_2}{\varthree_3\var_2}} \dots \esub{\varthree_n}{\la{\var_n} \consttwo \vartwo_n \esub{\vartwo_n}{\consttwo\var_n}}
\\
 ({\tobv}& \tolet) \,\, (\tobv \tolet^2) \,\cdots\, (\tobv \tolet^i) \\ 
&(\varthree_1 \var_0) \esub{\varthree_1}{\la{\var_1} \consttwo\vartwo_1 \esub{\vartwo_1}{\varthree_2\var_1}} \esub{\varthree_2}{\la{\var_2} \consttwo\vartwo_2 \esub{\vartwo_2}{\varthree_3\var_2}} \dots \\
&\dots \esub{\varthree_{n-i-1}}{\la{\var_{n-i-1}} \consttwo \vartwo_{n-i-1} \esub{\vartwo_{n-i-1}}{\varthree_{n-i}\var_{n-i-1}}} \esub{\varthree_{n-i}}{\la{\var_{n-i}} \consttwo \vartwo_{n-i} \esub{\vartwo_{n-i}}{\consttwo \vartwo_{n-i+1}} \esub{\vartwo_{n-i+1}}{\consttwo \vartwo_{n-i+2}} \\
&\dots \esub{\vartwo_n}{\consttwo \var_{n-i}}}
\\
\tobv \\
&(\varthree_1 \var_0) \esub{\varthree_1}{\la{\var_1} \consttwo\vartwo_1 \esub{\vartwo_1}{\varthree_2\var_1}} \esub{\varthree_2}{\la{\var_2} \consttwo\vartwo_2 \esub{\vartwo_2}{\varthree_3\var_2}} \dots \\
&\dots \esub{\varthree_{n-i-1}}{\la{\var_{n-i-1}} \consttwo \vartwo_{n-i-1} \esub{\vartwo_{n-i-1}}
{\consttwo \vartwo_{n-i} \esub{\vartwo_{n-i}}{\consttwo \vartwo_{n-i+1}} \esub{\vartwo_{n-i+1}}{\consttwo \vartwo_{n-i+2}} \dots \esub{\vartwo_n}{\consttwo \var_{n-i-1}}}}
\\
\tolet^{i+1} \\
&(\varthree_1 \var_0) \esub{\varthree_1}{\la{\var_1} \consttwo\vartwo_1 \esub{\vartwo_1}{\varthree_2\var_1}} \esub{\varthree_2}{\la{\var_2} \consttwo\vartwo_2 \esub{\vartwo_2}{\varthree_3\var_2}} \dots \\
&\dots \esub{\varthree_{n-i-1}}{\la{\var_{n-i-1}} \consttwo \vartwo_{n-i-1} \esub{\vartwo_{n-i-1}}{\consttwo \vartwo_{n-i}} \esub{\vartwo_{n-i}}{\consttwo \vartwo_{n-i+1}} \dots \\
&\dots \esub{\vartwo_n}{\consttwo \var_{n-i-1}}}
\end{align*}
\Description{Kennedy's example of evaluation in the monadic calculus where the number of commutation steps is quadratic in the number of $\betav$-steps.}
\caption{Kennedy's example of evaluation in the monadic calculus where the number of commutation steps is quadratic in the number of $\betav$-steps ($\to^i$ stands for the composition of $i$ $\to$-steps). 
The $i^\textup{th}$ $\betav$-step is immediately followed by 
$i$ commutation steps $\tolet$ that just append two lists of substitutions moving one substitution at a time. 
Therefore, to reach a normal form one needs $n$ $\betav$-steps and ${n(n+1)}/{2}$ $\text{let}$-steps.
In the 
\Crumble and \oaglam instead, the commutation steps are integrated in the multiplicative rule simply by appending the two lists in constant time.}
\label{fig:kennedy}
\end{figure*}

ANFs are just canonical shapes of monadic terms where the topmost term and the body of each abstraction is a \crumblep, \ie a term together with a list of ES that map variables to terms (instead of \crumble{s}). 
Kennedy rightly observes that ANFs are not preserved by standard $\betav$-reduction and thus, after each $\betav$-step, some commutative steps are required to reach the ANF shape. 
The quadratic example by Kennedy stands in the ANF fragment. 
So, Kennedy too hastily concludes that the quadratic blowup also affects the ANF calculus.

However, Kennedy misses the fact that the ES in ANFs form a list and that the commutations steps altogether just implement the append function of two lists. Since append can be implemented in constant time, the complexity of evaluation in the ANF calculus is just linear (and not quadratic) in the number of $\betav$-steps. 
This is the same complexity we achieved 
for the \Crumble and \oaglam.

\paragraph{Danvy and Filinski}
 In \citet{DBLP:journals/mscs/DanvyF92} the CPS transformation is shown to scale up to open \lat{s} (their Theorem 2). 
 On open \lat{s}, however, they consider Plotkin's \cbv operational semantics $\plotcalc$, which is \emph{not adequate} (it is adequate only for closed terms, see \citet{DBLP:conf/aplas/AccattoliG16,aplas18}). 
 When one considers one of the equivalent adequate \cbv semantics in \citet{DBLP:conf/aplas/AccattoliG16,aplas18} for the open case, for instance the fireball calculus $\firecalc$, then the properties of the CPS no longer hold, in particular it does not commute with evaluation, as the following example shows.
Take the following open \lat $\tm \defeq (\la{\var} {\la{\vartwo}\vartwo}) (\varthree \varthree) \val$, where $\val$ is a \lav, say a distinguished variable.
 In $\plotcalc$ the \lat $\tm$ is $\betav$-normal, but in $\firecalc$ we have:
\begin{align*}
	\tm \defeq (\la{\var} {\la{\vartwo}\vartwo}) (\varthree \varthree) \val \tof (\la{\vartwo}{\vartwo})\val \tof \val
\end{align*}

Now, consider the CPS translation $\mathtt{cps}(\tm)$ of $\tm$, according to the definition in \citet{DBLP:journals/mscs/DanvyF92}.
We use $\lambda$ for standard (``dynamic'', in Danvy's terminology) abstraction, and $\Lambda$ and $@$ for ``static'' abstraction and application, respectively.
If a generalized version of Theorem 2 in \citet{DBLP:journals/mscs/DanvyF92} held in the open case, one would expect that $@ (\mathtt{cps}(\tm)) I$ (where $I \defeq \la{\varthree}\varthree$) evaluates 
to $\val$, as $\val$ is a value.
But, even using an unrestricted $\beta$-reduction that goes under abstraction as evaluation, we obtain
(we reduce all static redexes first, followed by all dynamic redexes):
  \begin{align*}
   &@ (\mathtt{cps}(\tm)) I   \\
   = \ & (\Lambda k.
        @
         (\Lambda x.
           @
            (\Lambda y.
              @ y
               (\lambda w.
                 \lambda a.
                  @
                   (\Lambda b.
                     @ b
                      (\lambda c.
                        \lambda d.
                         @ (\Lambda e. @ e c) (\Lambda e. d e))) \\
                   &(\Lambda b. a b)))
            (\Lambda y.
              @
               (\Lambda j.
                 @ (\Lambda a. @ a z)
                  (\Lambda a.
                    @ (\Lambda b. @ b z)
                     (\Lambda b. (a b) (\lambda c. @ j c)))) \\
               &(\Lambda w. (y w) (\lambda a. @ x a))))
         (\Lambda x.
           @ (\Lambda y. @ y v)
            (\Lambda y. (x y) (\lambda w. @ K w)))) I\\
   \tob^*&
   (\varthree\varthree)
    (\lambda x.
      ((\lambda y. \lambda w. w (\lambda a. \lambda b. b a)) x)
       (\lambda y. y v (\lambda w. I w)))\\
  \tob^*&
  (\varthree\varthree) (\la \var \val)
\end{align*}

\noindent
where $ (\varthree\varthree) (\la \var \val)$ is not even $\beta$-equivalent to $\val$.
The CPS translation---like Plotkin's calculus---gets stuck trying to evaluate $z z$, whereas the term reduces to $v$ in the fireball calculus.

Summing up, we are not claiming that Theorem~2 in \citet{DBLP:journals/mscs/DanvyF92} is false, but just that it does not mean that their CPS transformation 
scale up to open \lat{s}:
to prove scalability, one should use an adequate \cbv evaluation for open \lat{s} (such as the one of the fireball calculus), instead of Plotkin's one.
Our counter-example shows that Danvy's and Filinski's CPS does not scale up to open \lat{s} with an adequate \cbv operational semantics for them.

It is worth noting that this problem affects also other CPS translations, such as the ones defined by \citet{DBLP:journals/tcs/Plotkin75} or by \citet{DBLP:conf/lics/Lassen05}. 
Likely, this is the reason why \citet{DBLP:conf/lics/Lassen05} states his Theorem 4.6 (the analogous of Theorem 2 in \citet{DBLP:journals/mscs/DanvyF92}) only for closed \lat{s}.

\section{Proofs of \refsect{plotkin} (\pif Calculus)}

\begin{proofof}[\refprop{pif-harmony}]
\Paste{prop:pif-harmony}
\end{proofof}
\begin{proof}\hfill
    \begin{description}
      \item [$(\Rightarrow)$] Proof by induction on the structure of $\tm$.
      \begin{itemize}
        \item Case $\tm$ value: trivial.
        \item Case $\tm = \tmtwo\tmthree$ for some terms $\tmtwo$ and $\tmthree$: we show that this case is not possible by deriving a contradiction. Since $\tm$ is $\topif$normal, then $\tmthree$ is $\topif$normal, and hence by \ih{} $\tmthree$ is a value. Since $\tm$ is $\topif$normal and $\tmthree$ is a value, then also $\tmtwo$ must be $\topif$normal, and hence a value.
        We proceed by cases on $\tmtwo$, showing that no case is possible. $\tmtwo$ cannot be a variable, because $\tm$ is closed; it cannot be an abstraction (because otherwise the rule $\tobv$ may be applied, contradicting the hypothesis that $\tm$ is $\topif$normal); it cannot be a boolean or $\err$ (because otherwise the rule $\toape$ may be applied).
        \item Case $\tm = \xite\tmtwo\tmthree\tmfour$ for some terms $\tmtwo,\tmthree,\tmfour$: we show that this case is not possible by deriving a contradiction. Since $\tm$ is $\topif$normal, then $\tmtwo$ is $\topif$normal, and hence by \ih{} $\tmtwo$ is a value.
        We proceed by cases on $\tmtwo$, showing that no case is possible. $\tmtwo$ cannot be a variable, because $\tm$ is closed; it cannot be an abstraction or $\err$ (because otherwise the rule $\toife$ may be applied, contradicting the hypothesis that $\tm$ is $\topif$normal); it cannot be a boolean (because otherwise one of the rules $\toift$ or $\toiff$ may be applied).
      \end{itemize}

      \item [$(\Leftarrow)$] By hypothesis, $\tm$ is a value. We proceed by cases on $\tm$. $\tm$ cannot be a variable, because by hypothesis $\tm$ is closed. If $\tm$ is an abstraction, then it is $\topif$normal since $\topif$ does not reduce under $\l$'s. Otherwise $\tm$ is a either a boolean or $\err$, which clearly are $\topif$normal.
    \end{description}
\end{proof}

\begin{lemma}[Composition of right \valuectx{s}]
	\label{l:r-ctxs-comp-closed}
  Let $\revctx$ and $\revctx'$ be right \valuectx{s}.
  Then their composition $\revctxp{\revctx'}$ is a right \valuectx.
\end{lemma}
\begin{proof}
	By induction on the right \valuectx $\revctx$.
	Cases:
	\begin{itemize}
		\item \emph{Hole}, \ie $\revctx \defeq \ctxhole$: then, $\revctxp{\revctxtwo} = \revctxtwo$ is a right \valuectx{} by hypothesis.

		\item  \emph{Right}, \ie $\revctx \defeq \tm\revctxthree$: then, $\revctxp{\revctxtwo} = \tm \revctxthreep{\revctxtwo}$ is a right \valuectx{} because $\revctxthreep{\revctxtwo}$ is a right \valuectx{} by \ih

		\item \emph{Left}, \ie $\revctx \defeq \revctxthree \val$: then, $\revctxp{\revctxtwo} = \revctxthreep{\revctxtwo} \val$ is a right \valuectx{} because $\revctxthreep{\revctxtwo}$ is a right \valuectx{} by \ih
    
    \item $\xitecompact$, \ie $\revctx \defeq (\xite \revctxthree \tm \tmtwo)$:
     then, $\revctxp{\revctxtwo} = (\xite{\revctxthreep{\revctxtwo}} \tm \tmtwo)$ is a right \valuectx{} because $\revctxthreep{\revctxtwo}$ is a right \valuectx{} by \ih
		\qedhere
	\end{itemize}
\end{proof}

\section{Proofs of \refsect{preliminaries} (preliminaries)}
\begin{lemma}
  Let $\env \defeq \esub{\var_1}{\mol_1} \dots \esub{\var_n}{\mol_n}$ a well-named environment.
If $\fv{\tm_i} \cap \{\var_1, \dots, \var_{i}\} = \emptyset$ for all $1 \leq i \leq n$, then $\fv{\env} = \bigcup_{i = 1}^n \fv{\mol_i} \smallsetminus \domain{\env}$ and $\fv{(\mol, \env)} = (\fv{\mol} \cup \fv{\env}) \smallsetminus \domain{\env}$.
\end{lemma}

\begin{definition}[Disjointedness]
	Let $P$ and $P'$ be two \crumblep{s} or environments:
	$P $ and $ P'$ are \emph{disjoint} (noted $P \disj P'$) if $\fv P \cap \domain {P'} = \emptyset$.
\end{definition}

\begin{definition}[Composition of \crumble contexts]
It is also possible to plug a \crumble context $\cctxtwo$ into another \crumble context $\cctx$, as follows:
\begin{align*}
	\cctxp{\cctxtwo} \defeq
	\begin{cases}
	 \cctx 		&\text{if } \cctxtwo = \ctxhole\\
	 \cctxtwo &\text{if } \cctx = \ctxhole\\
	 (\mol, \env \esub{\var}{\moltwo}\envtwo\esub{\vartwo}{\ctxhole}) 		&\text{if } \cctx = (\mol, \env \esub{\var}{\ctxhole}) \text{ and } \cctxtwo = (\moltwo, \envtwo \esub{\vartwo}{\ctxhole}) \\
	\end{cases}
\end{align*}

\end{definition}

The \emph{appending} of an environment context $\genv$ to a \crumble $(\mol, \envtwo)$ is defined as $\append{(\mol,\envtwo)}{\genv} \defeq (\mol, \envtwo \genv)$.

The notions of $\fv{\cdot}$ and $\domain{\cdot}$ are extended to \crumble contexts by:
  \begin{align*}
  \fv{\ctxhole} &\defeq \emptyset  & \fv{(\mol,\env\esub\var\ctxhole)} &\defeq \fv{(\mol,\env)} \smallsetminus \set\var \\
    \domain{\ctxhole} &\defeq \emptyset &
    \domain{(\mol,\env\esub\var\ctxhole)} &\defeq \domain{\env} \cup \set\var
\end{align*}

An environment context $\genv$ is \emph{well-named} if $\genvp{\var}$ is well-named (for $\var$ fresh) and a \crumble context $\cctx$ is \emph{well-named} if $\cctx \defeq \ctxhole$ or $\cctx \defeq (\mol, \genv)$ and $\genv$ is well-named.

\begin{lemma}[Decomposition of read back, auxiliary]
 \label{l:rb-comp-cell-0}
 Let $\env$ be an environment, and $\var$ be a variable such that
		$\var \notin \domain{\env} \cup \fv{\env}$.
    Then $\rbp{\append{\cell}{\esub\var\mol\env}} = \rbp{ \append{\cell}{\env}}\isub\var{\rb{(\mol,\env)}}$
	for every \crumble $\cell$ and \crumbledt $\mol$.
\end{lemma}

\begin{proof}
  Let $\cell = (\moltwo,\envtwo)$. We proceed by structural induction on $\env$:
  \begin{itemize}
    \item If $\env=\emptyenv$, then $\rb{(\mol,\env)} = \rb{\mol}$ and hence
  	$\rb{(\moltwo,\envtwo\esub\var\mol\env)} = \rb{(\moltwo,\envtwo\esub\var\mol)} = \rb{(\moltwo,\envtwo)}\isub\var{\rb\mol}%
  	= \rb{(\moltwo,\envtwo\env)}\esub\var{\rb{(\mol,\env)}}$.

	\item
	Otherwise $\env=\envthree\esub\varthree\molthree$ and then $\rb{(\mol,\env)} = \rb{(\mol,\envthree)}\isub\varthree{\rb\molthree}$, so
	\begin{align*}
	\rb{(\moltwo,\envtwo\esub\var\moltwo\env)} & = \rb{(\moltwo,\envtwo\esub\var\moltwo\envthree\esub\varthree\molthree)} \\
	& = \rb{(\moltwo,\envtwo\esub\var\moltwo\envthree)}\isub\varthree{\rb\molthree} & \\
	& = \rb{(\moltwo,\envtwo\envthree)}\isub\var{\rb{(\moltwo,\envthree)}}\isub\varthree{\rb\molthree} &  \text{by \ih{}} \\
	& = \rb{(\moltwo,\envtwo\envthree)}\isub\varthree{\rb\molthree}\isub\var{\rb{(\moltwo,\envthree)}\isub\varthree{\rb\molthree}} & \text{as $\var\not\in\fv{\rb\molthree}$} \\
	& = \rb{(\moltwo,\envtwo\env)}\isub\var{\rb{(\moltwo,\env)}} &
	\end{align*}
	where $\var \notin \fv{\rb\molthree} = \fv{\molthree}$ by \refrmk{preservation-fvs} and the hypothesis that $\var \notin \fv\env$.
\qedhere
\end{itemize}
\end{proof}

\begin{lemma}[Read-back vs. disjointedness]
	\label{l:disj-c-e}
  For every \crumble $\cell$ and environment $\env$:
		if $\cell\disj\env$, then $\rbp{\append\cell\env} = \rb\cell$.
\end{lemma}
\begin{proof}
	By structural induction on $\env$:
	\begin{itemize}
	\item If $\env \defeq \emptyenv$ then $\append{\cell}{\env} = \cell$ and hence $\rbp{\append{\cell}{\env}} = \rb{\cell}$.

	\item Otherwise $\env \defeq \envtwo \esub{\var}{\mol}$.
	By \ih (which can be applied since $\cell \disj \env$ implies $\cell \disj \envtwo$, because $\domain{\envtwo} \subseteq \domain{\env}$), $\rbp{\append{\cell}{\envtwo}} = \rb{\cell}$.
	From $\cell \disj \env$ it follows that $\var \notin \fv{\cell} = \fv{\rb{\cell}}$ (by \refrmk{preservation-fvs}), so $\rb{\cell}\isub{\var}{\rb{\mol}} = \rb{\cell}$.
	Therefore, $\rbp{\append{\cell}{\env}} = \rb{\append{\cell}{\envtwo}}\isub{\var}{\rb{\mol}} = \rb{\cell}\isub{\var}{\rb{\mol}} = \rb{\cell}$.
    \qedhere
  \end{itemize}
\end{proof}

\begin{lemma}[Decomposition of read back]
\label{l:rb-comp-cell}
		Let $\cell$ be a \crumble and $\env$ be an environment such that $\cell\disj\env$.
    Then $\rbp{\append\cell{\esub\var\mol\env}} \allowbreak= \rb{\cell}\isub\var{\rb{(\mol,\env)}}$
    for every \crumbledt $\mol$.
\end{lemma}

	Note that \reflemma{rb-comp-cell} does not hold without the hypothesis $\cell\disj\env$.
	Indeed, take $\cell \defeq (\vartwo,\emptyenv)$ and $\env \defeq \esub{\vartwo}{\varthree\varthree}$ with $\var \neq \vartwo$: for any term $\mol$, one has $\rbp{\append\cell{\esub\var\mol\env}} = \varthree\varthree \neq \vartwo =  \rb{\cell}\isub\var{\rb{(\mol,\env)}}$.

\begin{proof}
	According to \reflemma{rb-comp-cell-0},
	\begin{align*}
	\rbp{\append{\cell}{\esub\var\mol\env}} = \rbp{\append{\cell}{\env}}\isub\var{\rb{(\mol,\env)}} = \rb{\cell}\isub\var{\rb{(\mol,\env)}}
	\end{align*}
	where the last equality holds by \reflemma{disj-c-e}, since $\cell\disj\env$.
\end{proof}

\begin{proofof}[\refprop{read-back-inverse-transl}]
	\Paste{prop:read-back-inverse-transl}
\end{proofof}

\begin{proof}
	By mutual induction on the \lat $\tm$ and the \lav $\val$.
	Cases:
	\begin{itemize}
		\item \emph{Variable}, \ie $\tm \defeq \var \eqdef \val$;
		then, $\mytr{\tm} = (\var, \emptyenv)$ and $\auxtr{\val} = \var$, thus $\rb{\mytr{\tm}} = \var = \tm$ and $\rb{\auxtr{\val}} = \var = \val$.
    \item \emph{Error} or \emph{Boolean}: similar to the case \emph{Variable} above.
		\item \emph{Abstraction}, \ie $\tm \defeq \la{\var} \tmtwo \eqdef \val$;
		then, $\mytr{\tm} = (\la{\var}{\mytr{\tmtwo}}, \emptyenv)$ and $\auxtr{\val} = \la{\var}{\mytr{\tmtwo}}$;
		by \ih, $\rb{\mytr{\tmtwo}} = \tmtwo$, hence $\rb{\mytr{\tm}} = \la{\var}{\rb{\mytr{\tmtwo}}} = \tm$ and $\rb{\auxtr{\val}} = \la{\var}{\rb{\mytr{\tmtwo}}} = \val$.

    \item \emph{Conditional}, case $\tm\defeq (\xite\val\tmtwo\tmthree)$.
    Then $\mytr{\tm} = (\xite{\auxtr\val}{\mytr\tmtwo}{\mytr\tmthree},\emptyenv)$.
    Since $\rb{\mytr\tm} = (\xite{\rb{\auxtr\val}}{\rb{\mytr\tmtwo}}{\rb{\mytr\tmthree}})$, we can conclude by using the \ih{}

    \item \emph{Conditional}, case $\tm\defeq (\xite\tmtwo\tmthree\tmfour)$ with $\tmtwo$ not a \lav.
    Then $\mytr{\tm} = \append{(\xite\var{\mytr\tmthree}{\mytr\tmfour}, \emptyenv)}{\esub\var\tmfive\env}$ where $\mytr{\tmtwo} = (\tmfive, \env)$.
		By \ih, $\rb{\mytr{\tmthree}} = \tmthree$ and $\rb{\mytr{\tmfour}} = \tmfour$.
		We can suppose that $(\{\var\} \cup \fv\tmthree \cup \fv\tmfour) \cap \domain{\env} = \emptyset$ by the freshness condition in the definition of the transformation, and hence we can apply \reflemma{rb-comp-cell} so that
     $\rb{\mytr{\tm}} = (\xite\var{\rb{\mytr\tmthree}}{\rb{\mytr\tmfour}}) \isub{\var}{\rb{\mytr{\tmtwo}}} = \tm$.

		\item \emph{Application of two \lav{}s}, \ie $\tm = \val\valtwo$;
		then, $\mytr{\tm} = (\auxtr{\val} \auxtr{\valtwo}, \emptyenv)$;
		by \ih, $\rb{\auxtr{\val}} = \val$ and $\rb{\auxtr{\valtwo}} = \valtwo$, so $\rb{\mytr{\tm}} = \rb{\auxtr{\val}} \rb{\auxtr{\valtwo}} = \tm$.

		\item \emph{Application of a non-\lav to a \lav}, \ie $\tm \defeq \tmtwo \val$ where $\tmtwo$ is not a \lav;
		then, $\mytr{\tm} = (\var \auxtr{\val}, \esub{\var}{\tmthree}\env) = \append{(\var \auxtr{\val}, \emptyenv)}{\esub{\var}{\tmthree}\env}$ where $\mytr{\tmtwo} = (\tmthree, \env)$;
		we can suppose without loss of generality that $(\fv{\val} \cup \{\var\}) \cap \domain{\env} = \emptyset$.
		By \ih, $\rb{\mytr{\tmtwo}} = \tmtwo$ and $\rb{\auxtr{\val}} = \val$.
		By the freshness condition for $\var$, we can apply \reflemmas{rb-comp-cell-0}{disj-c-e} so that $\rb{\mytr{\tm}} = \rbp{\append{\var\auxtr{\val}}{\env}} \isub{\var}{\rb{(\tmthree,\env)}} = \var \rb{\auxtr{\val}} \isub{\var}{\rb{\mytr{\tmtwo}}} = \tm$.

		\item \emph{Application of a \lat to a non-\lav}, \ie $\tm \defeq \tmtwo\tmthree$ where $\tmthree$ is not a \lav;
		then, $\mytr{\tm} = \append{\mytr{\tmtwo\var}}{(\esub{\var}{\tmfour}\env)}$ where $\mytr{\tmthree} = (\tmfour, \env)$ and $\mytr{\tmtwo\var} = (\vartwo\var, \esub{\vartwo}{\tmfive}\envtwo) = \append{(\vartwo\var,\emptyenv)}{(\esub{\vartwo}{\tmfive}\envtwo)}$ with $\mytr{\tmtwo} = (\tmfive,\envtwo)$;
		we can suppose without loss of generality that $(\fv{\tmtwo} \cup \{\var\}) \cap \domain{\env} = \emptyset$ and $\{\var,\vartwo\} \cap \domain{\envtwo} = \emptyset$.
		By \ih, $\rb{\mytr{\tmtwo}} = \tmtwo$ and $\rb{\mytr{\tmthree}} = \tmthree$.
		By the freshness condition for $\var$ and $\vartwo$, we can apply \reflemmas{rb-comp-cell-0}{disj-c-e} so that $\rb{\mytr{\tmtwo\var}} = \rbp{\append{\vartwo\var}{\envtwo}}\isub{\vartwo}{\rbp{\tmfive, \envtwo}} = \vartwo\var\isub\vartwo{\rb{\mytr{\tmtwo}}} = \tmtwo\var$ and $\rb{\mytr{\tm}} = \rbp{\append {\mytr{\tmtwo\var}} {\env}} \isub{\var}{\rbp{\tmfour, \env}} = \rb{\mytr{\tmtwo\var}} \isub{\var}{\rb{\mytr{\tmthree}}} = \tmtwo\var \isub{\var}{\tmthree} = \tm$.
		\qedhere
	\end{itemize}
\end{proof}

\begin{lemma}\label{l:aux-aux-lookup-open}
  Let $\cell,\celltwo$ be \crumblep{s}, and $\env$ be an environment.
	If $\rb\cell = \rb\celltwo$, then $\rbp{\append\cell\env} = \rbp{\append\celltwo\env}$.
\end{lemma}
\begin{proof}
	By induction on $\env$:
	\begin{itemize}
		\item if $\env\defeq\emptyenv$, then we conclude because $\append\cell\env = \cell$ and $\append\celltwo\env = \celltwo$;
		\item if $\env\defeq\envtwo\esub\var\mol$, by \ih{}
		$\rbp{\append\cell\envtwo} = \rbp{\append\celltwo\envtwo}$,
		and we conclude because $\rbp{\append\cell\env} = \rbp{\append\cell\envtwo}\isub\var{\rb\mol}$ and  $\rbp{\append\celltwo\env} = \rbp{\append\celltwo\envtwo}\isub\var{\rb\mol}$.
		\qedhere
	\end{itemize}
\end{proof}

\begin{proposition}\label{prop:aux-rb-structural}
	Let $\env$ be a \crumbled environment. Then:
	\begin{enumerate}
		\item $\rbp{\var\,\molv,\env} = \rbp{\var,\env}\rbp{\molv,\env}$ for every variable $\var$ and \crumbled value $\molv$.
		\item $\rbp{\xite\var\cell\celltwo,\env} =
		 \xite{\rbp{\var,\env}}{\rbp{\append\cell\env}}{\rbp{\append\celltwo\env}}$ for every variable $\var$ and \crumblep{s} $\cell,\celltwo$.
	\end{enumerate}
\end{proposition}
\begin{proof}
	By induction on $\env$:
	\begin{itemize}
		\item Case $\env=\emptyenv$:
		\begin{enumerate}
			\item $\rbp{\var\,\molv,\emptyenv} = \rbp{\var\,\molv} = \rb\var\,\rb\molv = \rbp{\var,\emptyenv}\rbp{\molv,\emptyenv}$.
			\item $\rbp{\xite\var\cell\celltwo,\emptyenv} = \rbp{\xite\var\cell\celltwo} = (\xite{\rb\var}{\rb\cell}{\rb\celltwo}) = (\xite{\rbp{(\var,\emptyenv)}}{\rbp{(\append\cell\emptyenv)}}{\rbp{(\append\celltwo\emptyenv)}})$.
		\end{enumerate}
	  \item Case $\env=\envtwo\esub\vartwo\mol$ for some $\envtwo,\vartwo,\mol$:
		\begin{enumerate}
			\item By the definition of $\rbp\cdot$, $\rbp{\var\,\molv,\envtwo\esub\vartwo\mol} = \rbp{\var\,\molv,\envtwo}\isub\vartwo{\rb\mol}$.
			By \ih{} $\rbp{\var\,\molv,\envtwo}\isub\vartwo{\rb\mol} = (\rbp{\var,\envtwo}\rbp{\molv,\envtwo})\isub\vartwo{\rb\mol}$.
			By the definition of substitution, \sloppy $(\rbp{\var,\envtwo}\rbp{\molv,\envtwo})\isub\vartwo{\rb\mol} = (\rbp{\var,\envtwo}\isub\vartwo{\rb\mol})(\rbp{\molv,\envtwo}\isub\vartwo{\rb\mol})$.
			Again by the definition of $\rbp\cdot$, $(\rbp{\var,\envtwo}\isub\vartwo{\rb\mol})(\rbp{\molv,\envtwo}\isub\vartwo{\rb\mol})=\rbp{\var,\env}\rbp{\molv,\env}$ and we conclude.

			\item By the definition of $\rbp\cdot$, $\rbp{\xite\var\cell\celltwo,\envtwo\esub\vartwo\mol} = \rbp{\xite\var\cell\celltwo,\envtwo}\isub\vartwo{\rb\mol}$.
			By \ih{} $\rbp{\xite\var\cell\celltwo,\envtwo}\isub\vartwo{\rb\mol} = (\xite{\rbp{\var,\envtwo}}{\rbp{\append\cell\envtwo}}{\rbp{\append\celltwo\envtwo}})\isub\vartwo{\rb\mol}$.
			By the definition of substitution, \sloppy $(\xite{\rbp{\var,\envtwo}}{\rbp{\append\cell\envtwo}}{\rbp{\append\celltwo\envtwo}})\isub\vartwo{\rb\mol} =
			(\xite{\rbp{\var,\envtwo}\isub\vartwo{\rb\mol}}{\rbp{\append\cell\envtwo}\isub\vartwo{\rb\mol}}{\rbp{\append\celltwo\envtwo}\isub\vartwo{\rb\mol}})$.
			Again by the definition of $\rbp\cdot$, $(\xite{\rbp{\var,\envtwo}\isub\vartwo{\rb\mol}}{\rbp{\append\cell\envtwo}\isub\vartwo{\rb\mol}}{\rbp{\append\celltwo\envtwo}\isub\vartwo{\rb\mol}})=\xite{\rbp{\var,\env}}{\rbp{\append\cell\env}}{\rbp{\append\celltwo\env}}$ and we conclude.
			\qedhere
		\end{enumerate}
	\end{itemize}
\end{proof}

\begin{lemma}\label{l:aux-aux-lookup-open-two}
  Let $\env$ be an environment such that $\var\not\in\domain\env$.
  Then $\rbp{\var, \env \esub\var\mol \envtwo} = \rbp{\mol,\envtwo}$ for every $\mol,\envtwo$.
\end{lemma}
\begin{proof}
	By \reflemma{disj-c-e}, $\rbp{\var, \env} = \rbp{\var,\emptyenv}$ because $\var \notin \domain\env$. By \reflemma{aux-aux-lookup-open},
	$\rbp{\var, \env\esub\var\mol} = \rbp{\var,\esub\var\mol} = \var\isub\var{\rb\mol} = \rb\mol = \rbp{\mol,\emptyenv}$. Again by \reflemma{aux-aux-lookup-open},
	$\rbp{\var, \env \esub\var\mol \envtwo} = \rbp{\mol,\envtwo}$.
\end{proof}


\begin{lemma}\label{l:plug-append}
		For all \crumble context $\cctx$, \crumblep $\cell$, and environment $\env$, one has $\append{\cctxp\cell}\env = \cctxp{\append\cell\env}$.
\end{lemma}

\begin{proof}
	By cases according to the definition of the \crumble context $\cctx$.

	If $\cctx \defeq \ctxhole$, then $\append{\cctxp\cell}\env = \append{\cell}{\env} = \cctxp{\append\cell\env}$.

	Otherwise $\cctx \defeq (\mol, \envtwo\esub{\var}{\ctxhole})$;
	let $\cell \defeq (\moltwo,\envthree)$;
	then, $\append\cell\env = (\moltwo, \envthree\env)$ and hence $\append{\cctxp\cell}\env = \append{(\mol, \envtwo\esub{\var}{\moltwo}\envthree)}{\env} = (\mol, \envtwo\esub{\var}{\moltwo}\envthree\env) = \append{(\mol, \envtwo\esub{\var}{\moltwo})}{\envthree\env} = \cctxp{\append\cell\env}$.
\end{proof}

\begin{corollary}[Read back vs.~\crumble contexts]
	\label{c:rb-ctx}
			Let $\cell$ be a \crumblep and $\cctx, \cctxtwo$ be \crumble contexts.

			\begin{enumerate}
				\item\label{p:rb-ctx-cell}
				\emph{Plugging:}
				If $\cctx\disj\cell$ and 
				$\rb{\cctx}$ is a $\lambda$-context and $\cctxp{\cell}$ is well-named,	then $\rb{\cctxp\cell} = \rb{\cctx}\ctxholep{\rb\cell}$.
				\item\label{p:rb-ctx-comp}
				\emph{Composition:}
				If $\cctx\disj\cctxtwo$ and $\cctxp{\cctxtwo}$ is well-named, where $\rb{\cctx}$ and $\rb{\cctxtwo}$ are \lcontext{}s, then $\rb{\cctxp{\cctxtwo}} = \rb{\cctx}\ctxholep{\rb{\cctxtwo}}$, which is a \lcontext.
			\end{enumerate}
\end{corollary}

\begin{proof}
	\begin{enumerate}
		\item 	If $\cctx \defeq \ctxhole$, then $\rb{\cctx} =  \ctxhole$ and so $\rb{\cctxp\cell} = \rb{\cell} = \rb{\cctx}\ctxholep{\rb\cell}$.
		Otherwise $\cctx \defeq (\mol, \env\esub\var\ctxhole)$ with $\cell=(\moltwo,\envtwo)$;
		since $\cctx\ctxholep\cell = (\mol, \env\esub\var\moltwo\envtwo)$ is well-named, then $\var \notin \domain{\envtwo}$; therefore, from $\cctx\disj\cell$ it follows that $(\mol,\env)\disj\envtwo$;
		we have $\rb{\cctxp{\cell}} = \rb{(\mol,\env)}\isub\var{\rb{(\moltwo,\envtwo)}} \allowbreak= \rb{\cctx}\ctxholep{\rb\cell}$ by \reflemma{rb-comp-cell} and because by hypothesis $\cctx$ unfolds to a \lcontext.

		\item 	The composition of two \lcontext{s} is a \lcontext, thus $\rb{\cctx}\ctxholep{\rb{\cctxtwo}}$ is a \lcontext since $\rb{\cctx}$ and $\rb{\cctxtwo}$ are so.

		If $\cctx \defeq \ctxhole$, then $\rb{\cctx} =  \ctxhole$ and $\cctxp{\cctxtwo} = \cctxtwo$, thus $\rb{\cctxp\cctxtwo} = \rb{\cctxtwo} = \rb{\cctx}\ctxholep{\rb\cctxtwo}$.

		If $\cctxtwo \defeq \ctxhole$, then $\rb{\cctxtwo} = \ctxhole$ and $\cctxp{\cctxtwo} = \cctx$, so $\rb{\cctxp{\cctxtwo}} = \rb{\cctx} = \rb{\cctx}\ctxholep{\rb\cctxtwo}$.

		Finally, if $\cctx \defeq (\mol, \env\esub\var\ctxhole)$ and $\cctxtwo = (\moltwo, \envtwo\esub\vartwo\ctxhole)$, then $\cctxp{\cctxtwo} = (\mol, \env \esub{\var}{\moltwo}\envtwo\esub{\vartwo}{\ctxhole}) $ and so
		\begin{align*}
		\rb{\cctxp{\cctxtwo}} & =  \rb{(\mol,\env\esub\var\moltwo\envtwo)}\isub\vartwo{\ctxhole} & \\
		& = \rb{(\mol,\env)}\isub\var{\rb{(\moltwo,\envtwo)}}\isub\vartwo{\ctxhole} &  \text{by \reflemma{rb-comp-cell}} \\
		& = \rb{(\mol,\env)}\isub\var{\rb{(\moltwo,\envtwo)}\isub\vartwo{\ctxhole}} & \text{as $\vartwo \notin\fv{\rb{(\mol,\env)}}$} \\
		& = \rb{\cctx}\ctxholep{\rb{\cctxtwo}}
		\end{align*}
		where \reflemma{rb-comp-cell} can be applied because $(\mol, \env) \disj \envtwo$, since $\cctx \disj \cctxtwo$ and $\var \notin \domain{\envtwo}$ (as $\cctxp{\cctxtwo}$ is well-named); and $\vartwo \notin \fv{\rb{(\mol, \env)}}$ because $\cctx \disj \cctxtwo$.
		\qedhere
	\end{enumerate}
\end{proof}


\begin{lemma}
	\label{l:dec-c-fvs}
  Let $\cell=\cctxp{(\mol,\env)}$ be a \crumblep.
  Then $\fv\mol,\fv\cctx \subseteq \domain\env \cup \fv\cell $.
\end{lemma}

\begin{proof}
	By cases according to the definition of the \crumble context $\cctx$.

	If $\cctx \defeq \ctxhole$ then $\fv{\cctx} = \emptyset \subseteq \domain{\env} \cup \fv{\cell}$  and $\cell = (\mol,\env)$, so $\fv{\cell} = (\fv{\mol} \smallsetminus \domain{\env}) \cup \fv{\env}$ and hence $\fv{\mol} \subseteq (\fv{\cell} \smallsetminus \fv{\env}) \cup \domain{\env} \subseteq \fv{\cell} \cup \domain{\env}$.

	Otherwise $\cctx \defeq (\mol, \envtwo\esub{\var}{\ctxhole})$ and then $\fv{\cctx} = \fv{\moltwo} \cup (\domain{\envtwo} \smallsetminus \{\var\})$ and $\cell = (\tmtwo, \envtwo \esub{\var}{\mol} \env)$; therefore, $\fv{\cell} = \fv{\cctx} \cup (\fv{\mol} \smallsetminus \domain{\env}) \cup \fv{\env}$ and hence $\fv{\cctx} \subseteq \domain{\env}\cup \fv{\cell}$ and $\fv{\mol} \subseteq \domain{\env} \cup \fv{\cell}$.
\end{proof}

\begin{remark}
	\label{rmk:rb-append}
	For every \crumblep $\cell$, one has $\rb{\append{\cell}{\esub{\var}{\ctxhole}} } = \rb{\cell}\isub{\var}{\ctxhole} $.
	Indeed, let $\cell \defeq (\mol, \env)$: then, $\rb{\append{(\mol, \env)}{\esub{\var}{\ctxhole}} } = \rb{(\mol, \env\esub{\var}{\ctxhole}) } = \rb{(\mol, \env)}\isub{\var}{\ctxhole} $.
\end{remark}

\begin{proofof}[\reflemma{basecases-both}]
	\Paste{l:basecases-both}
\end{proofof}

\begin{proof}~
	\begin{enumerate}
		\item It follows immediately from the freshness condition in the definition of translation.

		\item By \refrmk{preservation-fvs}.

		\item It follows immediately from the freshness condition in the definition of translation.

		\item By induction on $\tm$ and by cases on the rules defining the translation.

		\item By induction on the size of $\tm$. Cases:
		\begin{itemize}
			\item \emph{$\l$-Value}, \ie $\tm = \val$.
			Then $\mytr{\tm} = (\auxtr{\val}, \emptyenv)$ and so  the only possible \crumble context $\cctx$ such that $\mytr{\tm} = \cctxp{\cell}$ for some \crumblep $\cell$ is $\cctx = \ctxhole$ 
			and so $\rb{\cctx} = \ctxhole$, which is a right \valuectx{}.

			\item \emph{$\l$-Value applied to \lav}, \ie $\tm = \val\valtwo$. As in the previous case.

			\item \emph{Application applied to a \lav}, \ie $\tm = \tmtwo \val$ with $\tmtwo$ not a $\lambda$-value.
			Then, $\mytr\tm = (\var \auxtr\val, \esub\var\tmthree\env)$ with $\mytr\tmtwo = (\tmthree,\env)$ and $\var$ fresh.
			Cases for $\cctx$:
			\begin{itemize}
				\item \emph{Empty}, \ie $\cctx = \ctxhole$: as for \lav{s} (see above).

				\item \emph{Non-empty}, \ie $\cctx = (\var \auxtr\val, \esub\var\ctxhole) \ctxholep\cctxtwo$ where $\cctxtwo$ is a \crumble context of 
				$\mytr{\tmtwo}$, \ie $\mytr{\tmtwo} = \cctxtwop{\cell}$.
				The read-back of the \crumble context $\cctxthree \defeq (\var \auxtr\val, \esub\var\ctxhole)$ is $\rb{\cctxthree} = \rb{\var \auxtr\val} \isub{\var}{\ctxhole} = \ctxhole \rb{\auxtr\val}$, which is a right \valuectx{} because $\rb{\auxtr{\val}}$ is a $\lambda$-value by \refrmk{preservation-fvs}{val-to-val}.
				By \ih, $\rb{\cctxtwo}$ is a right \valuectx{}.
				By the freshness conditions in the definition of translation, $\cctxthree\disj\cctxtwo$;
				according to \refcorollaryp{rb-ctx}{comp}, $\rb\cctx = \rb{\cctxthreep\cctxtwo} = \rb{\cctxthree}\ctxholep{\rb\cctxtwo}$, which is a right \valuectx{} since the composition of right \valuectx{} is a right \valuectx{} (\reflemma{r-ctxs-comp-closed}).
			\end{itemize}
			\item\emph{$\l$-Term applied to application}, \ie $\tm = \tmtwo\tmthree$ with $\tmthree$ not a $\lambda$-value.
			Then, $\mytr{\tm} = \append{\mytr{\tmtwo \var}}{\esub{\var}{\mol}}\env$ where $\mytr{\tmthree} = (\mol, \env)$ and $\var$ is fresh.
			Cases for $\cctx$:
			\begin{itemize}
				\item $\cctx$ is a \crumble context of $\mytr{\tmtwo\var}$, \ie~$\mytr{\tmtwo\var}=
				\cctxp{\cell}$, then $\rb{\cctx}$ is a right \valuectx{} by \ih (the size of the \lat $\tmtwo \var$ is strictly less than the size of $\tmtwo\tmthree$ because $\tmthree$ is not a \lav).
				\item $\cctx= \cctxthree\ctxholep{\cctxtwo}$ where $\cctxthree\defeq\append{\mytr{\tmtwo \var}}{\esub{\var}{\ctxhole}}$ and $\cctxtwo$ is a \crumble context of $\tmthree$ \ie~$\mytr\tmthree=
				\cctxtwop{\cell}$.
				By the freshness condition in the definition of translation and according to \refrmk{rb-append},  $\rb{\cctxthree} = \rb{\mytr{\tmtwo\var}}\isub{\var}{\ctxhole} = \rb{\mytr{\tmtwo}}\ctxhole$, which is a right \valuectx{}.
				By \ih, $\rb{\cctxtwo}$ is a right \valuectx{}.
				According to \refcorollaryp{rb-ctx}{comp}, $\rb\cctx = \rb{\cctxthreep\cctxtwo} = \rb{\cctxthree}\ctxholep{\rb\cctxtwo}$, which is a right \valuectx{} since the composition of right \valuectx{s} is a right \valuectx{} (\reflemma{r-ctxs-comp-closed}).
			\end{itemize}
      
      \item \emph{Conditional}, case $\tm\defeq(\xite\val\tmtwo\tmthree)$. Then $\mytr\tm=(\xite{\auxtr\val}{\mytr\tmtwo}{\mytr\tmthree},\emptyenv)$. Necessarily $\cctx=\ctxhole$, and hence $\rb\cctx$ is a right \valuectx{}.

      \item \emph{Conditional}, case $\tm\defeq(\xite\tmtwo\tmthree\tmfour)$ with $\tmtwo$ not a \lav. Then $\mytr\tm=(\xite\var{\mytr\tmthree}{\mytr\tmfour},\esub\var\tmfive\env)$ where $\mytr\tmtwo \eqdef (\tmfive,\env)$. We proceed by cases on $\cctx$:
      \begin{itemize}
        \item Case $\cctx=\ctxhole$. Then clearly $\rb\cctx$ is a right \valuectx{}.
        \item Case $\cctx=\cctxthreep\cctxtwo$, where $\cctxthree\defeq (\xite\var{\mytr\tmthree}{\mytr\tmfour},\esub\var\ctxhole)$ and $\cctxtwo$ is a \crumble context of $\mytr\tmtwo$.
        By the freshness condition in the definition of translation and according to \refrmk{rb-append},  $\rb\cctxthree = (\xite\ctxhole\tmthree\tmfour)$, which is a right \valuectx{}.
				By \ih, $\rb{\cctxtwo}$ is a right \valuectx{}.
				According to \refcorollaryp{rb-ctx}{comp}, $\rb\cctx = \rb{\cctxthreep\cctxtwo} = \rb{\cctxthree}\ctxholep{\rb\cctxtwo} = (\xite{\rb\cctxtwo}\tmthree\tmfour)$, which is a right \valuectx{}.
        \qedhere
      \end{itemize}
		\end{itemize}
	\end{enumerate}
\end{proof}

\section{Proofs of \refsect{closed} (closed case)}


\subsection{Proofs of \refsubsect{micro-step-closed}}
\begin{lemma}[Closure under substitution]
	\label{l:closure-substitution-closed}
	Let $\val$ and $\valtwo$ be \lav{s}.
	Then $\val\isub{\var}{\valtwo}$ is a \lav.
	If moreover $\val$ is a $\lambda$-abstraction,
	then $\val\isub{\var}{\valtwo}$ is so.
\end{lemma}

\begin{proof}
        By cases on the definition of \lav:
	\begin{itemize}
		\item \emph{Variable,} \ie either $\val \defeq \var$ and then $\val\isub{\var}{\valtwo} = \valtwo$ which is a \lav by hypothesis; 
		or $\val \defeq \vartwo \neq \var$ and then $\val\isub{\var}{\valtwo} = \vartwo$ which is a \lav.
		
		\item \emph{Abstraction,} \ie $\val \defeq \la{\vartwo}{\tmtwo}$ and 
		we can suppose without loss of generality that $\vartwo \notin \fv{\val} \cup \{\var\}$;
		therefore, $\val\isub{\var}{\valtwo} = \la\vartwo{(\tmtwo\isub{\var}{\valtwo})}$ which is a $\lambda$-abstraction and hence a \lav.
                \item \emph{Booleans and errors,} \ie $\val \in \set{\true,\false,\err}$. Trivial since $\val\isub{\var}{\valtwo} = \val$.
		\qedhere
	\end{itemize}
\end{proof}

\begin{lemma}[Read-back to \lav]
	\label{l:rb-val-f-closed}
	For every \crumbled value $\molv$ and \lenvironment{} $\aenv$, one has that $\rbp{\molv,\aenv}$ is a \lav.
	If moreover $\molv$ is an abstraction, then $\rbp{\molv,\aenv}$ is a $\lambda$-abstraction.
\end{lemma}
\begin{proof} By induction on the length of $\aenv$.
 We proceed by cases on the shape of $\molv$.
	\begin{itemize}
		\item \emph{Abstraction:} 
		If $\aenv \defeq \emptyenv$, then clearly $\rbp{\molv,\aenv}=\rb\molv$ is a $\lambda$-abstraction and hence a \lav.
		Otherwise $\aenv \defeq \aenvtwo\esub\var\mol$ where $\mol$ is a \crumbled value (and hence $\rb{\mol}$ is a \lav): thus, $\rbp{\molv,\aenv} = \rbp{\molv,\aenvtwo}\isub\var{\rb\mol}$;
		by \ih{}, $\rb{(\molv,\aenvtwo)}$ is a $\lambda$-abstraction, thus $\rbp{\molv,\aenv}$ is a $\lambda$-abstraction (and so a 
		\lav) by \reflemma{closure-substitution-closed}.
                \item \emph{Booleans and errors}: the proof is identical
                 to the previous case.
		\item \emph{Variable:} If $\molv \defeq\var \notin \domain{\aenv}$, then $\rbp{\molv,\aenv} = \var$ which is a \lav; 
		otherwise $\molv \defeq \var \in \domain{\aenv}$ with
		$\aenv \defeq \aenvtwo\esub\var{\la\vartwo\cell}\aenvthree$, and then $\rbp{\var,\aenv} = \rbp{\la\vartwo\cell,\aenvthree}$ (since $\var \notin \domain{\aenvtwo}$) is a \lav by \ih, according to the previous point.
		\qedhere
	\end{itemize}
\end{proof}

\begin{lemma}\label{l:harmony-aux1-closed}
  Let $\cell=(\mol,\aenv)$ be a well-named closed \crumble,
	and $\mol$ have the following property:
	$\mol$ is a \lav, or $\mol$ is $\var$ or $\var\molv$ or $\xite\var\cell\celltwo$ but $\var$ is not defined in $\aenv$. Then $\cell$ is a \lcell{}.
\end{lemma}
\begin{proof}
  Let $\cell=(\mol,\aenv)$ as above: it suffices to prove that 
  $\mol$ is a \lav. This follows easily from the hypothesis that $\cell$ is closed, since the cases where $\mol$ is $\var$ or $\var\molv$ or $\xite\var\cell\celltwo$ but $\var$ is not defined in $\aenv$ are impossible.
\end{proof}

\begin{corollary}
	\label{coro:harmony-aux1-coro-closed}
	If the well-named closed \crumble $(\mol,\aenv)$ is normal, then it is a \lcell{}.
\end{corollary}

\begin{proofof}[\refprop{harmony-closed}]
	\Paste{prop:harmony-closed}
\end{proofof}
\begin{proof}~
  \begin{itemize}
    \item[$(\Rightarrow)$]
     Let $\cell=(\mol,\env)$ be well-named, closed, and normal. We proceed by structural induction on $\env$:
     \begin{itemize}
       \item if $\env=\emptyenv$, then $(\mol,\emptyenv)$ is a \lcell{} by \refcoro{harmony-aux1-coro-closed};
       \item if $\env=\esub\var\moltwo\envtwo$, then also the \crumble $(\moltwo,\envtwo)$ is normal. 
       By \ih{} $(\moltwo,\envtwo)$ is a \lcell{}, and therefore $\env=\esub\var\moltwo\envtwo$ is a \lenvironment. By \refcoro{harmony-aux1-coro-closed}, $\cell=\mol,\env$ is a \lcell{}.
     \end{itemize}
     \item[$(\Leftarrow)$]
      Let $\cell=\acell$, we need to prove that $\acell$ is normal.
      Let $\acell=\cctxp{(\pval,\aenv)}$ for some $\cctx,\pval,\aenv$.
      Clearly no reduction rule is applicable, because $\pval$ is not a variable or an application.
      \qedhere
  \end{itemize}
\end{proof}

\subsection{Proofs of \refsubsect{the-implementation-theorem}}

\begin{proofof}[\reflemma{invariants-closed}]
	\Paste{l:invariants-closed}
\end{proofof}
\begin{proof}
	\newcommand{\tmpitem}[1]{\item[\ref{p:invariants-closed-#1}.] }
	By induction on the length of the reduction sequence leading to the \crumble. 
	The base cases hold by \reflemma{basecases-both} (by noting that for \refpoint{invariants-closed-ctx-decoding}, \reflemmap{basecases-both}{unfold-right} implies the weaker statement \reflemmap{invariants-closed}{ctx-decoding}).
	As for the inductive cases, we inspect each transition:
	\begin{itemize}
		\tmpitem{fresh}
		The substitution transitions $\evsym, \elsym, \eitesym$ do not change the set of variables occurring on the lhs of substitutions outside abstractions because they copy a value that does not contain any. Hence the claim follows from the \ih{}. 
		For transition $\msym$ the claim follows from the side condition. For the remaining rules $\iftsym,\iffsym,\ifesym,\apesym$ the claim follows from the fact that all substitutions outside abstractions in the rhs alredy occur in the lhs.
		\tmpitem{fvs}
		The substitution transitions $\evsym,\elsym,\eitesym$ do not change the domain of the \crumble and only copy to the left a value from the environment,
		and the claim follows from the \ih{}.
		
		Transition $\msym$ copies to the top level and renames the body of an abstraction.
		By the properties of $\alpha$-renaming $\fv{(\append\cell{\esub\var\molv})^\alpha} = \fv{\append\cell{\esub\var\molv}} = \fv{\la\var\cell}$,
		and since by \ih{} $\fv{\la\var\cell} \subseteq \domain\aenv$, we can conclude with $\fv{(\append\cell{\esub\var\molv})^\alpha} \subseteq \domain\aenv$.

                The terms in the rhs of the remaining transitions $\iftsym,\iffsym,\ifesym,\apesym$ already occur in the lhs under the same environment. Therefore the claim follows from the \ih{}.
		
		\tmpitem{abstractions}
		The rules $\evsym,\elsym,\eitesym$ may copy an abstraction, but the abstraction was already in the environment,
		and the claim follows from the \ih{}.
		The rule $\msym$ copies and renames the body of an abstraction that was already in the environment,
		and the claim follows from the \ih{} since the translation commutes with the renaming of free variables (\refrmkp{preservation-fvs}{commutation-fvs}).
                All the bodies in the rhs of the remaning rules $\iftsym,\iffsym,\ifesym,\apesym$ already occur in the lhs and therefore the claim follows from the \ih{}.

		\tmpitem{ctx-decoding}
		Let $\mytr\molthree \Rew{}^n \cctxtwop{(\moltwo, \aenvtwo)} \Rew{a} \cctx\ctxholep{(\mol, \aenv)}$ (where $\mol$ is not a \pvalue).
		Cases of the reduction step $\cctxtwop{(\moltwo, \aenvtwo)} \Rew{a} \cctx\ctxholep{(\mol, \aenv)}$:
		
		\begin{itemize}
			\item Case $\msym$: $\cctxtwop{((\la\var\cell)\molv, \aenv)} \tom \cctxtwop{\append{\cell^\alpha}{(\esub{\var^\alpha}\molv\aenv)}}$.
			
			Let $\cctxthree$ be a prefix of $\cctx$. There are two sub-cases:
			\begin{itemize}
				\item \emph{$\cctxthree$ is a prefix of $\cctxtwo$}: by \ih{} $\rb{\cctxthree}$ is a right \valuectx{}.
				
				\item \emph{$\cctxtwo$ is a prefix of $\cctxthree$},
				\ie{} $\cctxthree=\cctxtwop{\cctxfour}$ and $\cell^\alpha=\cctxfour\ctxholep{\cell'}$.
				By \reflemmap{basecases-both}{abstractions} and \reflemmap{invariants-closed}{abstractions} $\cell$ is the translation of a $\lambda$-term, by \refrmkp{preservation-fvs}{commutation-fvs} $\cell^\alpha$ is so, and thus by \reflemmap{basecases-both}{unfold-right} $\rb{\cctxfour}$ is a right \valuectx{}.
				By \ih, $\rb\cctxtwo$ is a right \valuectx{} as well.
				Since $\rb{\cctxthree} = \rb{\cctxtwo}\ctxholep{\rb{\cctxfour}}$ according to \refcorollaryp{rb-ctx}{comp},  we obtain that $\rb{\cctxthree}$ is a right \valuectx{} as composition of right \valuectx{s} (\reflemma{r-ctxs-comp-closed}).
			\end{itemize}

			\item Case $\iftsym$: $\cctxtwop{(\xite\true\cell\celltwo, \aenv)} \tom \cctxtwop{\append{\cell}{\aenv}}$.
			Let $\cctxthree$ be a prefix of $\cctx$. There are two sub-cases:
			\begin{itemize}
				\item \emph{$\cctxthree$ is a prefix of $\cctxtwo$}: by \ih{} $\rb{\cctxthree}$ is a right \valuectx{}.
				
				\item \emph{$\cctxtwo$ is a prefix of $\cctxthree$},
				\ie{} $\cctxthree=\cctxtwop{\cctxfour}$ and $\cell=\cctxfour\ctxholep{\cell'}$.
				By \reflemmap{basecases-both}{abstractions} and \reflemmap{invariants-closed}{abstractions} $\cell$ is the translation of a $\lambda$-term and thus by \reflemmap{basecases-both}{unfold-right} $\rb{\cctxfour}$ is a right \valuectx{}.
				By \ih, $\rb\cctxtwo$ is a right \valuectx{} as well.
				Since $\rb{\cctxthree} = \rb{\cctxtwo}\ctxholep{\rb{\cctxfour}}$ according to \refcorollaryp{rb-ctx}{comp},  we obtain that $\rb{\cctxthree}$ is a right \valuectx{} as composition of right \valuectx{s} (\reflemma{r-ctxs-comp-closed}).
			\end{itemize}

			\item Case $\iffsym$: identical to the previous case.
			
			\item Cases $\evsym,\ifesym,\apesym$: they follow from the \ih{} since $\cctx$ is necessarily a prefix of $\cctxtwo$ because $\mol$ is a \pvalue.
			
			\item Cases $\elsym$ and $\eitesym$: they follow from the \ih{},
			since $\aenvtwo=\aenv$ and $\cctx=\cctxtwo$.
			\qedhere
		\end{itemize}
	\end{itemize}
\end{proof}

\begin{lemma}[Determinism]
  \label{l:determinism-closed}
  $\tocrumble$ is deterministic.
\end{lemma}
\begin{proof}
  Assume that there exists a \crumble that may be
  decomposed in two ways $\cctxp{(\mol, \aenv)} = \cctxtwop{\moltwo, \aenvtwo}$ such that they reduce respectively $\cctxp{(\mol, \aenv)} \Rew a \cctxp\cell$ and $\cctxtwop{(\moltwo, \aenvtwo)} \Rew b \cctxtwop\celltwo$ with rules $a,b\in\set{\msym, \iftsym, \iffsym, \ifesym, \apesym, \evsym, \elsym, \eitesym}$.
  
  We prove that it must necessarily be $a=b$, $\cctx=\cctxtwo$, and $\cell=\celltwo$ (up to alpha).
  Three cases:
  \begin{itemize}
    \item $\cctx$ strict initial segment of $\cctxtwo$, \ie{} $\cctxtwo = \cctxp\cctxthree$ for some $\cctxthree\neq\ctxhole$.
    We show that this case is not possible: in fact,
    it follows that $ \aenv = \envctxp{(\moltwo, \aenvtwo)} $ for some $\envctx$, thus
    $(\moltwo, \aenvtwo)$ is a \lcell{}, and by \refprop{harmony-closed} it must be normal, contradicting the hypothesis that $(\moltwo,\aenvtwo)$ and $\cell$ reduce with rule $b$.
    \item $\cctx = \cctxtwo$.
     By inspection of the reduction rules, $a=b$: in fact the rule $\msym$ applies only when $\mol$ is the application of an abstraction to a \crumbled value, the rule $\evsym$ only when $\mol$ is a variable, and the rule $\elsym$ only when $\mol$ is the application of a variable to a \crumbled value, etc.
     It remains to show that $\cell=\celltwo$ (up to alpha): this follows from the determinism of the lookup in the environment during $\evsym,\elsym$ and $\eitesym$ reductions.
     
    \item $\cctxtwo$ initial segment of $\cctx$, \ie{} $\cctx = \cctxtwop\cctxthree$. Symmetric to the first case.
    \qedhere
  \end{itemize}
\end{proof}

%
%


\begin{lemma}
	\label{l:fvt-domc-closed}
	In every reachable \crumble $\cctxp{(\mol,\env)}$ one has $\fv{\mol} \cap \domain\cctx = \emptyset$.
\end{lemma}
\begin{proof}
	By \reflemma{dec-c-fvs} and \reflemmasps{invariants-closed}{fresh}{fvs}.
\end{proof}

\begin{proposition}[Overhead transparency]
	\label{prop:transparency-closed}
  Let $\cell$ be a reachable \crumble, and let $a \in \set{\evsym,\elsym,\eitesym}$.
  If $ \cell \Rew{a} \celltwo $ then $\rb\cell = \rb\celltwo$.
\end{proposition}
\begin{proof}  
	Let $\cell\defeq\cctxp{(\mol,\aenv)} \Rew a \cctxp{(\moltwo, \aenv)} \eqdef \celltwo $, and let $\aenvtwo,\aenvthree$ such that $\aenv = \aenvtwo\esub\var{\aenv(\var)}\aenvthree$, noting that $\var$ does not occur in $\aenvthree$ by \reflemmap{invariants-closed}{fresh} and \reflemma{fvt-domc-closed}.
	We first prove that $\rbp{\mol,\aenv} = \rbp{\moltwo,\aenv}$:
	\begin{itemize}
		\item Case $\evsym$, \ie{} $\mol \defeq \var$ and $\moltwo = \aenv(\var)$. 
		By \reflemma{rb-comp-cell-0}, $\rbp{\var,\aenvtwo\esub\var{\aenv(\var)}\aenvthree} = \rbp{\var,\aenvtwo\aenvthree}\isub\var{\rbp{\aenv(\var),\aenvthree}} = \rbp{\aenv(\var),\aenvthree}$
		as $\cell$ is well-named (\reflemmap{invariants-closed}{fresh}).
		By \reflemma{fvt-domc-closed}, $\fv{\aenv(\var)}\cap\domain{\aenvtwo\esub\var{\aenv(\var)}} = \emptyset$, therefore $\rbp{\aenv(\var),\aenvthree} = \rbp{\aenv(\var),\aenv}$, and we conclude with $\rbp{\var,\aenv}= \rbp{\aenv(\var),\aenv}$.
		
		\item Case $\elsym$, \ie{} $\mol \defeq \var\,\molv$ and $\moltwo = \aenv(\var)\,\molv$. 
		Since $\rbp{\var\,\molv,\aenv} = \rbp{\var,\aenv}\rbp{\molv,\aenv}$ (\refprop{aux-rb-structural}), we can use the point above to conclude.
		\item Case $\eitesym$, \ie{} $\mol \defeq \xite\var\cell\celltwo$ and $\moltwo = \xite{\aenv(\var)}\cell\celltwo$. 
		Since $\rbp{\xite\var\cell\celltwo,\aenv} = \xite{\rbp{\var,\aenv}}{\rbp{\append\cell\aenv}}{\rbp{\append\celltwo\aenv}}$ (\refprop{aux-rb-structural}), we can use the point above to conclude.
	\end{itemize}
	
	We now prove that $\rb{\cctxp{(\mol,\aenv)}} = \rb{\cctxp{(\moltwo, \aenv)}} $ under the hypothesis that $\rbp{\mol,\aenv} = \rbp{\moltwo,\aenv}$. 
	By cases on $\cctx$: if $\cctx \defeq \ctxhole$ just use the hypothesis.
	Otherwise $\cctx \defeq (\molthree, \env\esub\var\ctxhole)$ and so
	$\rbp{\molthree, \env\esub\var\mol \aenv} = \rbp{\molthree,\env\aenv} \isub\var{\rbp{\mol,\aenv}} = \rbp{\molthree,\env\aenv} \isub\var{\rbp{\moltwo,\aenv}} \allowbreak= \rbp{\molthree, \env\esub\var\moltwo \aenv} $ by \reflemma{rb-comp-cell-0}.
\end{proof}

\begin{lemma}[Substitution]
	\label{l:substitution-closed}
	Let $\tm$ and $\tmtwo$ be \lat{s}, and $\val$ be a \lav.
	If $\tm \topif \tmtwo$ then $\tm\isub{\var}{\val} \allowbreak
	\topif \tmtwo\isub{\var}{\val}$.
\end{lemma}

\begin{proof}
	By induction on the definition of $\tm \topif \tmtwo$.
	Cases:
	\begin{itemize}
		\item \emph{Root-step}.
          \begin{itemize}
           \item \emph{$\betavsym$-step},
           \ie $\tm \defeq (\la{\vartwo}{\tmthree})\valtwo \rtobv \tmthree \isub{\vartwo}{\valtwo} \eqdef \tmtwo$ and we can suppose without loss of generality that $\vartwo \notin \fv{\val} \cup \{\var\}$. 
		According to \reflemma{closure-substitution-closed}, $\valtwo\isub{\var}{\val}$ is a \lav.
		As a consequence, $\tm\isub{\var}{\val} =\allowbreak (\la{\vartwo}{\tmthree\isub{\var}{\val}})(\valtwo\isub{\var}{\val}) \tobv \tmthree\isub\vartwo{\valtwo\isub{\var}{\val}} = 
		\tmthree\isub\vartwo\valtwo\isub{\var}{\val}
		= \tmtwo\isub{\var}{\val}$.
           \item the \emph{$\iftsym,\iffsym,\ifesym,\apesym$ steps} are similar to the $\betavsym$-step. \reflemma{closure-substitution-closed} is used in the proof of the $\ifesym$-step to prove that a $\ifesym$-redex where the guard is an abstraction is mapped to a $\ifesym$-redex of the same kind.
           \end{itemize}
		
		\item \emph{Application right}, \ie $\tm \defeq \tmthree\tmfour \topif \tmthree\tmfive \eqdef \tmtwo$ with $\tmfour \topif \tmfive$;
		by \ih $\tmfour\isub{\var}{\val} \topif \tmfive\isub{\var}{\val}$, and therefore $\tm\isub{\var}{\val} = \tmthree\isub{\var}{\val} (\tmfour\isub{\var}{\val}) \topif \tmthree\isub{\var}{\val} (\tmfive\isub{\var}{\val}) = \tmtwo\isub{\var}{\val}$.
		
		\item \emph{Application left}, \ie $\tm \defeq \tmthree\valtwo \topif \tmfour\valtwo \eqdef \tmtwo$ with $\tmthree \topif \tmfour$; 
		by \ih, $\tmthree\isub{\var}{\val} \topif \tmfour\isub{\var}{\val}$;
		according to \reflemma{closure-substitution-closed}, $\valtwo\isub{\var}{\val}$ is a \lav and hence $\tm\isub{\var}{\val} = \tmthree\isub{\var}{\val}(\valtwo\isub{\var}{\val}) \topif \tmfour\isub{\var}{\val} (\valtwo\isub{\var}{\val}) = \tmtwo\isub{\var}{\val}$.

		\item \emph{If-then-else guard}, \ie $\tm \defeq \xite\tmfour\tmthree\tmsix \topif \xite\tmfive\tmthree\tmsix \eqdef \tmtwo$ with $\tmfour \topif \tmfive$;
		by \ih $\tmfour\isub{\var}{\val} \topif \tmfive\isub{\var}{\val}$, and therefore $\tm\isub{\var}{\val} = \xite{\tmfour\isub\var\val}{\tmthree\isub\var\val}{\tmsix\isub\var\val} \topif \xite{\tmfive\isub\var\val}{\tmthree\isub\var\val}{\tmsix\isub\var\val} = \tmtwo\isub{\var}{\val}$.
		\qedhere
	\end{itemize}
\end{proof}

%
%
%
%

\begin{lemma}
	\label{l:clos-sigma-abs-closed}
   Let $\cell,\celltwo$ be \crumble{s}, and let $\aenv$ be a \lenvironment{}.
   If $\rb\cell \topif{} \rb\celltwo$, then 
   $\rbp{\append\cell\aenv} \topif \rbp{\append\celltwo\aenv}$.
\end{lemma}
\begin{proof}
	By induction on the length of $\aenv$.
	 If $\aenv \defeq \emptyenv$ then $\rbp{\append\cell\aenv} = \rb{\cell} \topif \rb{\celltwo} = \rbp{\append\celltwo\aenv}$.
	 Otherwise $\aenv \defeq \aenvtwo \esub{\var}{\molv}$ where $\molv$ is a \pvalue (and hence $\rb{\molv}$ is a \lav); 
	 by \ih, $\rbp{\append\cell\aenvtwo} \topif \rbp{\append\celltwo\aenvtwo}$ and hence $\rbp{\append\cell\aenv} = \rbp{\append\cell\aenvtwo}\isub{\var}{\rb\molv} \topif \rbp{\append\celltwo\aenvtwo} \isub{\var}{\rb\molv} = \rbp{\append\celltwo\aenv}$ according to \reflemma{substitution-closed}.
\end{proof}

\begin{proposition}[Principal projection]
	\label{prop:beta-projection-closed}
  Let $\cell$ be a reachable \crumble. If $\cell \Rew{a} \celltwo$ for $a \in \set{\msym,\iftsym,\iffsym,\ifesym,\apesym}$ then $\rb\cell \Rew{a} \rb\celltwo$.
\end{proposition}
\begin{proof}
 Note that for every $\mol,\aenv$,  $(\mol,\aenv) = \append{(\mol,\epsilon)}{\aenv}$. Therefore all steps can be written in the form
 $\cctxp{(\mol,\aenv)} \Rew{a} \cctxp{\append\cell\aenv}$ where $\mol$ is not a
\crumbled value. The \crumble context $\cctx$ unfolds to a right \valuectx{} by \reflemmap{invariants-closed}{ctx-decoding}. We need to prove that
$\rb{\cctxp{(\mol,\aenv)}} \Rew{a} \rbp{\cctxp{\append\cell\aenv}}$. By Lemma~\ref{l:plug-append} and \reflemma{clos-sigma-abs-closed}, it suffices to prove that 
$\rb{\cctxp{(\mol,\epsilon)}} \Rew{a} \rb{\cctxp{\cell}}$.

We proceed by cases on the rule $\Rew{a}$.
\begin{itemize}
 \item Rule $\betavsym$: we need to prove that
	$\rb{\cctxp{((\la\var\cell)\,\molv,\emptyenv)}} \tobv{} \rb{\cctx\ctxholep{\append\cellthree{\esub\vartwo\molv}}}$ where
	$\append\cellthree{\esub\vartwo\molv} \defeq (\append\cell{\esub\var\molv})^\alpha$.
	\begin{align*}
	\rb{\cctxp{((\la\var\cell)\,\molv,\emptyenv)}} & = \rb\cctx\ctxholep{(\la\var\rb\cell)\,\rb\molv} & \text{by \refcorollaryp{rb-ctx}{cell}}\\
	& \alphaequiv{}
	\rb\cctx\ctxholep{(\la{\vartwo}\rb{\cellthree})\,\rb\molv} & \\
	& \tobv \rb\cctx\ctxholep{\rb{\cellthree}\isub{\vartwo}{\rb\molv}} & \\
	& = \rb\cctx\ctxholep{\rb{\append{\cellthree}{\esub{\vartwo}\molv}}} \\
	& = \rb{\cctx\ctxholep{\append{\cellthree}{\esub{\vartwo}\molv}}} & \text{by \refcorollaryp{rb-ctx}{cell}}
	\\
	& = \rb{\cctx\ctxholep{(\append{\cell}{\esub\var\molv})^\alpha}}. &
	\end{align*}
	Note that the second use of \refcorollaryp{rb-ctx}{cell} requires that
	$\cctx\disj(\append\cellthree{\esub\vartwo\molv})$ \ie that
	$\fv{\cctx} \cap \domain{\append\cellthree{\esub\vartwo\molv}} = \emptyset$,
	which follows from the side condition about $\alpha$-renaming in the $\msym$ rule.
 \item Rules $\iftsym,\iffsym,\ifesym,\apesym$:
  a quick check by cases over $(\mol,\aenv) \rootRew{a} \append\cell\aenv$ shows that $\rb\mol \Rew{a} \rb\cell$.

  For example,
  $(\xite\true\cell\celltwo,\aenv) \rootRew{\iftsym} \append\cell\aenv$ and\\
  $\rbp{\xite\true\cell\celltwo} = \xite\true{\rb\cell}{\rb\celltwo}
   \toift \rb\cell$. The other cases are all similar.

  Thus
	\begin{align*}
	  \rb{\cctxp{(\mol,\epsilon)}} & = \rb\cctx\ctxholep{\rb\mol} & \text{by \refcorollaryp{rb-ctx}{cell}}\\
        & \Rew{a} \rb\cctx\ctxholep{\rb\cell} \\
        & = \rb{\cctx\ctxholep{\cell}} & \text{by \refcorollaryp{rb-ctx}{cell}}\\
        \end{align*}
  Note that the second use of \refcorollaryp{rb-ctx}{cell} requires that
  $\cctx\disj \cell$. The property holds because all substitutions outside
  abstractions in the rhs of the rules $\Rew{a}$ under consideration were such in the lhs,
  and because reachable \crumblep{s} are well-named (\reflemmap{invariants-closed}{fresh}).
  \qedhere
\end{itemize}
\end{proof}

\begin{lemma}[Halt]
  \label{l:progress-closed}
   Let $\cell$ be a closed \crumble.
   If $\cell$ is $\crumblesym$\!-normal then $\rb\cell$ is $\pif$-normal.
\end{lemma}
\begin{proof}
  By \refprop{harmony-closed}, if $\cell$ is normal then it is a \lcell{} \ie{} $\cell=\acell$.
  By \reflemma{rb-val-f-closed}, $\rb\cell$ is a \lav.
	By \pif harmony (\refprop{pif-harmony}), $\rb\cell$ is $\topif$-normal.
\end{proof}

\begin{proofof}[\refthm{implementation-closed}]
\Paste{thm:implementation-closed}
\end{proofof}
\begin{proof}
	\begin{enumerate}
		\item See \refprop{read-back-inverse-transl}.
		\item See \refprop{beta-projection-closed}.
		\item See \refprop{transparency-closed}.
		\item See \reflemma{determinism-closed}.
		\item See \reflemma{progress-closed}, since $\cell$ is closed by \reflemmap{invariants-closed}{fvs}.
		\item Immediate consequence of \reflemma{linear-bounds} (proved independently).
		\qedhere
	\end{enumerate} 
\end{proof}

\subsection{Proofs of \refsubsect{complexity-closed}}

%

\begin{proofof}[\refprop{mytr_constants_new}]
	\Paste{prop:mytr_constants_new}
\end{proofof}
\begin{proof}
By mutual induction on $\tm$ and $\val$:
	\begin{itemize}
		\item \emph{Variable}, \ie $\val = \tm \defeq \var$\,: then $\auxtr{\val} = \var$ and $\mytr{\tm} = (\var, \emptyenv)$, hence $\lenv{\auxtr\val}=\lenv{\mytr\tm}=1$.

		\item \emph{Boolean or error}, \ie $\val = \tm \in \set{\true,\false,\err}$: then $\auxtr{\val} = \val$ and $\mytr{\tm} = (\val, \emptyenv)$, hence $\lenv{\auxtr\val}=\lenv{\mytr\tm}=0$.
		
		\item \emph{Abstraction}, \ie $\val \defeq \la{\var}{\tmtwo} \eqdef \tm$\,: then $\auxtr{\val} = \la{\var}{\mytr{\tmtwo}}$ and $\mytr{\tm} = (\la{\var}{\mytr{\tmtwo}}, \emptyenv)$;
		by \ih, $\lenv{\mytr\tmtwo} \leq 1$. We have $\lenv{\auxtr\val}=\lenv{\mytr\tm}=0$ by definition.
		
		\item \emph{Application of two $\lambda$-values}, \ie $\tm \defeq \valtwo\valthree$\,: then, $\mytr{\tm} = (\auxtr\valtwo\,\auxtr\valthree,\emptyenv)$.
		We have $\lenv{\mytr\tm}=0$ by definition.

		\item \emph{Conditional on a \lav}, \ie $\tm \defeq \xite\valtwo\tmtwo\tmthree$\,: then, $\mytr{\tm} = (\xite{\auxtr\valtwo}{\mytr\tmtwo}{\mytr\tmthree},\emptyenv)$.
		We have $\lenv{\mytr\tm}=0$ by definition.
				
		\item \emph{Application of a non-$\lambda$-value to a $\lambda$-value}, \ie $\tm \defeq \tmtwo\val$ where $\tmtwo$ is not a \lav : then, $\mytr{\tm} = (\var\auxtr\val, \esub\var\mol \env)$ where $\mytr{\tmtwo} = (\mol, \env)$;
		by \ih, $\lenv{\mytr{\tmtwo}}=0$.
		Note that $\lenv{\mytr{\tm}} = \lenv{(\var\auxtr\val, \esub\var\mol \env)} = \lenv\mol + \lenv \env = \lenv{(\mol, \env)} = \lenv{\mytr{\tmtwo}} =_{\ih} 0$.
		
		\item \emph{Application of a \lat to a non-\lav}, \ie $\tm \defeq \tmtwo\tmthree$ where $\tmthree$ is not a \lav:
		then, $\mytr{\tm} = \append{\mytr{\tmtwo\var}}{(\esub{\var}{\mol}\env)} = (\vartwo\var, \esub{\vartwo}{\moltwo}\envtwo\esub{\var}{\mol}\env)$ where $\mytr{\tmthree} = (\mol, \env)$ and $\mytr{\tmtwo\var} = (\vartwo\var, \esub{\vartwo}{\moltwo}\envtwo)$ with $\mytr{\tmtwo} = (\moltwo, \envtwo)$.
		By \ih, $\lenv{\mytr{\tmtwo \var}} = \lenv{\mytr{\tmthree}} = 0$.
		Since $\lenv{\mytr{\tm}} = \lenv{\mytr{\tmtwo}} + \lenv{\mytr\tmthree}$, we have $\lenv{\mytr{\tm}} = 0$.
		
				\item \emph{Conditional on a non-\lav}, \ie $\tm \defeq \xite\tmtwo\tmthree\tmfour$ where $\tmtwo$ is not a \lav\,: then, $\mytr{\tm} = (\xite{\var}{\mytr\tmthree}{\mytr\tmfour},\esub\var\mol\env)$ with $\mytr\tmtwo = (\mol,\env)$. By \ih, $\lenv{\mytr\tmtwo}=0$. Then 
				$$\lenv{\mytr\tm}=\lenv{(\xite{\var}{\mytr\tmthree}{\mytr\tmfour},\esub\var\mol\env)} = \lenv\mol + \lenv\env = \lenv{(\mol,\env)} = \lenv{\mytr\tmtwo} =_{\ih} 0$$

		\qedhere
	\end{itemize}
\end{proof}

We define the size $\len\cdot$ of \crumbled forms, \crumble{s} and environments as expected:
\begin{align*}
\len{(\mol,\env)} & \defeq \len\mol + \len\env &
\len\emptyenv & \defeq 0 &
\len{\env\esub\var\mol} & \defeq 1 + \len\env + \len\mol \\
\len\var & \defeq 1 &
\len{\la\var\cell} & \defeq 1 + \len\cell &
\len{\molv\molvtwo} & \defeq 1 + \len\molv+\len\molvtwo \\
\len{\true} & \defeq 1 & \len{\false} & \defeq 1 & \len{\err} & \defeq 1
\end{align*}

\vspace{-2\baselineskip}
\begin{align*}
\len{\xite\val\cell\celltwo} & \defeq 1 + \len{\val} + \len{\cell} + \len{\celltwo}
\end{align*}

\begin{proofof}[\reflemma{size-mytr}]
	\Paste{l:size-mytr}
\end{proofof}

\begin{proof}
	By mutual induction on  $\tm$ and $\val$:
	\begin{itemize}
		\item \emph{Variable}, \ie $\val \defeq \var \eqdef \tm$\,: then $\auxtr{\val} = \var$ and $\mytr{\tm} = (\var, \emptyenv)$, hence $\len{\auxtr\val} = 1 \leq 5\len{\val}$ and $\len{\mytr\tm} = \len{\var} + \len{\emptyenv} = 1 \leq 5 \, \size{\tm}$.
    
    \item \emph{Error} or \emph{Boolean}: similar to the case above.

		\item \emph{Abstraction}, \ie $\val \defeq \la{\var}{\tmtwo} \eqdef \tm$\,: then $\auxtr{\val} = \la{\var}{\mytr{\tmtwo}}$ and $\mytr{\tm} = (\la{\var}{\mytr{\tmtwo}}, \emptyenv)$;
		by \ih, $\len{\mytr{\tmtwo}} \leq 5\len{\tmtwo}$ and hence $\len{\auxtr{\val}} = \len{\mytr{\tmtwo}} + 1 \leq 5\len{\tmtwo} + 1 \leq 5(\len{\tmtwo} + 1) = 5\len{\val}$ and $\len{\mytr{\tm}} = \len{\la\var\mytr\tmtwo} + \len{\emptyenv} = \len{\mytr\tmtwo} + 1 \leq 5\size{\tmtwo} + 1 \leq 5(\size{\tmtwo} + 1) = 5 \, \size{\tm}$.

		\item \emph{Application of two $\lambda$-values}, \ie $\tm \defeq \val\valtwo$\,: then, $\mytr{\tm} = (\auxtr\val\,\auxtr\valtwo,\emptyenv)$;
		by \ih, $\len{\auxtr{\val}} \leq 5\len{\val}$ and $\len{\auxtr{\valtwo}} \leq 5\len{\valtwo}$;
		hence $\len{\mytr{\tm}} = \len{\auxtr\val\,\auxtr\valtwo} + \len{\emptyenv} = \len{\auxtr\val} + \len{\auxtr\valtwo} + 1 \leq 5 \len\val + 5 \cdot \len\valtwo + 1 \leq 5 (\size{\val} + \size{\valtwo} + 1) = 5 \, \size{\tm}$.

		\item \emph{Application of a non-$\lambda$-value to a $\lambda$-value}, \ie $\tm \defeq \tmtwo\val$ where $\tmtwo$ is not a \lav : then, $\mytr{\tm} = (\var\auxtr\val, \esub\var\mol \env)$ where $\mytr{\tmtwo} = (\mol, \env)$;
		by \ih, $\len{\auxtr{\val}} \leq 5\len{\val}$ and $\len{\mytr{\tmtwo}} = \len{\mol} + \len{\env} \leq 5\len{\tmtwo}$;
		hence $\len{\mytr{\tm}} = \len{\var} + \len{\auxtr\val} + 1 + \len{\mol} + \len{\env} + 1 = \len{\auxtr\val} + \len{\mytr\tmtwo} + 3 \leq 5\size{\val} + 5\size{\tmtwo} + 3 \leq 5(\size{\val} + \size{\tmtwo} + 1) = 5 \, \size{\tm}$.

		\item \emph{Application of a \lat to a non-\lav}, \ie $\tm \defeq \tmtwo\tmthree$ where $\tmthree$ is not a \lav:
		then, $\mytr{\tm} = \append{\mytr{\tmtwo\var}}{(\esub{\var}{\mol}\env)} = (\vartwo\var, \esub{\vartwo}{\moltwo}\envtwo\esub{\var}{\mol}\env)$ where $\mytr{\tmthree} = (\mol, \env)$ and $\mytr{\tmtwo\var} = (\vartwo\var, \esub{\vartwo}{\moltwo}\envtwo)$ with $\mytr{\tmtwo} = (\moltwo, \envtwo)$.
	  By \ih, $\len{\mytr{\tmthree}} = \len{\mol} + \len{\env} \leq 5 \len{\tmthree}$ and 
	  $\len{\mytr{\tmtwo}} = \len{\moltwo} + \len{\envtwo} \leq 5\size{\tmtwo}$.
		Therefore, $\len{\mytr{\tm}} = \len{\vartwo\var} + \len{\moltwo} + \len{\envtwo} + 1 + \len{\mol} + \len{\env} + 1 = \len{\mytr\tmtwo} + \len{\mytr\tmthree} + 5 \leq 5\len{\tmtwo} + 5\len{\tmthree} + 5 \leq 5(\size{\tmtwo} + \size{\tmthree} + 1) = 5\size{\tm}$.

  \item \emph{Conditional}, case $\tm\defeq(\xite\val\tmtwo\tmthree)$. Then $\mytr\tm=(\xite{\auxtr\val}{\mytr\tmtwo}{\mytr\tmthree},\emptyenv)$ and $\len{\mytr\tm} = 1 + \len{\auxtr\val} + \len{\mytr\tmtwo} + \len{\mytr\tmthree}$. Conclude by \ih{}

  \item \emph{Conditional}, case $\tm\defeq(\xite\tmtwo\tmthree\tmfour)$ with $\tmtwo$ not a \lav. Then $\mytr\tm=(\xite\var{\mytr\tmthree}{\mytr\tmfour},\esub\var\tmfive\env)$ where $\mytr\tmtwo \eqdef (\tmfive,\env)$.
  Then $\len{\mytr\tm} = 1 + 1 + \len{\mytr\tmthree} + \len{\mytr\tmfour} + 1 + \len{\mytr\tmtwo}$. Conclude by \ih{}
		\qedhere
	\end{itemize}
\end{proof}

\begin{proofof}[\refthm{compl-closed}]
	\Paste{thm:compl-closed}
\end{proofof}
\begin{proof}
  By \reflemma{linear-bounds} and the discussion in the body about costs.
\end{proof}

\subsection{The \apglam}
\label{app:towards-impl-closed-pre}

As discussed at the beginning of \refsect{closed}, we call the \aglam{} a machine even though it does not satisfy the usual requirements for abstract machines. 
One of these requirements is the presence of rules that guide evaluation to the next redex: in the \aglam{} the search for the next redex is instead left implicit in the evaluation rules, because it corresponds to going through the environment from right to left.

In order to implement the \aglam while respecting the complexity analysis that we just presented in \refsubsect{complexity-closed}, we introduce a variant of the \aglam called \apglam, where the search for the next redex is decomposed into $O(1)$ steps of a new \emph{search} transition. The other steps of the machine are in one-to-one correspondence with the ones of the \aglam. 
Finally, we prove that the overall number of the search transitions is bilinear in the number of principal transitions and the size of the initial term, establishing bilinearity of the \apglam.
To improve the readability of this section, all proofs are moved to \Cref{app:towimpl-closed}.

\paragraph{\Pcell{s} and \penvironment{}s.}
The key idea to turn the \aglam into the \apglam is to avoid plugging and unplugging in the rules by letting them act on \emph{\pcell{s}}, \ie on \crumblep{s} where the beginning of the evaluated coda is explicitly marked using a pointer. 
For example, the crumble $(\mol,\env \aenv)$ could be represented as $(\mol,\env \sep \aenv)$ where ``$\sep$'' is the explicit separator that must be followed by \lenvironment{s} only. A \pcell{}
$(\mol,\env \esub\var\moltwo \sep \aenv)$ is the machine state that is attempting to evaluate the 
\crumblep $\ctxp{(\moltwo,\aenv)}$, where
$\ctx = (\mol,\env\esub\var\ctxhole)$. If $(\moltwo,\aenv)$ is a \aglam $a$-redex, the \apglam will evaluate according to the corresponding $a$-transition that also takes care of setting (in $O(1)$) the pointer to the rightmost unevaluated \crumblep. 
Otherwise, by harmony (\refprop{harmony-closed}), $\moltwo$ must be a \crumbled value $\molv$ and therefore the pointer is moved (in $O(1)$) one step to the left, looking for the next redex: $(\mol,\env \esub\var\molv \sep \aenv) \toc (\mol,\env \sep \esub\var\molv \aenv)$.

Not all \pcell{} configurations are of the form $(\mol,\esub\var\moltwo \sep \aenv)$: the configurations $(\mol, \sep \aenv)$ must be also taken into account and evaluated if $\mol$ is not a \crumbled value. 
However, there is no simple way to describe machine transitions that act uniformly on configurations $(\mol,\sep \aenv)$ and $(\mol, \env \esub\var\moltwo \sep \aenv)$ without duplicating the rules or without re-introducing a notion of contextual closure. To solve the issue, we abandon \pcell{}s and adopt \penvironment{}s instead.

A \penvironment $(\esub\var\mol \env \sep \aenv)$ is just a representation of a \pcell{} $(\mol, \env \sep \aenv)$. The leftmost
variable $\var$ in a \penvironment{} can be understood as the name given to the machine output. 
It plays a role similar to the outermost $\lambda$-abstraction introduced by CPS transformations, that binds the continuation that is fed with the output of the evaluation. 
In particular, a normal \penvironment $(\sep \esub \var \molv \aenv)$ represents the normal \crumblep $(\molv, \aenv)$.

\paragraph*{Formal definition of \penvironment{}s and read-back.}
\Penvironment{}s are defined as 
$ \benv \defeq \env \sep \envtwo$ for any non-pointed environments $\env$ and $\envtwo$ such that either  $\env$ or $\envtwo$ is non-empty.
The environment on the left of the cursor $\sep$ is the \emph{unevaluated} environment; the one on the right is the
\emph{evaluated} environment.

The \emph{translation} $\myiota\cdot$ embeds 
\crumblep{s} into \penvironment{}s:
\begin{align*} 
	\myiota{\mol,\env} \defeq \esub\var\mol \env \sep \emptyenv
\end{align*}
where $\var$ is any variable name fresh in $\mol$ and $\env$.

The left inverse of $\myiota\cdot$ is the \emph{read-back} function $\atoi\cdot$ (from \penvironment{s} to \crumblep{s}):
\begin{align*}
  \atoi{\emptyenv \sep \esub\var\mol\env} & \defeq (\mol,\env) &
  \atoi{\esub\var\mol\env \sep \envtwo} & \defeq (\mol,\env\envtwo)
\end{align*}

\paragraph*{Evaluation.} The transition rules of the \apglam are (the adaptation of) those of the \aglam plus the new \emph{search} transition $\toc$, and they are all defined only at top level, without a contextual closure---their union is noted $\topcrumble$:
\begin{align*}
  \env \esub\var{(\la\vartwo\cell)\,\molv} \sep \aenv
  &  \tom 
  \env \esub\var\mol\envtwo\esub\varthree\molv
  \sep \aenv
  && (i)
  \\%
  \\[-.7\baselineskip]
    \env \esub\var{\xite\true\cell\celltwo} \sep \aenv
		    & \toift 
		      \env \esub\var\mol \envtwo \sep \aenv & & (ii) \\
    \env \esub\var{\xite\false\cell\celltwo} \sep \aenv
		    & \toiff 
		         \env \esub\var\mol \envtwo \sep \aenv & & (iii)
		     \\
		    \env \esub\var{\xite\molv \cell \celltwo} \sep \aenv
		    & \toife 
		    \env \esub\var{\err} \sep \aenv & & (iv)
		    \\
		    \env \esub\var{\molv \molvtwo} \sep \aenv
		    & \toape 
		    \env \esub\var{\err} \sep \aenv& &(v)
		    \\
		      \\[-.7\baselineskip]
  \env \esub\vartwo\var \sep \aenv
  & \toev
  \env\esub\vartwo{\aenv(\var)} \sep \aenv
  & & (vi)
  \\
  \env \esub\vartwo{\var\molv} \sep \aenv
  & \toel 
  \env \esub\vartwo{\aenv(\var)\molv} \sep \aenv
  & & (vi)
  \\
  \env \esub\vartwo{\xite\var\cell\celltwo} \sep \aenv
  & \toeite 
  \env \esub\vartwo{\xite{\aenv(\var)}\cell\celltwo} \sep \aenv
    & & (vi)
  \\%
  \\[-.7\baselineskip]
  \env \esub\var\mol \sep \aenv
  & \toc 
  \env \sep \esub\var\mol \aenv
  & & (vii)
\end{align*}

\noindent where 
\begin{enumerate}[i.]
 \item $\la\varthree{(\mol,\envtwo)} \defeq (\la\vartwo\cell)^\alpha$ such that $(\env \esub\var\mol\envtwo\esub\varthree\molv
\sep \aenv)$ is well-named.
\item where $\cell \eqdef (\mol,\envtwo)$
\item where $\celltwo \eqdef (\mol,\envtwo)$
\item if $\molv=\la\var\cellthree$ or $\molv = \err$.
\item if $\molv\in \set{\true,\false,\err}$.
 \item if $\var \in \domain{\aenv}$.
 \item if none of the other rules is applicable, \ie{} when $\mol$ is a practical value $\pval$ or when $\mol$ is $\var$, $\var\molv$, or $\xite\var\cell\celltwo$  but $\var$ is not defined in $\aenv$.
\end{enumerate}

A \emph{principal transition} of the \apglam is a transition $\Rew{a}$ for any rule $a\in\set{\betav,\iftsym,\iffsym,\ifesym,\apesym}$.

\begin{proposition}[Harmony for the \apglam]
  \label{prop:zip-harmony-closed}
\Copy{prop:zip-harmony-closed}{
  A reachable \penvironment{} $\benv$ is normal iff it has the form $(\emptyenv \sep \aenv)$ for some non-empty \lenvironment{} $\aenv$.
}
\end{proposition}

\paragraph*{Implementation.} We prove that that the \apglam{} simulates the \aglam{} following the standard schema
introduced in \refsubsect{the-implementation-theorem} and that we have already employed for the \aglam{} in \refsubsect{complexity-closed}.
 We only state the machine invariants and the statement of the implementation theorem; the details of the proof can be found in the \refapp{towimpl-closed}.

\begin{definition}[Reachable state]
  A \penvironment{} is said to be \emph{reachable} if it is obtained from evaluation steps starting from the translation $\myiota\cell$ of a well-named closed \crumblep $\cell$.
\end{definition}

\begin{lemma}[Invariants for the \apglam] 
  \label{l:zip-qualitative-invariants-closed}
  \Copy{l:zip-qualitative-invariants-closed}{
  Let $\benv$ be a reachable \penvironment in the \apglam:
  \begin{enumerate}
    \item\label{p:zip-qualitative-invariants-closed-freshness} \emph{Freshness:} $\benv$ is well-named.
    \item \emph{Closure:} $\benv$ is closed.
    \item \emph{Rightmost:} $\benv = (\env \sep \aenv)$ for some environment $\env$ and some \lenvironment{} $\aenv$.
  \end{enumerate}}
\end{lemma}

\begin{theorem}[Implementation]
  \label{thm:zip-implementation-closed}
\Copy{thm:zip-implementation-closed}{
  Let $\benv$ be a reachable \penvironment in the \apglam.
  \begin{enumerate}
    \item \emph{Initialisation: }
     $\atoi{\myiota\cell}=\cell$ for every \crumblep $\cell$.
    \item \emph{Principal projection: }
     if $\benv \to_a \benvtwo$
      then $\atoi\benv \to_a \atoi\benvtwo$ for any rule $a\neq \csym$.
    \item \emph{Overhead transparency: }
     if $\benv \toc \benvtwo$
      then $\atoi\benv = \atoi\benvtwo$.
    \item \emph{Determinism: }
     the transition function $\topcrumble$ is deterministic.
    \item \emph{Halt: }
     if $\benv$ is normal then $\atoi\benv$ is normal.    
    \item \emph{Overhead termination}:  $\toc$ terminates.
  \end{enumerate}

	Therefore, the \apglam (with its transition function $\topcrumble$), the \aglam  (with $\tocrumble$), and the read-back $\atoi\cdot$ form an implementation system.
}
\end{theorem}

\paragraph*{Complexity}
We reuse the measures introduced in \refsubsect{complexity-closed}. The only difference
here is that we need to bound also the number of $\csym$ steps, which intuitively depends on the length of the pointed environment that is being evaluated.
We define the new measure $\length\cdot$ on environments and \crumble{s}, that simply counts the number of entries:
\begin{align*}
\length{\mol,\env} & \defeq 1 + \length\env & & &
\length\emptyenv & \defeq 0 &
\length{\env\esub\var\mol} & \defeq 1 + \length\env
\end{align*}

After each $\msym/\iftsym/\iffsym$ step, the measure increases by the length
of the body that is being concatenated. For every environment $\env$, we define the constant $L(\env)$ that bounds the length of the bodies occurring anywhere in $\env$:
\begin{align*} L(\env) \defeq & \sup\set{ \length\cell : \cell   \mbox{ body in } \env}. 
\end{align*}

We extend the definitions above to pointed environments and \crumblep{s} in the expected way:
\begin{align*}
\length{\env \sep \envtwo} &\defeq \length{\env\envtwo} & L(\env \sep \envtwo) &\defeq L(\env\envtwo) & L(\cell) &\defeq L(\myiota\cell).
\end{align*}

\begin{remark}
  For any \crumblep $\cell$ and \lat{} $\tm$:
  $\length\cell = \length{\myiota\cell}$,
  $L(\mytr\tm)\leq \len\tm$ and $\length{\mytr\tm} \leq \len\tm$.
\end{remark}

As usual, an execution $\exec$ is a sequence of evaluation steps, and we use $\no{a}\exec$ to count the number of $a$-evaluation steps in $\exec$.

By \refthm{zip-implementation-closed}, all transition steps but $\csym$ steps are mapped one-to-one to corresponding transition steps in the \aglam{}. As for the number of $\csym$ transitions:

\begin{lemma}[Number of $\csym$-transitions]
	\label{l:number-comm-pointed-closed}
	\Copy{l:number-comm-pointed-closed}{
  Let $\cell$ be a well-named \crumblep, and let $\exec \colon \myiota\cell \topcrumble^* \benv = (\env\sep\aenv)$ an execution in the \apglam. 
  Then $\length\env \leq \length\cell + (\no\msym\exec+\no\iftsym\exec+\no\iffsym\exec) \cdot L(\cell) - \no\csym\exec$.}
\end{lemma}

\begin{corollary}\label{coro:zip-aux-r-e-closed}
  Let $\tm$ be a \lat{}. 
  For a normalizing execution $\exec$ in the \apglam starting from $\myiota{\mytr{\tm}}$, we have $\no\csym\exec \leq (\sizepr\exec + 1) \cdot \len\tm$.
\end{corollary}

\paragraph*{Cost of evaluation.}
 The cost of all transitions but $\csym$ was already discussed in \refsubsect{complexity-closed}. The cost of a $\csym$ transition is clearly $O(1)$.
  
\begin{theorem}[The \apglam is bilinear]
  \label{thm:efficient-pointed-closed}
  \Copy{thm:efficient-pointed-closed}{
   For any closed \lat{} $\tm$ and any \apglam execution $\exec\colon\myiota{\mytr\tm} \topcrumble^* \benv$, the cost of implementing $\exec$ on a RAM is  $O((\sizepr\exec + 1) \cdot \len{\tm})$.
  }
\end{theorem}

The previous theorem also has a consequence for the non-pointed case. Since the theorem shows that the cost of searching for redexes and the (un)plugging operations can indeed be realised in bilinear time, it is also true that the \aglam can be implemented in bilinear time, improving \refthm{compl-closed} by removing the \emph{up to search and (un)plugging} side condition.
\begin{corollary}[The \aglam is bilinear]\label{cor:compl-closed}
 		For any closed \lat{} $\tm$ and any \aglam execution $\exec\colon\mytr\tm\tocrumble^*\cell$, the cost of implementing $\exec$ on a RAM is $O((\sizepr\exec + 1) \cdot \len{\tm})$.
\end{corollary}

\subsection{Proofs of \refapp{towards-impl-closed-pre}}
\label{app:towimpl-closed}

\begin{lemma}\label{l:zip-aux-cell-rrb-closed}
  Let $\mol$ be a \crumbledt and $\env$ an environment.
  $\esub\var\mol\env$ is an \lenvironment if and only if $(\mol,\env)$ is a \lcell.
\end{lemma}
\begin{proof}
  Obvious from the definition of \lenvironment{} and \lcell{}.
\end{proof}

\begin{proofof}[\reflemma{zip-qualitative-invariants-closed}] 
  \Paste{l:zip-qualitative-invariants-closed}
\end{proofof}
\begin{proof}~
  By induction on the length of the evaluation sequence leading to $\benv$. The base cases hold by the definition of reachability and by the definition of $\myiota\cdot$. As for the inductive cases, we proceed by cases on the transition rules:
  \begin{enumerate}
    \item
    For $\msym$ the claim follows from the side condition.
    The rules in $\set{\iftsym,\iffsym,\ifesym,\apesym,\csym}$ do not increase the number of explicit substitutions occuring in $\benv$ outside of abstractions, hence the claim follows from the \ih{}. The rules in $\set{\evsym,\elsym,\eitesym}$ copy a \pvalue from the environment $\aenv$: note that $\aenv(\var)$ is either an abstraction (which does not influence well-namedness) or a boolean or an error (which do not contain explicit substitutions).
    
    \item Similar to the discussion in \reflemma{invariants-closed}.

    \item 
    The rules in $\set{\msym,\iftsym,\iffsym,\ifesym,\apesym,\evsym,\elsym,\eitesym}$ do not change the evaluated part, hence the claim follows from the \ih{}. 
    As for the $\csym$ rule, by \reflemma{zip-aux-cell-rrb-closed} it suffices to prove that $(\mol,\aenv)$ is a \lcell{}, knowing that the other transition rules cannot be applied. This follows from the side condition of the $\csym$ rule: the \crumble{} $\mol$ is necessarily a \pvalue, as the other cases in which the free variable $\var$ occurs are not possible because $\benv$ is a closed \penvironment.
    \qedhere
  \end{enumerate}
\end{proof}

\begin{proofof}[\refprop{zip-harmony-closed}]
  \Paste{prop:zip-harmony-closed}
\end{proofof}
\begin{proof}
  The proof of the implication from right to left is trivial.
  Let us now prove the other direction. Let $\benv$ be a reachable normal \penvironment. 
  By \reflemma{zip-qualitative-invariants-closed}, $\benv$ has the form $(\env \sep \aenv)$. $\env$ cannot be non-empty, because otherwise one of the transition rules in $\set{\msym,\iftsym,\iffsym,\ifesym,\apesym,\evsym,\elsym,\eitesym,\csym}$ could be applied, contradicting the hypothesis that $\benv$ is normal.
\end{proof}

\begin{lemma}
  \label{l:aux-zip-implementation-closed}
  For every \penvironment $(\env\esub\var\mol\sep\envtwo)$:
  $\atoi{\env\esub\var\mol\sep\envtwo} = \atoi{\cctxp{\mol,\envtwo}}$ where $\cctx$ is:
  \[\cctx \defeq \begin{cases}
    \ctxhole & \mbox{if } \env=\emptyenv \\
    (\moltwo, \envthree\esub\var\ctxhole) & \mbox{if } \env\defeq\esub\vartwo\moltwo\envthree 
  \end{cases}. \]
\end{lemma}
\begin{proof}
  Easy by the definition of $\atoi\cdot$.
\end{proof}

\begin{proofof}[\refthm{zip-implementation-closed}]
  \Paste{thm:zip-implementation-closed}
\end{proofof}
\begin{proof}~
  \begin{enumerate}
    \item Let $\cell=(\mol,\env)$: by the definitions,
     $\atoi{\myiota\cell}=\atoi{\esub\var\mol\env\sep\emptyenv} = (\mol,\env\emptyenv) = \cell$.
    \item There is a clear one-to-one correspondence between the transitions of the \apglam and the \aglam (apart from $\csym$). In order to prove it, one may  proceed by cases on each transition rule, but we just show the case of $\msym$ as the others are similar.
    
    Suppose $\env \esub\var{(\la\vartwo\cell)\,\molv} \sep \aenv
    \tom 
    \env \esub\var\mol\envtwo\esub\varthree\molv
    \sep \aenv$.
     By \reflemma{aux-zip-implementation-closed},
      $\atoi{\env \esub\var{(\la\vartwo\cell)\,\molv} \sep \aenv} = \cctxp{((\la\vartwo\cell)\,\molv, \aenv)}$ and $\atoi{\env \esub\var\mol\envtwo\esub\varthree\molv
      \sep \aenv} = \cctxp{\mol,\envtwo\esub\varthree\molv\aenv}$ for some  \crumble context $\cctx$.
      Clearly also $\cctxp{((\la\vartwo\cell)\,\molv, \aenv)} \tom \cctxp{(\mol,\envtwo\esub\varthree\molv\aenv)}$.
    
    \item By inspection of the rule $\csym$. We need to prove that 
     $\atoi{\env \esub\var\mol \sep \aenv} =
     \atoi{\env \sep \esub\var\mol \aenv}$. By cases on the structure of $\env$:
      if $\env=\emptyenv$, then $\atoi{\emptyenv \esub\var\mol \sep \aenv} = (\mol, \aenv) =
      \atoi{\emptyenv \sep \esub\var\mol \aenv}$.
      If instead $\env=\esub\vartwo\moltwo\envtwo$, then 
      $\atoi{\env \esub\var\mol \sep \aenv} = (\moltwo, \envtwo\esub\var\mol\aenv) =
      \atoi{\env \sep \esub\var\mol \aenv}$.
    \item The rule $\csym$ can be applied by definition only when the other rules cannot be applied. The remaining rules apply to a \penvironment of the form $(\env\esub\var\mol\sep\aenv)$, for distinct shapes of $\mol$:
    \begin{itemize}
    \item The rules $\set{\msym,\apesym,\elsym}$ apply when $\mol$ is an application, and respectively the application of an abstraction to a \crumbled value, of a boolean/error to a \crumbled value, and of a variable to a \crumbled value.
    \item The rule $\evsym$ applies when $\mol$ is a variable.
    
    \item The rules in $\set{\iftsym,\iffsym,\ifesym,\eitesym}$ apply when $\mol$ is a $\xitecompact$, and in clearly disjoint cases according to the structure of the condition: respectively, when the condition is $\true$, $\false$, an abstraction or an error, and a variable.

    \item Finally, note that no single rule in $\set{\evsym,\elsym,\eitesym}$ can transition to different \penvironment{s} due to the lookup of a variable which has multiple occurrences in the environment: the lookup is deterministic because the environment is well-named (\reflemmap{zip-qualitative-invariants-closed}{freshness}).
    \end{itemize}
    
    \item By \refprop{zip-harmony-closed}, $\benv$ is normal iff it has the form $(\emptyenv \sep \aenv)$ for some non-empty \lenvironment{} $\aenv$.
    Then $\atoi{\benv}=\atoi{\emptyenv\sep\aenv}$ which is a \lcell{} by \reflemma{zip-aux-cell-rrb-closed}. 
    By \refprop{harmony-closed}, $\atoi{\benv}$ is normal.
    
    \item Immediate consequence of forthcoming \refcoro{zip-aux-r-e-closed} (proved independently). 
    \qedhere
  \end{enumerate}
\end{proof}

\begin{proofof}[\reflemma{number-comm-pointed-closed}]
	\Paste{l:number-comm-pointed-closed}
\end{proofof}
\begin{proof}
	By induction on the length of $\exec$. In the base case $\length\env = \length\cell$ and $\no\msym\exec=\no\iftsym\exec=\no\iffsym\exec=\no\csym\exec=0$. In the inductive case, use the \ih{} and proceed by cases on the transition rules. 
	The transitions $\evsym,\elsym,\eitesym$ do not change the measure of the unevaluated environment because $\aenv(\var)$ is a value, whose measure is 0 by definition. A $\csym$ transition decreases by 1 the measure of the unevaluated part. A $\msym/\iftsym$ transition increases the measure of the unevaluated part by a number bound by $L(\cell)$. A $\iffsym$ transition increases the measure of the unevaluated part by a number bound by $L(\celltwo)$. All remaining transitions do not increase the measure of the unevaluated environment.
\end{proof}

\begin{proofof}[\refthm{efficient-pointed-closed}]
	\Paste{thm:efficient-pointed-closed}
\end{proofof}
\begin{proof}
	The cost $\len\exec$ of $\exec$ is the the total cost of principal transitions (which was proved in \refsubsect{complexity-closed} to be bilinear in the number of principal steps (plus one) and the size of the initial \crumblep{} when starting from \lat{s}) plus the total cost of $\csym$ transitions.
	The cost of $\csym$ transitions in $\exec$ is linear in the number of $\csym$ transitions, and therefore by \refcoro{zip-aux-r-e-closed}, again bilinear in the number of principal steps (plus one) and the size of the initial term.
\end{proof}

\section{Proofs of \refsect{open} (open case)}

\subsection{Proofs of \refsubsect{fireball}}

\begin{lemma}[Composition of right \firectx{s}]
	\label{l:r-ctxs-comp-open}
	Let $\revctx$ and $\revctx'$ be right \firectx{s}. 
	Then their composition $\revctxp{\revctx'}$ is a right \firectx.
\end{lemma}
\begin{proof}
	By induction on the right \firectx $\revctx$.
	Cases:
	\begin{itemize}
		\item \emph{Hole}, \ie $\revctx \defeq \ctxhole$: then, $\revctxp{\revctxtwo} = \revctxtwo$ is a right \firectx{} by hypothesis.
		
		\item  \emph{Right}, \ie $\revctx \defeq \tm\revctxthree$: then, $\revctxp{\revctxtwo} = \tm (\revctxthreep{\revctxtwo})$ is a right \firectx{} because $\revctxthreep{\revctxtwo}$ is a right \firectx{} by \ih
		
		\item \emph{Left}, \ie $\revctx \defeq \revctxthree \fire$: then, $\revctxp{\revctxtwo} = \revctxthreep{\revctxtwo} \fire$ is a right \firectx{} because $\revctxthreep{\revctxtwo}$ is a right \firectx{} by \ih
		
    \item $\xitecompact$, \ie $\revctx \defeq (\xite \revctxthree \tm \tmtwo)$:
     then, $\revctxp{\revctxtwo} \allowbreak= \allowbreak (\xite{\revctxthreep{\revctxtwo}} \tm \tmtwo)$ is a right \firectx{} because $\revctxthreep{\revctxtwo}$ is a right \firectx{} by \ih
		\qedhere
	\end{itemize}
\end{proof}

\begin{proofof}[\refprop{distinctive-fireball-new}]
	\Paste{prop:distinctive-fireball-new}
\end{proofof}
\begin{proof}
	\begin{enumerate}
		\item We show separately the two implications:
	\begin{description}
		\item [$(\Rightarrow)$] Proof by induction on the structure of $\tm$.
		\begin{itemize}
			\item Case $\tm$ value: trivial.
			\item Case $\tm = \tmtwo\tmthree$ for some terms $\tmtwo$ and $\tmthree$: we show that then $\tm$ is an inert term (and hence a fireball). 
			Since $\tm$ is $\cbetafsym$-normal, then $\tmthree$ is $\cbetafsym$-normal, and hence by \ih{} $\tmthree$ is a fireball. 
			Since $\tm$ is $\cbetafsym$-normal and $\tmthree$ is a fireball, then also $\tmtwo$ must be $\cbetafsym$-normal, and hence a fireball by \ih
			We proceed by cases on $\tmtwo$, showing that the only possible cases are when $\tmtwo$ is a variable or an inert term (and so $\tm$ is a fireball). Indeed, $\tmtwo$ cannot be an abstraction (because otherwise the rule $\tof$ may be applied, contradicting the hypothesis that $\tm$ is $\cbetafsym$-normal); it cannot be a boolean or $\err$ (because otherwise the rule $\toape$ may be applied).
			\item Case $\tm = \xite\tmtwo\tmthree\tmfour$ for some terms $\tmtwo,\tmthree,\tmfour$: we show that the only possible cases are when $\tmtwo$ is a variable or a inert term (and hence $\tm$ is a fireball).
			 Since $\tm$ is $\cbetafsym$-normal, then $\tmtwo$ is $\cbetafsym$-normal, and hence by \ih{} $\tmtwo$ is a fireball.
			$\tmtwo$ it cannot be an abstraction or $\err$ (because otherwise the rule $\toife$ may be applied, contradicting the hypothesis that $\tm$ is $\cbetafsym$-normal); it cannot be a boolean (because otherwise one of the rules $\toift$ or $\toiff$ may be applied).
		\end{itemize}
		
		\item [$(\Leftarrow)$] By hypothesis, $\tm$ is a fireball, \ie either a value or an inert term. 
		\begin{itemize}
			\item \emph{Value:} We proceed by cases on the definition of value. 
			If $\tm$ is a variable, then it is clearly $\cbetafsym$-normal. 
			If $\tm$ is an abstraction, then it is $\cbetafsym$-normal since $\tocf$ does not reduce under $\l$'s. 
			Otherwise $\tm$ is a either a boolean or $\err$, which clearly are $\cbetafsym$-normal.
			
			\item \emph{Inert term:} We proceed by cases on the definition of inert term.
			If $\tm = \var \fire$ for some variable $\var$ and some fireball $\fire$, then $\fire$ is $\cbetafsym$-normal by \ih, and $\var$ is clearly $\cbetafsym$-normal and different from abstractions, booleans and $\err$; therefore, no reduction rule can be applied to $\tm$, \ie $\tm$ is $\cbetafsym$-normal. 
			If $\tm = \var \fire$ or $\tm = \itm \fire$ for some inert term $\itm$ and some fireball $\fire$, then $\itm$ and $\fire$ are $\cbetafsym$-normal by \ih; moreover, $\itm$ is not an abstraction or boolean or $\err$; therefore, no reduction rule can be applied to $\tm$, \ie $\tm$ is $\cbetafsym$-normal.
			Otherwise, $\tm = \xite\tmtwo\tmthree\tmfour$ for some terms $\tmtwo,\tmthree,\tmfour$ where $\tmtwo$ is either a variable or a inert term;
			since $\tmtwo$ is $\cbetafsym$-normal (if $\tmtwo$ is a inert term, this holds by \ih), then no reduction rule can be applied to $\tm$, \ie $\tm$ is $\cbetafsym$-normal.
			\qedhere
		\end{itemize}
	\end{description}
	\item The direction ($\Leftarrow$) is exactly \reflemma{not-creation}.
  The direction ($\Rightarrow$) is proved by induction on the definition of $\tm \tocf \tmtwo$, in a way that is analogue to the proof of \reflemma{not-creation}.
	\qedhere
\end{enumerate}
\end{proof}

\begin{lemma}[Fireballs are closed under substitution and anti-substitution of inert terms]
	\label{l:closure-substitution}
	Let $\tm$ be a term and $\itm$ be an inert term.
	Then, $\tm \isub{\var}{\itm}$ is a fireball (resp.~an abstraction) if and only if $\tm$ is a fireball (resp.~an abstraction).
	Moreover, $\tm \isub{\var}{\itm} = \true$ (resp.~$\tm \isub{\var}{\itm} = \false$; $\tm \isub{\var}{\itm} = \err$) if and only if $\tm = \true$ (resp.~$\tm = \false$; $\tm = \err$).
\end{lemma}

\begin{proof}
	The left-to-right direction ($\Rightarrow$) is proved by a simple
	induction on the fireball structure of $\tm\isub{\var}{\itm}$. 
	Conversely, the right-to-left direction
	($\Leftarrow$) is proved by a simple induction on the inert structure of $\tm$.
\end{proof}

\begin{lemma}[Substitution of inert terms does not create redexes]
	\label{l:not-creation}
	Let $\tm, \tmtwo$ be terms and $\itm$ be an inert term. 
	There is a term $\tmthree$ such that, if $\tm\isub{\var}{\itm} \tocf \tmtwo$ then $\tm \tocf \tmthree$ and $\tmtwo = \tmthree \isub{\var}{\itm}$.  
\end{lemma}

\begin{proof}
	By induction on the right fireball context closing the redex. Cases:
	\begin{itemize}
		\item \emph{Step at the root:} Sub-cases:
		\begin{enumerate}
			\item \emph{Beta-step}, \ie $\tm\isub{\var}{\itm} = (\la{\vartwo}\tmfour\isub{\var}{\itm})\tmfive\isub{\var}{\itm} \tof \tmfour\isub{\var}{\itm}\isub{\vartwo}{\tmfive\isub{\var}{\itm}} \allowbreak\eqdef \tmtwo$ where $\tm \defeq (\la{\var}\tmfour)\tmfive$ and $\tmfive\isub{\var}{\itm}$ is a fireball.
			By \reflemma{closure-substitution}, $\tmfive$ is a fireball and hence $\tm = (\la{\var}\tmfour)\tmfive \tocf \tmfour\isub{\vartwo}{\tmfive}$.
			So, $\tmthree \defeq \tmfour\isub{\vartwo}{\tmfive}$ satisfies the statement, as $\tmthree \isub{\var}{\itm} = \tmfour\isub{\vartwo}{\tmfive}\isub{\var}{\itm} = \tmtwo$. 
			
			\item \emph{Conditional step with $\true$}, \ie $\tm\isub{\var}{\itm} = \xite{\allowbreak\true}{\tmthree\isub{\var}{\itm}}{\tmfive\isub{\var}{\itm}} \allowbreak\toift \tmthree\isub{\var}{\itm} \eqdef \tmtwo$ where $\tm \defeq \xite{\tmfour}{\tmthree}{\tmfive}$ for some term $\tmfour$ such that $\tmfour\isub{\var}{\itm} = \true$.
			By \reflemma{closure-substitution}, $\tmfour = \true$ and hence $\tm = \xite{\true}{\tmthree}{\tmfive} \toift \tmthree$.

			\item \emph{Conditional step with $\false$}, \ie $\tm\isub{\var}{\itm} = \xite{\allowbreak\true}{\tmfive\isub{\var}{\itm}}{\tmthree\isub{\var}{\itm}} \allowbreak\toiff \tmthree\isub{\var}{\itm} \eqdef \tmtwo$ where $\tm \defeq \xite{\tmfour}{\tmfive}{\tmthree}$ for some term $\tmfour$ such that $\tmfour\isub{\var}{\itm} = \false$.
			By \reflemma{closure-substitution}, $\tmfour = \false$ and hence $\tm = \xite{\false}{\tmfive}{\tmthree} \toiff \tmthree$.
			
			\item \emph{Conditional step to error}, \ie $\tm \isub{\var}{\itm} = \xite{\tmfour}{\tmfive\isub{\var}{\itm}}{\tmseven\isub{\var}{\itm}} \allowbreak\toife \err \eqdef \tmtwo$ where $\tmfour \defeq \la{\vartwo}{\tmfour'}$ or $\tmfour = \err$, and $\tm \defeq \xite {\tmsix}{\tmfive}{\tmseven}$ for some term $\tmsix$ such that $\tmsix\isub{\var}{\itm} = \tmfour$.
			By \reflemma{closure-substitution}, $\tmsix$ is either an abstraction or $\err$ and hence $\tm = \xite{\tmfour}{\tmfive}{\tmseven} \allowbreak\toife \err$.
			So, $\tmthree \defeq \err$ satisfies the statement, as $\tmthree\isub{\var}{\itm} = \err = \tmtwo$. 
			
			\item \emph{Application step to error}, \ie $\tm\isub{\var}{\itm} = \tmfour\isub{\var}{\itm} \tmfive\isub{\var}{\itm} \toape \err \eqdef \tmtwo$ where $\tmfour\isub{\var}{\itm} \in \{\true,\false,\err\}$ and $\tm \defeq \tmfour\tmfive$. 
			By \reflemma{closure-substitution}, $\tmfour \in \{\true, \false, \err\}$ and hence $\tm = \tmfour\tmfive \toape \err$. 
			So, $\tmthree \defeq \err$ satisfies the statement, as $\tmthree\isub{\var}{\itm} = \err = \tmtwo$. 
		\end{enumerate}
	
		\item \emph{Application left}, \ie $\tm\isub{\var}{\itm} = \tmfour\isub{\var}{\itm}\fire\isub{\var}{\itm} \tocf \tmsix\fire\isub{\var}{\itm} \eqdef \tmtwo$ where $\tm \defeq \tmfour\fire$ and $\tmfour\isub{\var}{\itm} \tocf \tmsix$.
		By \ih there is a term $\tmthree'$ such that $\tmsix = \tmthree'\isub{\var}{\itm}$ and $\tmfour \tocf \tmthree'$.
		Then $\tmthree \defeq \tmthree'\fire$ satisfies the statement, as $\tmthree\isub{\var}{\itm} = \tmthree'\isub{\var}{\itm}\fire{\var}{\itm} = \tmtwo$.
		\item \emph{Application right}, \ie $\tm\isub{\var}{\itm} = \tmfive\isub{\var}{\itm}\tmfour\isub{\var}{\itm} \tocf \tmfive\isub{\var}{\itm}\tmsix \eqdef \tmtwo$ where $\tm \defeq \tmfive\tmfour$ and $\tmfour\isub{\var}{\itm} \tocf \tmsix$.
		Analogous to the application left case, just switch left and right.
		\item \emph{If-then-else}, \ie $\tm\isub{\var}{\itm} = \xite{\tmfour\isub{\var}{\itm}}{\tmfive\isub{\var}{\itm}}{\tmsix\isub{\var}{\itm}} \tocf \xite{\tmseven}{\tmfive\isub{\var}{\itm}}{\tmsix\isub{\var}{\itm}} \eqdef \tmtwo$ where $\tm \defeq \xite\tmfour\tmfive\tmsix$ and $\tmfour\isub{\var}{\itm} \tocf \tmseven$.
		By \ih there is a term $\tmthree'$ such that $\tmseven = \tmthree'\isub{\var}{\itm}$ and $\tmfour \tocf \tmthree'$.
		Then $\tmthree \defeq \xite{\tmthree'}{\tmfive}{\tmsix}$ satisfies the statement, as $\tmthree\isub{\var}{\itm} = \xite {\tmthree'\isub{\var}{\itm}}{\tmfive\isub{\var}{\itm}}{\tmsix\isub{\var}{\itm}} = \tmtwo$.
		\qedhere
	\end{itemize}
\end{proof}

\subsection{Proofs of \refsubsect{micro-step-open}}

\subsubsection{Evaluated environments.}\hfill

Given the \lat{}s $\tm_1, \dots, \tm_n$ and pairwise distinct variables $\var_1, \dots, \var_n$, let $\sub \defeq \subs{\var_1}{\tm_1}{\var_n}{\tm_n}$ be the \emph{simultaneous} capture-avoiding substitution of $\tm_i$ for $\var_i$, for all $1 \leq i \leq n$.
In particular, if $n = 0$ then $\sub = \{\}$, the identity (or empty) substitution.
If $\tmtwo$ is a \lat and $\vartwo \neq \var_i$ for all $1 \leq i \leq n$, we set $\isub{\vartwo}{\tmtwo} \cup \sigma \defeq \{\replace{\vartwo}{\tmtwo}, \replace{\var_1}{\tm_1}, \dots, \replace{\var_n}{\tm_n}\}$.

Given an environment $\env$, the simultaneous substitution $\subunf{\env}$ associated with $\env$ is defined by induction on the length of $\env$:
\begin{align*}
\subunf{\emptyenv} &\defeq \{ \} & \subunf{\esub{\var}{\mol}\env} &\defeq \isub{\var}{\rbp{\mol, \env}} \cup \subunf{\env} \,.
\end{align*}

\begin{lemma}[Semi-closure under substitution]
\label{l:closure-substitution-open}
	Let $\val$ be a \lav, $\var_1, \dots, \var_n$ be pairwise distinct variables, 
	and $\fire_1, \dots, \fire_n$ be fireballs.
	Then $\val\subs{\var_1}{\fire_1}{\var_n}{\fire_n}$ is a fireball.
	If moreover $\val$ is a $\lambda$-abstraction,
	then $\val\subs{\var_1}{\fire_1}{\var_n}{\fire_n}$ is a $\lambda$-abstraction.
\end{lemma}

\begin{proof}
	Let $\sigma \defeq \subs{\var_1}{\fire_1}{\var_n}{\fire_n}$.
        By cases on the definition of \lav:
	\begin{itemize}
		\item \emph{Variable,} \ie either $\val \defeq \var_i$ for some $1 \leq i \leq n$ and then $\val\sigma = \fire_i$ which is a fireball by hypothesis; 
		or $\val \defeq \vartwo \neq \var_i$ for all $1 \leq i \leq n$ and then $\val\sigma = \vartwo$ which is a fireball.
		
		\item \emph{Abstraction,} \ie $\val \defeq \la{\vartwo}{\tmtwo}$ and 
		we can suppose without loss of generality that $\vartwo \notin \bigcup_{i=1}^n\fv{\fire_i} \cup \{\var_1, \dots, \var_n\}$;
		therefore, $\val\sigma = \la\vartwo{(\tmtwo\sigma)}$ which is a $\lambda$-abstraction and hence a fireball.
                \item \emph{Booleans and errors,} \ie $\val \in \set{\true,\false,\err}$. Trivial since $\val\sigma = \val$.
		\qedhere
	\end{itemize}
\end{proof}

\begin{lemma}[Read-back to fireball]
	\label{l:rb-val-f-open}
	For every \crumbled value $\molv$ and \fireenvironment $\fenv$, one has that $\rbp{\molv,\fenv}$ is a fireball.
	If moreover $\molv$ is a \pvalue, then $\rbp{\molv,\fenv}$ is a \pvalue.
\end{lemma}
\begin{proof} By induction on the length of $\fenv$. 
        We proceed by cases on the shape of $\molv$.
	\begin{itemize}
		\item \emph{Abstraction:} 
		If $\fenv \defeq \emptyenv$, then clearly $\rbp{\molv,\fenv}=\rb\molv$ is a $\lambda$-abstraction and hence a fireball.
		Otherwise $\fenv \defeq \fenvtwo\esub\var\mol$ where $\mol$ is a \crumbledt such that $\rb{\mol}$ is a fireball: thus, $\rbp{\molv,\fenv} = \rbp{\molv,\fenvtwo}\isub\var{\rb\mol}$;
		by \ih{}, $\rb{(\molv,\fenvtwo)}$ is a $\lambda$-abstraction, thus $\rbp{\molv,\fenv}$ is a $\lambda$-abstraction (and so a 
		fireball) by \reflemma{closure-substitution-open}.
		
                \item \emph{Booleans and errors}: the proof is identical
                 to the previous case.
		\item \emph{Variable:} If $\molv \defeq\var \notin \domain{\fenv}$, then $\rbp{\molv,\fenv} = \var$ which is a fireball; 
		otherwise $\molv \defeq \var \in \domain{\fenv}$ with
		$\fenv \defeq \fenvtwo\esub\var{\mol}\fenvthree$, and then $\rbp{\var,\fenv} = \rbp{\mol,\fenvthree}$ (since $\var \notin \domain{\fenvtwo}$) is a fireball by definition of \fireenvironment, as the \fireenvironment $\esub{\var}{\mol} \fenvthree = \genvp{(\mol, \fenvthree )}$ with $\genv \defeq \esub{\var}{\ctxhole}$.
		\qedhere
	\end{itemize}
\end{proof}

\begin{lemma}\label{l:aux-lookup-open}
	Let $\fenv$ well-named. If $\rbp{\var,\fenv}$ is a \pvalue, then $\fenv(\var)$ is a \pvalue.
\end{lemma}
\begin{proof}
	Assume by contradiction that $\var$ is not defined in $\fenv$: then by \reflemma{disj-c-e} $\rbp{\var,\fenv}=\rbp{\var,\emptyenv}=\var$, contradicting the hypothesis that $\rbp{\var,\fenv}$ is a \pvalue. Therefore
	$\var$ must be defined in $\fenv$, \ie{} $\fenv \defeq \fenvtwo \esub\var\mol \fenvthree$ with $\mol\defeq\fenv(\var)$.
	By the hypothesis that $\fenv$ is well-named, $\var\not\in\domain\fenvtwo$; therefore $\rbp{\var,\fenv} = \rbp{\mol,\fenvthree}$ by \reflemma{aux-aux-lookup-open-two}.
	By the definition of $\fenv$, since $\rbp{\mol,\fenvthree}$ is a \pvalue, then also $\mol$ is a \pvalue, and we conclude.
\end{proof}

\begin{lemma}\label{l:harmony-aux1-open}
  Let $\cell=(\mol,\fenv)$ be a well-named \crumble,
	and let $\mol$ have the following property:
	either $\mol$ is a \pvalue, or $\mol$ is $\var$ or $\var\molv$ or $\xite\var\cell\celltwo$ but $\var$ is not defined in $\fenv$ or $\fenv(\var)$ is not a \pvalue. 
	Then $\cell$ is a \firecell.
\end{lemma}
\begin{proof}
  Let $\cell=(\mol,\fenv)$ as above: it suffices to prove that 
  $\rbp{\mol,\fenv}$ is a fireball, and that if $\rbp{\mol,\fenv}$ is a \pvalue, then also $\mol$ is a \pvalue. By cases on the property about $\mol$ in the hypothesis:
  \begin{itemize}
    \item Abstraction: nothing to prove because $\rbp{\mol,\fenv}$ is an abstraction by \reflemma{rb-val-f-open} and thus a fireball.
    \item Booleans and errors: nothing to prove because $\rbp{\mol,\fenv}$ is a boolean or an error and thus a fireball.
    \item Variable, \ie{} $\mol=\var$ for some $\var$ when $\var$ is not defined in $\fenv$ or $\fenv(\var)$ is not a \pvalue.
     $\rbp{\mol,\fenv}$ is a fireball by \reflemma{rb-val-f-open}. Let us now assume that $\rbp{\mol,\fenv}$ is a \pvalue, and show that it is not possible: in fact by \reflemma{aux-lookup-open} $\fenv(\var)$ must then be defined and a \pvalue, contradicting the hypothesis.
    \item Application of a variable to a \crumbled value, \ie{} $\mol = \var\molv$ when $\var$ is not defined in $\fenv$ or $\fenv(\var)$ is not a \pvalue.
    Note that $\rbp{\mol,\fenv} = \rbp{\var,\fenv}\rbp{\molv,\fenv}$,
     where both $\rbp{\var,\fenv}$ and $\rbp{\molv,\fenv}$ are fireballs by \reflemma{rb-val-f-open}. If $\rbp{\var,\fenv}$ is inert there is nothing else to prove, because then $\rbp{\mol,\fenv}$ is a fireball, and clearly not a \pvalue.
     The case when $\rbp{\var,\fenv}$ is a \pvalue is not possible: 
		 again by \reflemma{aux-lookup-open} $\fenv(\var)$ should be defined and a \pvalue, contradicting the hypothesis.
    \item $\xitecompact$ where the guard is a variable, \ie{} $\mol = \xite\var\cell\celltwo$ when $\var$ is not defined in $\fenv$ or $\fenv(\var)$ is not a \pvalue.
    Note that $\rbp{\mol,\fenv} = \xite{\rbp{\var,\fenv}}{\rbp{\append\cell\fenv}}{\rbp{\append\celltwo\fenv}}$,
     where $\rbp{\var,\fenv}$ is a fireball by \reflemma{rb-val-f-open}. If $\rbp{\var,\fenv}$ is inert there is nothing else to prove, because then $\rbp{\mol,\fenv}$ is a fireball, and clearly not a \pvalue.
     The case when $\rbp{\var,\fenv}$ is a \pvalue is not possible: 
		 again by \reflemma{aux-lookup-open} $\fenv(\var)$ should be defined and a \pvalue, contradicting the hypothesis.
		 \qedhere
  \end{itemize}
\end{proof}

\begin{corollary}
	\label{coro:harmony-aux1-open-coro}
	If the well-named \crumble $(\mol,\fenv)$ is normal, then it is a \firecell.
\end{corollary}

\begin{proofof}[\refprop{harmony-open}]
	\Paste{prop:harmony-open}
\end{proofof}
\begin{proof}~
  \begin{itemize}
    \item[$(\Rightarrow)$]
     Let $\cell=(\mol,\env)$ be well-named and normal. We proceed by structural induction on $\env$:
     \begin{itemize}
       \item if $\env=\emptyenv$, then $(\mol,\emptyenv)$ is a \firecell by \refcoro{harmony-aux1-open-coro};
       \item if $\env=\esub\var\moltwo\envtwo$, then also the \crumble $(\moltwo,\envtwo)$ is normal. 
       By \ih{} $(\moltwo,\envtwo)$ is a \firecell, and therefore $\env=\esub\var\moltwo\envtwo$ is a \fireenvironment. By \refcoro{harmony-aux1-open-coro}, $\cell=(\mol,\env)$ is a \firecell.
     \end{itemize}
     \item[$(\Leftarrow)$]
      Let $\cell=\acell$, we need to prove that $\acell$ is normal.
      Let $\acell=\cctxp{(\mol,\fenv)}$ for some $\cctx,\mol,\fenv$.
      First of all, note that by the definition of $\acell$,
      $\rbp{\mol,\fenv}$ is a fireball, and that if $\rbp{\mol,\fenv}$ is a \pvalue, then also $\mol$ is a \pvalue.
      We prove that no reduction rule is applicable to $\cctxp{(\mol,\fenv)}$:
      \begin{itemize}
        \item Rule $\betafsym$ can be applied only if $\mol=(\la\var\moltwo)\,\molv$, but this contradicts the hypothesis that $\rbp{\mol,\fenv}$ is a fireball, since $\rbp{(\la\var\moltwo)\,\molv, \fenv} = \rbp{\la\var\moltwo,\fenv}\,\rbp{\molv,\fenv}$ and $\rbp{\la\var\moltwo, \fenv}$ is an abstraction by \reflemma{rb-val-f-open}.
        \item The analysis for the rules $\iftsym,\iffsym,\ifesym,\apesym$ is similar to the previous case: in all cases the read-back of the lhs of the rule is not a fireball, contradicting the hypothesis.
        \item Rule $\evsym$ can be applied only if $\mol$ is some variable $\var$ and $\fenv(\var)$ is defined and a \pvalue. This contradicts the hypothesis that if $\rbp{\mol,\fenv}$ is a \pvalue, then also $\mol$ is a \pvalue.
        \item Rule $\elsym$ can be applied only if $\mol = \var\molv$ for some $\var,\molv$, and $\fenv(\var)$ is defined and a \pvalue.
        This contradicts the hypothesis that $\rbp{\mol,\fenv}$ is a fireball, since $\rbp{\var\molv, \fenv} = \rbp{\var,\fenv}\rbp{\molv,\fenv}$ and $\rbp{\var, \fenv}$ is a \pvalue by \reflemma{rb-val-f-open} because $\fenv(\var)$ is a \pvalue.
        \item Rule $\eitesym$ can be applied only if $\mol = \xite\var\cell\celltwo$ for some $\var,\cell,\celltwo$, and $\fenv(\var)$ is defined and a \pvalue.
        This contradicts the hypothesis that $\rbp{\mol,\fenv}$ is a fireball, since $\rbp{\xite\var\cell\celltwo, \fenv} = \xite{\rbp{\var,\fenv}}{\rbp{\append\cell\fenv}}{\rbp{\append\celltwo\fenv}}$ and $\rbp{\var, \fenv}$ is a \pvalue by \reflemma{rb-val-f-open} because $\fenv(\var)$ is a \pvalue.
        \qedhere
      \end{itemize}
  \end{itemize}
\end{proof}

\subsubsection{The implementation theorem for the \oaglam}

\begin{lemma}[Invariants for the \oaglam]
	\label{l:invariants-open}
  For every reachable \crumblep $\cell$:
  \begin{enumerate}
    \item\mylabel{p:invariants-open-fresh}
     \emph{Freshness:} $\cell$ is well-named.
    \item\mylabel{p:invariants-open-fvs}
     \emph{Disjointedness}: if $\cell = \cctxp{(\mol,\env)}$ then $\domain\cctx \cap \fv\mol = \emptyset$.
    \item\mylabel{p:invariants-open-abstractions}
    \emph{Bodies:} any body in $\cell$ is a subterm (up to renaming) of the initial \crumblep.
    \item\mylabel{p:invariants-open-ctx-decoding}
		\emph{Weak contextual decoding:}
		for every decomposition $\cctxp{(\mol, \fenv)}$ where $\rbp{\mol,\fenv}$ is not a fireball, if $\cctxthree$ is a prefix of $\cctx$ then $\rb{\cctxthree}$ is a right \valuectx.
  \end{enumerate}
\end{lemma}
\begin{proof}
	\newcommand{\tmpitem}[1]{\item[\ref{p:invariants-open-#1}.] }
	By induction on the length of the reduction sequence leading to the \crumble $\cell$. 
	The base cases hold by \reflemma{basecases-both} (by noting that for \refpoint{invariants-open-ctx-decoding}, \reflemmap{basecases-both}{unfold-right} implies the weaker statement \reflemmap{invariants-open}{ctx-decoding}).
	As for the inductive cases, we inspect each reduction rule:
	\begin{itemize}
		\tmpitem{fresh}
		The substitution transitions $\evsym, \elsym, \eitesym$ do not change the set of variables occurring on the lhs of substitutions outside abstractions because they copy a value that does not contain any. Hence the claim follows from the \ih{}. 
		For the rule $\betafsym$ the claim follows from the side condition. For the remaining rules $\iftsym,\iffsym,\ifesym,\apesym$ the claim follows from the fact that all substitutions outside abstractions in the rhs alredy occur in the lhs.
		\tmpitem{fvs}
		The substitution transitions $\evsym,\elsym,\eitesym$ do not change the domain of the \crumble and only copy to the left a value $\val$ such that
$\cell = \cctxtwop{(\val,\envtwo)}$ and $\cctx$ is a prefix of $\cctxtwo$.
Thus the claim follows from the \ih{} because $\domain\cctx \subseteq \domain\cctxtwo$ and the free variables of the rhs are a subset of the union of the free variables of the lhs with $\fv\val$.
				
		Transition $\betafsym$ copies to the toplevel and renames the body of an abstraction.
		By the properties of $\alpha$-renaming $\fv{(\append\cell{\esub\var\molv})^\alpha} = \fv{\append\cell{\esub\var\molv}} = \fv{\la\var\cell}$.
		If the $\mol$ is chosen to be in the \crumble context (say $\cctxthree$) or in $\fenv$ of the reduction rule, then the claim follows from the \ih{}. 
		Let instead $(\append\cell{\esub\var\molv})^\alpha \eqdef \cctxtwop{(\mol,\envtwo)}$ with $\cctx = \cctxthreep\cctxtwo$: then $\fv{\cctxtwop{(\mol,\envtwo)}} = \fv{\la\var\cell}$ and $\fv{\mol} \subseteq \domain\envtwo \cup \fv{\la\var\cell}$.
		$\domain\envtwo \cap \domain\cctx = \emptyset$ by the side condition of $\betafsym$, and $\fv{\la\var\cell} \cap \domain\cctx = \emptyset$ by \ih{}, therefore we conclude with $\fv\mol \cap \domain\cctx = \emptyset$.
		
		\tmpitem{abstractions}
		Transitions $\evsym,\elsym,\eitesym$ may copy an abstraction, but the abstraction was already in the environment,
		and the claim follows from the \ih{}.
		Transition $\betafsym$ copies and renames the body of an abstraction that was already in the environment,
		and the claim follows from the \ih{} since the translation commutes with the renaming of free variables (\refrmkp{preservation-fvs}{commutation-fvs}).
                All the bodies in the rhs of the remaning transitions $\iftsym,\iffsym,\ifesym,\apesym$ already occur in the lhs and therefore the claim follows from the \ih{}.

		\tmpitem{ctx-decoding}
		Let $\mytr\molthree \Rew{}^n \cctxtwop{(\moltwo, \aenvtwo)} \Rew{a} \cctx\ctxholep{(\mol, \aenv)}$ (where $\mol$ is not a \pvalue).
		Cases of the transition $\cctxtwop{(\moltwo, \aenvtwo)} \Rew{a} \cctx\ctxholep{(\mol, \aenv)}$:
		
		\begin{itemize}
			\item Case $\betafsym$: $\cctxtwop{((\la\var\cell)\molv, \aenv)} \tom \cctxtwop{\append{\cell^\alpha}{(\esub{\var^\alpha}\molv\aenv)}}$.
			
			Let $\cctxthree$ be a prefix of $\cctx$. There are two sub-cases:
			\begin{itemize}
				\item \emph{$\cctxthree$ is a prefix of $\cctxtwo$}: by \ih{} $\rb{\cctxthree}$ is a right \valuectx{}.
				
				\item \emph{$\cctxtwo$ is a prefix of $\cctxthree$},
				\ie{} $\cctxthree=\cctxtwop{\cctxfour}$ and $\cell^\alpha=\cctxfour\ctxholep{\cell'}$.
				By \reflemmap{basecases-both}{abstractions} and \reflemmap{invariants-open}{abstractions} $\cell$ is the translation of a $\lambda$-term, by \refrmkp{preservation-fvs}{commutation-fvs} $\cell^\alpha$ is so, and thus by \reflemmap{basecases-both}{unfold-right} $\rb{\cctxfour}$ is a right \valuectx{}.
				By \ih, $\rb\cctxtwo$ is a right \valuectx{} as well.
				Since $\rb{\cctxthree} = \rb{\cctxtwo}\ctxholep{\rb{\cctxfour}}$ according to \refcorollaryp{rb-ctx}{comp},  we obtain that $\rb{\cctxthree}$ is a right \valuectx{} as composition of right \valuectx{s} (\reflemma{r-ctxs-comp-closed}).
			\end{itemize}

			\item Case $\iftsym$: $\cctxtwop{(\xite\true\cell\celltwo, \aenv)} \tom \cctxtwop{\append{\cell}{\aenv}}$.
			Let $\cctxthree$ be a prefix of $\cctx$. There are two sub-cases:
			\begin{itemize}
				\item \emph{$\cctxthree$ is a prefix of $\cctxtwo$}: by \ih{} $\rb{\cctxthree}$ is a right \valuectx{}.
				
				\item \emph{$\cctxtwo$ is a prefix of $\cctxthree$},
				\ie{} $\cctxthree=\cctxtwop{\cctxfour}$ and $\cell=\cctxfour\ctxholep{\cell'}$.
				By \reflemmap{basecases-both}{abstractions} and \reflemmap{invariants-open}{abstractions} $\cell$ is the translation of a $\lambda$-term and thus by \reflemmap{basecases-both}{unfold-right} $\rb{\cctxfour}$ is a right \valuectx{}.
				By \ih, $\rb\cctxtwo$ is a right \valuectx{} as well.
				Since $\rb{\cctxthree} = \rb{\cctxtwo}\ctxholep{\rb{\cctxfour}}$ according to \refcorollaryp{rb-ctx}{comp},  we obtain that $\rb{\cctxthree}$ is a right \valuectx{} as composition of right \valuectx{s} (\reflemma{r-ctxs-comp-closed}).
			\end{itemize}

			\item Case $\iffsym$: identical to the previous case.
			
			\item Cases $\evsym,\ifesym,\apesym$: they follow from the \ih{} since $\cctx$ is necessarily a prefix of $\cctxtwo$ because $\mol$ is a \pvalue.
			
			\item Cases $\elsym$ and $\eitesym$: they follow from the \ih{},
			since $\aenvtwo=\aenv$ and $\cctx=\cctxtwo$.
			\qedhere
		\end{itemize}
	\end{itemize}
\end{proof}

\begin{lemma}[Determinism]
  \label{l:determinism-open}
  $\toocrumble$ is deterministic.
\end{lemma}
\begin{proof}
  Assume that there exists a \crumble that may be
  decomposed in two ways $\cctxp{(\mol, \fenv)} = \cctxtwop{(\moltwo, \fenvtwo)}$ such that they reduce respectively $\cctxp{(\mol, \fenv)} \Rew a \cctxp\cell$ and $\cctxtwop{(\moltwo, \fenvtwo)} \Rew b \cctxtwop\celltwo$ with rules $a,b\in\set{\betafsym, \iftsym, \iffsym, \ifesym, \apesym, \evsym, \elsym, \eitesym}$.
  
  We prove that it must necessarily be $a=b$, $\cctx=\cctxtwo$, and $\cell=\celltwo$ (up to $\alpha$-equivalence).
  Three cases:
  \begin{itemize}
    \item $\cctx$ strict initial segment of $\cctxtwo$, \ie{} $\cctxtwo = \cctxp\cctxthree$ for some $\cctxthree\neq\ctxhole$.
    We show that this case is not possible: in fact,
    it follows that $ \fenv = \envctxp{(\moltwo, \fenvtwo)} $ for some $\envctx$, thus
    $(\moltwo, \fenvtwo)$ is a \firecell, and by \refprop{harmony-open} it must be normal, contradicting the hypothesis that $(\moltwo,\fenvtwo)$ and $\cell$ reduce with rule $b$.
    \item $\cctx = \cctxtwo$.
     By inspection of the reduction rules, $a=b$: in fact the rule $\betafsym$ applies only when $\mol$ is the application of an abstraction to a \crumbled value, the rule $\evsym$ only when $\mol$ is a variable, and the rule $\elsym$ only when $\mol$ is the application of a variable to a \crumbled value, etc.
     It remains to show that $\cell=\celltwo$ (up to alpha): this follows from the determinism of the lookup in the environment during $\evsym,\elsym$ and $\eitesym$ reductions.
    \item $\cctxtwo$ initial segment of $\cctx$, \ie{} $\cctx = \cctxtwop\cctxthree$. Symmetric to the first case.
    \qedhere
  \end{itemize}
\end{proof}

\begin{proposition}[Overhead transparency]
	\label{prop:transparency-open}
  Let $\cell$ be a reachable \crumble, and let $a \in \set{\evsym,\elsym,\eitesym}$.
  If $ \cell \Rew{a} \celltwo $ then $\rb\cell = \rb\celltwo$.
\end{proposition}
\begin{proof}  
	Let $\cell\defeq\cctxp{(\mol,\fenv)} \Rew a \cctxp{(\moltwo, \fenv)} \eqdef \celltwo $, and let $\fenvtwo,\fenvthree$ be such that $\fenv = \fenvtwo\esub\var{\fenv(\var)}\fenvthree$, noting that $\var$ does not occur in $\fenvthree$ by \reflemmap{invariants-open}{fresh} and \reflemmap{invariants-open}{fvs}.
	We first prove that $\rbp{\mol,\fenv} = \rbp{\moltwo,\fenv}$:
	\begin{itemize}
		\item Case $\evsym$, \ie{} $\mol \defeq \var$ and $\moltwo = \fenv(\var)$. 
		By \reflemma{rb-comp-cell-0}, $\rbp{\var,\fenvtwo\esub\var{\fenv(\var)}\fenvthree} = \rbp{\var,\fenvtwo\fenvthree}\isub\var{\rbp{\fenv(\var),\fenvthree}} = \rbp{\fenv(\var),\fenvthree}$
		as $\cell$ is well-named (\reflemmap{invariants-open}{fresh}).
		By \reflemmap{invariants-open}{fvs}, $\fv{\fenv(\var)}\cap\domain{\fenvtwo\esub\var{\fenv(\var)}} = \emptyset$, therefore $\rbp{\fenv(\var),\fenvthree} = \rbp{\fenv(\var),\fenv}$, and we conclude with $\rbp{\var,\fenv}= \rbp{\fenv(\var),\fenv}$.
		
		\item Case $\elsym$, \ie{} $\mol \defeq \var\molv$ and $\mol = \fenv(\var)\,\molv$. 
		Since $\rbp{\var\molv,\fenv} = \rbp{\var,\fenv}\rbp{\molv,\fenv}$ (\refprop{aux-rb-structural}), we can use the point above to conclude.
		\item Case $\eitesym$, \ie{} $\mol \defeq \xite\var\cell\celltwo$ and $\moltwo = \xite{\aenv(\var)}\cell\celltwo$. 
		Since $\rbp{\xite\var\cell\celltwo,\aenv} = \xite{\rbp{\var,\aenv}}{\rbp{\append\cell\aenv}}{\rbp{\append\celltwo\aenv}}$ (\refprop{aux-rb-structural}), we can use the point above to conclude.
	\end{itemize}
	
	We now prove that $\rb{\cctxp{(\mol,\fenv)}} = \rb{\cctxp{(\moltwo, \fenv)}} $ under the hypothesis that $\rbp{\mol,\fenv} = \rbp{\moltwo,\fenv}$. 
	By cases on $\cctx$: if $\cctx \defeq \ctxhole$ just use the hypothesis.
	Otherwise $\cctx \defeq (\molthree, \env\esub\var\ctxhole)$ and so
	$\rbp{\molthree, \env\esub\var\mol \fenv} = \rbp{\molthree,\env\fenv} \isub\var{\rbp{\mol,\fenv}} = \rbp{\molthree,\env\fenv} \isub\var{\rbp{\moltwo,\fenv}} \allowbreak= \rbp{\molthree, \env\esub\var\moltwo \fenv} $ by \reflemma{rb-comp-cell-0}.
\end{proof}

\begin{lemma}[Substitution]
\label{l:substitution-open}
	Let $\tm$ and $\tmtwo$ be \lat{s}, $\var_1, \dots, \var_n$ pairwise distinct variables,
	and $\fire_1, \dots, \fire_n$ be fireballs.
	If $\tm \Rew{a} \tmtwo$
        for $a \in \set{\betavsym,\iftsym,\iffsym,\ifesym,\apesym}$,
        then $\tm\subs{\var_1}{\fire_1}{\var_n}{\fire_n} \allowbreak
	\tocf \tmtwo\subs{\var_1}{\fire_1}{\var_n}{\fire_n}$.
\end{lemma}

\begin{proof}
	Let $\sigma \defeq \subs{\var_1}{\fire_1}{\var_n}{\fire_n}$.
	By induction on the definition of $\tm \tobv \tmtwo$.
	Cases:
	\begin{itemize}
		\item \emph{Top level:}.
          \begin{itemize}
                \item \emph{$\betavsym$-step},
		\ie $\tm \defeq (\la{\vartwo}{\tmthree})\valtwo \rtobv \tmthree \isub{\vartwo}{\valtwo} \eqdef \tmtwo$ and we can suppose without loss of generality that $\vartwo \notin \bigcup_{i=1}\fv{\fire_i} \cup \{\var_1, \dots, \var_n\}$. 
		According to \reflemma{closure-substitution-open}, $\valtwo\sigma$ is a fireball.
		As a consequence, $\tm\sigma=\allowbreak (\la{\vartwo}{\tmthree\sigma})(\valtwo\sigma) \tof \tmthree\sigma\isub\vartwo{\valtwo\sigma} = 
		\tmthree\isub\vartwo\valtwo\sigma
		=
		\tmtwo\sigma$.
           \item the \emph{$\iftsym,\iffsym,\ifesym,\apesym$ steps} are similar to the $\betavsym$-step. \reflemma{closure-substitution-open} is used in the proof of the $\ifesym$-step to prove that a $\ifesym$-redex where the guard is an abstraction is mapped to a $\ifesym$-redex of the same kind.
           \end{itemize}
		
		\item \emph{Application right}, \ie $\tm \defeq \tmthree\tmfour \Rew{a} \tmthree\tmfive \eqdef \tmtwo$ with $\tmfour \Rew{a} \tmfive$;
		by \ih $\tmfour\sigma \tocf \tmfive\sigma$, and therefore $\tm\sigma=\tmthree\sigma (\tmfour\sigma) \tocf \tmthree\sigma (\tmfive\sigma) = \tmtwo\sigma$.
		
		\item \emph{Application left}, \ie $\tm \defeq \tmthree\valtwo \Rew{a} \tmfour\valtwo \eqdef \tmtwo$ with $\tmthree \Rew{a} \tmfour$; 
		by \ih, $\tmthree\sigma \tocf \tmfour\sigma$;
		according to \reflemma{closure-substitution-open}, $\valtwo\sigma$ is a fireball and hence $\tm\sigma = \tmthree\sigma(\valtwo\sigma) \tocf \tmfour\sigma (\valtwo\sigma) = \tmtwo\sigma$.

		\item \emph{If-then-else guard}, \ie $\tm \defeq \xite\tmfour\tmthree\tmsix \Rew{a} \xite\tmfive\tmthree\tmsix \eqdef \tmtwo$ with $\tmfour \Rew{a} \tmfive$;
		by \ih $\tmfour\sigma \tocf \tmfive\sigma$, and therefore $\tm\sigma = \xite{\tmfour\sigma}{\tmthree\sigma}{\tmsix\sigma} \tocf \xite{\tmfive\sigma}{\tmthree\sigma}{\tmsix\sigma} = \tmtwo\sigma$.	
		\qedhere
	\end{itemize}
\end{proof}

%
%
%
%

\begin{lemma}[Substitution of fireballs]
	\label{l:subs-fireballs-open}
	If $\fenv \defeq \esub{\var_1}{\mol_1} \ldots \esub{\var_n}{\mol_n}$ is a \fireenvironment, then $\subunf{\fenv} = \subs{\var_1}{\tmtwo_1}{\var_n}{\tmtwo_n}$ where all the $\tmtwo_i$\!'s are fireballs.
\end{lemma}

\begin{proof}
	By induction on the length of $\env$.
	If $\fenv \defeq \emptyenv$, then $\subunf{\fenv} = \{\}$ and the statement is vacuously true.
	Otherwise $\fenv \defeq \esub{\var}{\mol}\fenvtwo$ with $\fenvtwo \defeq  \esub{\var_1}{\mol_1}\ldots\esub{\var_n}{\mol_n}$ and then $\subunf{\fenv} = \isub{\var}{\rbp{\mol, \fenvtwo}} \cup \subunf{\fenvtwo}$;
	by \ih (since $\fenvtwo$ is a \fireenvironment), $\subunf{\fenvtwo} = \subs{\var_1}{\tmtwo_1}{\var_n}{\tmtwo_n}$ where all the $\tmtwo_i$\!'s are fireballs;
	also $\rbp{\mol, \fenvtwo}$ is a fireball by definition of \fireenvironment, as $\fenv = \genvp{(\mol, \envtwo)}$ with $\genv \defeq \esub{\var}{\ctxhole}$;
	therefore, $\subunf{\fenv} = \{\replace{\var}{\rbp{\mol,\fenvtwo}}, \replace{\var_1}{\tmtwo_1}, \dots, \replace{\var_n}{\tmtwo_n}\}$ satisfies the statement.
\end{proof}

\begin{lemma}[Read-back vs. append]
\label{l:rb-append-open}
	For any \crumble $\cell$ and \fireenvironment $\fenv$,  $\rbp{\append{\cell}{\fenv}} = \rb{\cell} \subunf{\fenv}$.
\end{lemma}

\begin{proof}
	By induction on the length of $\fenv$. 
	If $\fenv \defeq \emptyenv$ then $\subunf{\fenv} = \{\}$ and hence $\rbp{\append{\cell}{\fenv}} = \rb{\cell} = \rb{\cell} \{\} = \rb{\cell} \subunf{\fenv}$.
	Otherwise $\fenv \defeq \esub{\var}{\mol}\fenvtwo$ where $\var \notin \domain{\fenvtwo}$;
	by \ih (since $\fenvtwo$ is a fireball environment), $\rbp{\append{(\append{\cell}{\esub{\var}{\mol}})}{\fenvtwo}} = \rbp{\append{\cell}{\esub{\var}{\mol}}} \subunf{\fenvtwo}$ and $\rbp{\mol, \fenvtwo} = \rbp{\append{(\mol,\emptyenv)}{\fenvtwo}} = \rbp{\mol, \emptyenv} \subunf{\fenvtwo} = \rb{\mol} \subunf{\fenvtwo}$;
	by the definitions of append and read-back, $\rbp{\append{\cell}{\esub{\var}{\mol}}} = \rb{\cell}\isub{\var}{\rb{\mol}}$;
	therefore, $\rbp{\append{\cell}{\fenv}} = \rbp{\append{(\append{\cell}{\esub{\var}{\mol}})}{\fenvtwo}} = \rbp{\append{\cell}{\esub{\var}{\mol}}} \subunf{\fenvtwo} = (\rb{\cell}\isub{\var}{\rb{\mol}}) \subunf{\fenvtwo} = \rb{\cell} (\isub{\var}{\rb{\mol}\subunf{\fenvtwo}} \cup \subunf{\fenvtwo}) = \rb{\cell} \isub{\var}{\rbp{\mol,\fenvtwo}} \cup \subunf{\fenvtwo} = \rb{\cell}\subunf{\fenv}$.
\end{proof}

\begin{lemma}
	\label{l:clos-sigma-abs-open}
   If $\rb\cell \Rew{a} \rb\celltwo$
   for $a \in \set{\betavsym,\iftsym,\iffsym,\ifesym,\apesym}$,
   then $\rbp{\append\cell\fenv} \tof{} \rbp{\append\celltwo\fenv}$.
\end{lemma}
\begin{proof}
	According to \reflemma{subs-fireballs-open}, $\subunf{\fenv} = \subs{\var_1}{\fire_1}{\var_n}{\fire_n}$ where $\fire_1, \dots, \fire_n$ are fireballs.
	By \reflemma{rb-append-open} and \reflemma{substitution-open}, $\rbp{\append\cell\fenv} = \rb{\cell} \subunf{\fenv} \tof \rb{\celltwo} \subunf{\fenv} = \rbp{\append\celltwo\fenv}$
\end{proof}

Note that \reflemma{clos-sigma-abs-open} does not hold if we let $\Rew{a}$ be $\toin$ in the hypothesis.
Indeed, take $\cell \defeq ((\la{\var}{(\var, \emptyenv)})\vartwo, \esub{\vartwo}{(\varthree\varthree, \emptyenv)})$ and $\celltwo \defeq (\vartwo, \esub{\vartwo}{\varthree\varthree})$ and $\fenv \defeq \esub{\varthree}{\la{\var}{(\var\var, \emptyenv)}}$: then, $\rb{\cell} = (\la{\var}\var)(\varthree\varthree) \toin \varthree \varthree = \rb{\celltwo}$ but $\rbp{\append\cell\fenv} = (\la{\var}\var)((\la{\var}\var\var)\la{\var}\var\var) \not\tof (\la{\var}\var\var)\la{\var}\var\var = \rbp{\append\celltwo\fenv}$.
The problem is essentially due to the fact that fireballs, contrary to \lav{s}, are not closed by substitution: this a notable difference
between the closed case (where the normal forms coincide with closed \lav{}s) and the open case (where the normal forms coincide with fireballs).

\begin{proposition}[Principal projection]
	\label{prop:beta-projection-open}
  Let $\cell$ be a reachable \crumble. If $\cell \Rew{a} \celltwo$ for $a \in \set{\betafsym,\iftsym,\iffsym,\ifesym,\apesym}$ then $\rb\cell \Rew{a} \rb\celltwo$.
\end{proposition}
\begin{proof}
 Note that for every $\mol,\aenv$,  $(\mol,\aenv) = \append{(\mol,\epsilon)}{\aenv}$. Therefore all steps can be written in the form
 $\cctxp{(\mol,\aenv)} \Rew{a} \cctxp{\append\cell\aenv}$ where $\mol$ is not a
\crumbled value. The \crumble context $\cctx$ unfolds to a right \valuectx{} by \reflemmap{invariants-open}{ctx-decoding}. We need to prove that
$\rb{\cctxp{(\mol,\aenv)}} \Rew{a} \rbp{\cctxp{\append\cell\aenv}}$. By Lemma~\ref{l:plug-append} and \reflemma{clos-sigma-abs-open}, it suffices to prove that $\rb{\cctxp{(\mol,\epsilon)}} \Rew{b} \rb{\cctxp{\cell}}$ for all
$a \neq \betafsym$ and $b = a$ or for $a = \betafsym$ and $b = \betav$.

We proceed by cases on the rule $\Rew{a}$.
\begin{itemize}
 \item Rule $\betafsym$: we need to prove that
	$\rb{\cctxp{((\la\var\cell)\,\molv,\emptyenv)}} \tobv{} \rb{\cctx\ctxholep{\append\cellthree{\esub\vartwo\molv}}}$ where
	$\append\cellthree{\esub\vartwo\molv} \defeq (\append\cell{\esub\var\molv})^\alpha$.
	\begin{align*}
	\rb{\cctxp{((\la\var\cell)\,\molv,\emptyenv)}} & = \rb\cctx\ctxholep{(\la\var\rb\cell)\,\rb\molv} & \text{by \refcorollaryp{rb-ctx}{cell}}\\
	& \alphaequiv{}
	\rb\cctx\ctxholep{(\la{\vartwo}\rb{\cellthree})\,\rb\molv} & \\
	& \tobv \rb\cctx\ctxholep{\rb{\cellthree}\isub{\vartwo}{\rb\molv}} & \\
	& = \rb\cctx\ctxholep{\rb{\append{\cellthree}{\esub{\vartwo}\molv}}} \\
	& = \rb{\cctx\ctxholep{\append{\cellthree}{\esub{\vartwo}\molv}}} & \text{by \refcorollaryp{rb-ctx}{cell}}
	\\
	& = \rb{\cctx\ctxholep{(\append{\cell}{\esub\var\molv})^\alpha}}. &
	\end{align*}
	Note that the second use of \refcorollaryp{rb-ctx}{cell} requires that
	$\cctx\disj(\append\cellthree{\esub\vartwo\molv})$ \ie that
	$\fv{\cctx} \cap \domain{\append\cellthree{\esub\vartwo\molv}} = \emptyset$,
	which follows from the side condition about $\alpha$-renaming in the $\betafsym$ rule.
 \item Rules $\iftsym,\iffsym,\ifesym,\apesym$:
  a quick check by cases over $(\mol,\aenv) \rootRew{a} \append\cell\aenv$ shows that $\rb\mol \Rew{a} \rb\cell$.

  For example,
  $(\xite\true\cell\celltwo,\aenv) \rootRew{\iftsym} \append\cell\aenv$ and\\
  $\rbp{\xite\true\cell\celltwo} = \xite\true{\rb\cell}{\rb\celltwo}
   \toift \rb\cell$. The other cases are all similar.

  Thus
	\begin{align*}
	  \rb{\cctxp{(\mol,\epsilon)}} & = \rb\cctx\ctxholep{\rb\mol} & \text{by \refcorollaryp{rb-ctx}{cell}}\\
        & \Rew{a} \rb\cctx\ctxholep{\rb\cell} \\
        & = \rb{\cctx\ctxholep{\cell}} & \text{by \refcorollaryp{rb-ctx}{cell}}\\
        \end{align*}
  Note that the second use of \refcorollaryp{rb-ctx}{cell} requires that
  $\cctx\disj \cell$. The property holds because all substitutions outside
  abstractions in the rhs of the rules $\Rew{a}$ under consideration were such in the lhs,
	and because reachable \crumblep{s} are well-named (\reflemmap{invariants-open}{fresh}).
	\qedhere
\end{itemize}
\end{proof}

\begin{lemma}[Halt]
  \label{l:progress-open}
   If $\cell$ is $\ocrumblesym$-normal then $\rb\cell$ is $\tocf$-normal.
\end{lemma}
\begin{proof}
  By \refprop{harmony-open}, if $\cell$ is normal then it is a \firecell \ie{} $\cell=\fcell$.
  By definition of $\fcell$, $\rb\cell$ is a fireball.
	By harmony for $\cfirecalc$, $\rb\cell$ is $\cbetafsym$-normal.
\end{proof}

\begin{proofof}[\refthm{implementation-open}]
\Paste{thm:implementation-open}
\end{proofof}
\begin{proof}
	\begin{enumerate}
		\item See \refprop{read-back-inverse-transl}.
		\item See \refprop{beta-projection-open}.
		\item See \refprop{transparency-open}.
		\item See \reflemma{determinism-open}.
		\item See \reflemma{progress-open}.
		\item Immediate consequence of the equivalent of \reflemma{linear-bounds} for the \oaglam.
		\qedhere
	\end{enumerate}
\end{proof}

\subsection{The \oapglam}\label{app:towards-impl-open-pre}

The design of a machine, called \emph{\oapglam}, for the open case that decomposes the search for the next redex and (un)plugging in $O(1)$ transitions follows 
the same pattern as for the \apglam  in \refapp{towards-impl-closed-pre} for the closed case. 
In particular, the definitions of pointed environment, translation $\myiota\cdot$ (from \crumblep{s} to \penvironment{s}) and read-back $\atoi\cdot$ (from \penvironment{s} to \crumblep{s}) are the same.
The transitions of the \oapglam (whose union is noted $\topocrumble$) differ from the ones of the \apglam exactly as the transitions of \oaglam differ from the ones of the \aglam.



\subsubsection{Implementation.}

\paragraph*{\Penvironment{}s.}
%
%
%
For the \oapglam, the definitions of pointed environment, translation $\myiota\cdot$ (from \crumblep{s} to \penvironment{s}) and read-back $\atoi\cdot$ (from \penvironment{s} to \crumblep{s}) are the same as for the \apglam, see \refapp{towards-impl-closed-pre}.

\paragraph*{Evaluation.}
The transition rules of the \oapglam are:
\begin{align*}
  \env \esub\var{(\la\vartwo\cell)\,\molv} \sep \fenv
  &  \tof 
  \env \esub\var\mol\envtwo\esub\varthree\molv
  \sep \fenv
  && (i)
  \\%
  \\[-.5\baselineskip]
    \env \esub\var{\xite\true\cell\celltwo} \sep \fenv
		    & \toift 
		      \env \esub\var\mol \envtwo \sep \fenv & & (ii) \\
    \env \esub\var{\xite\false\cell\celltwo} \sep \fenv
		    & \toiff 
		         \env \esub\var\mol \envtwo \sep \fenv & & (iii)
		     \\
		    \env \esub\var{\xite\molv \cell \celltwo} \sep \fenv
		    & \toife 
		    \env \esub\var{\err} \sep \fenv & & (iv)
		    \\
		    \env \esub\var{\molv \molvtwo} \sep \fenv
		    & \toape 
		    \env \esub\var{\err} \sep \fenv& &(v)
		    \\
		      \\[-.5\baselineskip]
  \env \esub\vartwo\var \sep \fenv
  & \toev
  \env\esub\vartwo{\fenv(\var)} \sep \fenv
  & & (vi)
  \\
  \env \esub\vartwo{\var\molv} \sep \fenv
  & \toel 
  \env \esub\vartwo{\fenv(\var)\molv} \sep \fenv
  & & (vi)
  \\
  \env \esub\vartwo{\xite\var\cell\celltwo} \sep \fenv
  & \toeite 
  \env \esub\vartwo{\xite{\fenv(\var)}\cell\celltwo} \sep \fenv
    & & (vi)
  \\%
  \\[-.5\baselineskip]
  \env \esub\var\mol \sep \fenv
  & \toc 
  \env \sep \esub\var\mol \fenv
  & & (vii)
\end{align*}

\noindent where 
\begin{enumerate}[i.]
 \item $\la\varthree{(\mol,\envtwo)} \defeq (\la\vartwo\cell)^\alpha$ such that $(\env \esub\var\mol\envtwo\esub\varthree\molv
\sep \fenv)$ is well-named.
\item where $\cell \eqdef (\mol,\envtwo)$
\item where $\celltwo \eqdef (\mol,\envtwo)$
\item if $\molv=\la\var\cellthree$ or $\molv = \err$.
\item if $\molv\in \set{\true,\false,\err}$.
 \item if $\var \in \domain{\fenv}$.
 \item if none of the other rules is applicable, \ie{} when $\mol$ is a practical value $\pval$ or when $\mol$ is $\var$, $\var\molv$, or $\xite\var\cell\celltwo$  but $\var$ is not defined in $\fenv$.
\end{enumerate}

The transition function $\topocrumble$ of the \oapglam is then defined as the union of the rules above.
A \emph{principal transition} of the \oapglam is a transition $\Rew{a}$ for any rule $a\in\set{\betaf,\iftsym,\iffsym,\ifesym,\apesym}$.

\begin{definition}[Reachable state]
  A \penvironment is said to be \emph{reachable} (in the \oapglam) if it is obtained from evaluation steps starting from the translation $\myiota\cell$ of a well-named \crumblep $\cell$.
\end{definition}

\begin{lemma}[Invariants] 
  \label{l:zip-qualitative-invariants-open}
  Let $\benv$ be a reachable \penvironment:
  \begin{enumerate}
    \item \label{p:zip-qualitative-invariants-open-freshness}\emph{Freshness:} $\benv$ is well-named;
    \item \emph{Rightmost:} $\benv = (\env \sep \fenv)$ for some $\env$ and some \fireenvironment $\fenv$.
  \end{enumerate}
\end{lemma}
\begin{proof}
  By induction on the length of the evaluation sequence leading to $\benv$. The base cases hold by the definition of reachability and by the definition of $\myiota\cdot$. As for the inductive step, we proceed by cases on the transitions:
  \begin{enumerate}
    \item
      For $\betafsym$ the claim follows from the side condition.
      The rules in $\set{\iftsym,\iffsym,\ifesym,\apesym,\csym}$ do not increase the number of explicit substitutions occuring in $\benv$ outside of abstractions, hence the claim follows from the \ih{}. The rules in $\set{\evsym,\elsym,\eitesym}$ copy a \pvalue from the environment $\fenv$: note that $\fenv(\var)$ is either an abstraction (which does not influence well-namedness) or a boolean or an error (which do not contain explicit substitutions).
    \item The rules in $\set{\betafsym,\iftsym,\iffsym,\ifesym,\apesym,\evsym,\elsym,\eitesym}$ do not change the evaluated part, hence the claim follows from the \ih{}. 
    As for the $\csym$ rule, by \reflemma{zip-aux-cell-rrb-open} it suffices to prove that $(\mol,\fenv)$ is a \firecell{}, knowing that the other transition rules cannot be applied: this follows from \reflemma{harmony-aux1-open}.
    \qedhere
  \end{enumerate}
\end{proof}

\begin{lemma}[Harmony for the \oapglam]
  \label{l:zip-harmony-open}
  Let $\benv$ be a reachable \penvironment in the \oapglam:
  $\benv$ is normal if and only if it has the form $(\emptyenv \sep \fenv)$ for some non-empty \fireenvironment $\fenv$.
\end{lemma}
\begin{proof}
  The proof of the implication from right to left is trivial.
  As for the other direction, let $\benv$ be a reachable normal \penvironment. 
  By \reflemma{zip-qualitative-invariants-open}, $\benv$ has the form $(\env \sep \fenv)$. Note that $\env$ must be empty; otherwise one of the transitions in $\set{\betafsym,\iftsym,\iffsym,\ifesym,\apesym,\evsym,\elsym,\eitesym,\csym}$ could be applied, contradicting the hypothesis that $\benv$ is normal.
\end{proof}

\begin{lemma}\label{l:zip-aux-cell-rrb-open} 
  If $\esub\var\mol\env$ is a \fireenvironment,
  then $(\mol,\env)$ is a \firecell.
\end{lemma}
\begin{proof}
  Assume that $\esub\var\mol\env$ is a \fireenvironment, and
  let $(\mol,\env) = \cctxp{(\moltwo,\envtwo)}$ for some $\cctx,\moltwo,\envtwo$. Then $\esub\var\mol\env = \envctxp{\cctxp{\moltwo,\envtwo}}$ for $\envctx\defeq\esub\var\ctxhole$.
  The requirements for a \firecell follow from the definition of \fireenvironment for $\esub\var\mol\env$.
\end{proof}

%

\begin{theorem}[Implementation]
	\label{thm:zip-implementation-open}
		Let $\benv$ a \penvironment that is reachable by the \oapglam.
		\begin{enumerate}
			\item \emph{Initialization: }
			$\atoi{\myiota\cell}=\cell$ for every \crumblep $\cell$.
			\item \emph{Principal Projection: }
			if $\benv \to_a \benvtwo$
			then $\atoi\benv \to_a \atoi\benvtwo$ for any rule $a\neq \csym$.
			\item \emph{Overhead Transparency: }
			if $\benv \toc \benvtwo$,
			then $\atoi\benv = \atoi\benvtwo$.
			\item \emph{Determinism: }
			the transition function $\topocrumble$ is deterministic.
			\item \emph{Halt: }
			if $\benv$ is normal, then $\atoi\benv$ is normal. 
			\item \emph{Overhead Termination: } $\toc$ terminates.   
		\end{enumerate}
		Therefore, the \oapglam (with its transition function $\topocrumble$), the \oaglam  (with $\toocrumble$), and the read-back $\atoi\cdot$ form an implementation system.
\end{theorem}
\begin{proof}~
  \begin{enumerate}
    \item Let $\cell=(\mol,\env)$: by the definitions,
     $\atoi{\myiota\cell}=\atoi{\esub\var\mol\env\sep\emptyenv} = (\mol,\env\emptyenv) = \cell$.
    \item There is a clear one-to-one correspondence between the transitions of the \oapglam and the \oaglam (apart from $\csym$). The proof is similar to the one of \refthm{zip-implementation-closed}.

    \item By inspection of the rule $\csym$. We need to prove that 
     $\atoi{\env \esub\var\mol \sep \fenv} =
     \atoi{\env \sep \esub\var\mol \fenv}$. By cases on the structure of $\env$:
      if $\env=\emptyenv$, then $\atoi{\emptyenv \esub\var\mol \sep \fenv} = (\mol, \fenv) =
      \atoi{\emptyenv \sep \esub\var\mol \fenv}$.
      If instead $\env=\esub\vartwo\moltwo\envtwo$, then 
      $\atoi{\env \esub\var\mol \sep \fenv} = (\moltwo, \envtwo\esub\var\mol\fenv) =
      \atoi{\env \sep \esub\var\mol \fenv}$.

    \item The rule $\csym$ can be applied by definition only when the other rules cannot be applied. The remaining rules apply to a \penvironment of the form $(\env\esub\var\mol\sep\fenv)$, for distinct shapes of $\mol$:
    \begin{itemize}
    \item The rules $\set{\betafsym,\apesym,\elsym}$ apply when $\mol$ is an application, and respectively the application of an abstraction to a \crumbled value, of a boolean/error to a \crumbled value, and of a variable to a \crumbled value.
    \item The rule $\evsym$ applies when $\mol$ is a variable.
    
    \item The rules in $\set{\iftsym,\iffsym,\ifesym,\eitesym}$ apply when $\mol$ is a $\xitecompact$, and in clearly disjoint cases according to the structure of the condition: respectively, when the condition is $\true$, $\false$, an abstraction or an error, and a variable.

    \item Finally, note that no single rule in $\set{\evsym,\elsym,\eitesym}$ can transition to different \penvironment{s} due to the lookup of a variable which has multiple occurrences in the environment: the lookup is deterministic because the environment is well-named (\reflemmap{zip-qualitative-invariants-open}{freshness}).
    \end{itemize}
    
    \item By \reflemma{zip-harmony-open}, $\benv$ is normal iff it has the form $(\emptyenv \sep \fenv)$ for some non-empty \fireenvironment $\fenv$.
     Then $\atoi{\benv}=\atoi{\emptyenv\sep\fenv}$ which is a \firecell by \reflemma{zip-aux-cell-rrb-open}. 
     By \refprop{harmony-open}, $\atoi{\benv}$ is normal.
     
     \item Immediate consequence of forthcoming \refcoro{zip-aux-r-e-open} (proved independently).
     \qedhere
  \end{enumerate}
\end{proof}

\subsubsection{Complexity}
The complexity analysis for the \oapglam is analogue to the one for the \apglam.
More precisely, following the same approach and notations as in \Cref{app:towards-impl-closed-pre} and in \Cref{app:towimpl-closed}, we can prove:

\begin{lemma}[Number of $\csym$-transitions]
	\label{l:number-comm-pointed-open}
		Let $\cell$ be a well-named \crumblep, and let $\exec \colon \myiota\cell \topocrumble^* \benv = (\env\sep\aenv)$ an execution in the \oapglam. 
		Then $\length\env \leq \length\cell + (\no{\betaf}\exec+\no\iftsym\exec+\no\iffsym\exec) \cdot L(\cell) - \no\csym\exec$.
\end{lemma}

\begin{corollary}\label{coro:zip-aux-r-e-open}
	Let $\tm$ be a \lat{}. 
	For a normalizing execution $\exec$ in the \oapglam starting from $\myiota{\mytr{\tm}}$, we have $\no\csym\exec \leq (\sizepr\exec + 1) \cdot \len\tm$.
\end{corollary}

\begin{theorem}[The \oapglam is bilinear]
	\label{thm:efficient-pointed-open}
	For any \lat{} $\tm$ and any \oapglam execution $\exec\colon\myiota{\mytr\tm} \topocrumble^* \benv$, the cost of implementing $\exec$ on a RAM is  $O((\sizepr\exec + 1) \cdot \len{\tm})$.
\end{theorem}
\begin{proof}
	Essentially the same proof as the one of \refthm{efficient-pointed-closed}.
\end{proof}

Analogously to the closed case, the previous theorem also has a consequence for the non-pointed case. Since the theorem shows that the cost of searching for redexes and the (un)plugging operations can indeed be realised in bilinear time, it is also true that the \oaglam can be implemented in bilinear time, improving \refthm{compl-open} by removing the \emph{up to search and (un)plugging} side condition.
\begin{corollary}[The \oaglam is bilinear]\label{cor:compl-open}
	For any \lat{} $\tm$ and any \oaglam execution $\exec\colon\mytr\tm\toocrumble^*\cell$, the cost of implementing $\exec$ on a RAM is $O((\sizepr\exec + 1) \cdot \len{\tm})$.
\end{corollary}

\section{Implementation in OCaml}
\label{sect:ocaml}
\lstset{language=[Objective]{Caml}}
\label{app:ocamlimpl}

\lstset{
	basicstyle=\small\ttfamily,          
	stringstyle=\ttfamily
}


The goal of this section is implementing in OCaml the \apglam and \oapglam 
presented in \refapp{towards-impl-closed-pre} and \refapp{towards-impl-open-pre}. 
We are going to describe the abstract requirements of the data structures, before picking a concrete implementation. 
The two machines share the same data structures and auxiliary functions, and only differ by five lines of code.

\paragraph*{Data structures.}

The machine works on \penvironment{}s of the form $(\env \sep \aenv)$.
The unevaluated part of the environment $\env$ and the evaluated part $\aenv$ are subject to different requirements:
\begin{itemize}
  \item The unevaluated part $\env$ is extended only on the right by the transitions $\msym,\iftsym,\iffsym$ by concatenating an unevaluated environment.
  Only the rightmost entry is inspected by every reduction rule. The $\csym$ transition removes the rightmost entry, moving it to the evaluated part. Therefore the unevaluated environment must implement the \emph{catenable stack} interface, allowing to perform catenation, pop, and topmost inspection in constant time.
  \item As for the evaluated part $\aenv$, transitions never exploit the sequential structure of $\aenv$. On the contrary, the $\evsym,\elsym,\eitesym$ transitions need to access the entry associated with a variable $\var$ in constant time. The only other operation required (by the $\csym$ rule) is to add an entry to it in constant time. 
\end{itemize}

To satisfy lookup in constant time for $\aenv$, we implement $\aenv$ as a store, thus ignoring its list structure. In turn, this choice impacts on the data structure for terms, because it forces (occurrences of) variables to be implemented as pointers to memory locations. More explicitly, an entry $\esub\var\mol$ is represented as a node $n$ which is a record containing a field \verb+content+ holding $\mol$ together with additional fields soon to be described. If $\alpha$ is the address of $n$, occurrences of $\var$ are represented in OCaml as $\mathtt{Shared}(n)$, which means a memory cell tagged as \verb+Shared+ and holding the pointer $\alpha$.

Occurrences of $\lambda$-bound variables are instead presented as $\texttt{Var}(v)$ where $v$ is a unique identifier for that variable. Thus the data structure for terms is:

\begin{lstlisting}
...
and term =
 | Err
 | True
 | False
 | Var of var
 | Lam of var * crumblep
 | App of term * term
 | IfThenElse of term * crumblep * crumblep
 | Shared of node
...
\end{lstlisting}

A \crumblep is a term coupled with an unevaluated environment. As discussed above, the unevaluated environment implements a catenable stack interface. The simplest implementation, which we adopted, is a linked list of nodes, referenced by two pointers to the first and the last entries of the stack, or a special value to represent the empty stack. Concretely, in OCaml we use an option type for that.


\begin{lstlisting}
...
and env = (node * node) option
and crumblep = term * env
...
\end{lstlisting}

Each node has a field \verb+prev+ used to point to the previous entry in the environment. An additional field \verb+copying+ and mutability of all fields are required to implement $\alpha$-renaming in linear time; we explain their use later. Therefore a node is:

\begin{lstlisting}
type node = 
 { mutable content : term
 ; mutable copying : bool
 ; mutable prev : node option }
...
\end{lstlisting}

Unreachable nodes can be garbage-collected by the runtime of OCaml. Because the evaluator holds a pointer to the unevaluated environment, only evaluated nodes can be garbage-collected.

Finally, we implement the datatype of unique $\lambda$-bound variable identifiers \verb+var+ as the address of an OCaml record that holds no useful information. Thus comparing variables can be achieved using pointer equality \verb|==|. Concrete implementations can add fields to the record, for example to associate the name of the variable as a string.

\begin{lstlisting}
type var = {dummy: unit} (*no empty records in OCaml*)
\end{lstlisting}

A summary of the data structures can be found in \reftab{data-structures}.

\begin{table}
\begin{tabular}{|c|}\hline
\begin{lstlisting}
type var = { dummy : unit }
type node = 
 { mutable content : term
 ; mutable copying : bool
 ; mutable prev : node option }
and env = (node * node) option
and crumblep = term * env
and term =
 | Err
 | True
 | False
 | Var of var
 | Lam of var * crumblep
 | App of term * term
 | IfThenElse of term * crumblep * crumblep
 | Shared of node

let mk_node content = 
 { content ; copying = false ; prev = None }

let push n e =
 n.prev <- Some e
\end{lstlisting}\\
\hline\end{tabular}
\caption{Data structures}
\label{tab:data-structures}
\end{table}

\paragraph*{Implementation of transitions.}
The code that implements evaluation in the closed and open cases can be found in \reftab{abnormal-evaluation}. For the same reason as it is discussed in \refapp{towards-impl-closed-pre}, the input of the evaluation functions \verb+eval_c+/\verb+eval_o+ is not a \crumblep, but a \crumbled unevaluated environment $\myiota{\mol,\env}=\esub\var\mol\env$ where $\var$ is a fresh variable. More precisely, the evaluation functions take in input just a node \verb+n+, which is the rightmost entry of the \crumbled unevaluated environment that has to be evaluated.

The code that implements the transition $\msym$ for $(\la\vartwo\cell)\molv$ is the most complex because it must:
\begin{enumerate}
  \item $\alpha$-rename $\la\vartwo\cell$ to
    $\la{\vartwo'}{(\mol', \env')}$;
  \item 
   change the top of the unevaluated environment (stack) from $\esub{\texttt n}{(\la\vartwo\cell)\molv}$ to
   $\esub{\texttt{n}}{\mol'}$ and append $\env'$ to it;
  \item push $\esub{\vartwo'}{\molv}$ on top of the unevaluated environment.
\end{enumerate}

To implement the previous steps efficiently, the code creates the node $\vartwo'$ containing $\molv$ and then calls a function \verb+copy_crumbp+ $y \, y' \, \cell$ that performs step 1 in linear time, returning the new unevaluated \crumblep $\cell'$.
Then $\append n {\cell'}$ performs step 2 in constant time by concatenating $\cell'$ to the unevaluated environment (and therefore to $n$, its topmost element). The code of $\_ \append \_$ is given in Table~\ref{tab:copy_env}.

Finally, \verb+push+ $y' \, (\append n {\cell'})$ performs step 3 in constant time by pushing $y'$ on the new top of the unevaluated environment.

The transitions $\elsym,\evsym,\eitesym,\apesym,\ifesym$ rules just update the content of the top of the unevaluated environment in the required way.

The transitions $\iftsym$ and $\iffsym$ perform in constant time the plugging of the \crumblep $\cell$ to the unevaluted environment, then call the evaluation function on the new topmost entry of the unevaluated environment.

The transition $\csym$ is implemented by the function \texttt{pop} that pops the top of the unevaluated environment and calls evaluation on the new top (if present):
\begin{lstlisting}
...
and pop n =
 match n.prev with
 | None -> n.content
 | Some p ->
    n.prev <- None; 
    eval_c/o p
...
\end{lstlisting}
When a node is popped, its \verb+prev+ pointer is unset to facilitate garbage-collection of unreferenced nodes.

Otherwise a normal form is reached, and evaluation returns the term that, pointing to the evaluated environment, consists of
the normal \crumblep. 

Let us remark that the implementation is tail-recursive\footnote{when the \texttt{pop} function is inlined}; since OCaml
optimizes tail-recursion, the machine only consumes constant space on
the process execution stack.

As a minor optimization to the expected code, our implementation
merges execution of rule $\evsym$ with that of the $\csym$ step which always follows the former. The merging is obtained calling \verb+pop+ in place of \verb+eval_c+/\verb+eval_o+. 

\begin{figure*}
  \hspace{-1em}
\begin{tabular}{|l|l|}\hline
\begin{lstlisting}[mathescape=true]
let rec eval_c n =
 match n.content with
 | App(Lam(y,c), t) -> 
    (* $\msym$ *)
    let y' = mk_node t in
    let c' = copy_crumbp y y' c in
    push y' (n @ c') ;
    eval_c y'
 | App
    (Shared{content=t1},

     t2) ->
    (* $\elsym$ *)
    n.content <- App(t1, t2);
    eval_c n
 | App((True|False|Err), _) ->
    (* $\apesym$ *)
    n.content <- Err ;
    eval_c n
 | IfThenElse(True,c,_)  (* $\iftsym$ *)
 | IfThenElse(False,_,c) (* $\iffsym$ *)
    -> eval_c (n @ c)
 | IfThenElse
    (Shared {content=t1},

     t2,t3) ->
    (* $\eitesym$ *)
    n.content <-
     IfThenElse(t1,t2,t3) ;
    eval_c n
 | IfThenElse((Lam _|Err),_,_) ->
    (* $\ifesym$ *)
    n.content <- Err ;
    eval_c n
 | Shared {content}
   ->
    (* $\evsym$ *)
    n.content <- content ;
    pop n
 | Lam _ | Err | True | False ->
    (* $\csym$ *)
    pop n
 | Var _ | App(Var _, _)

 | IfThenElse(Var _,_,_) ->

    failwith "Open term"
 | _ -> assert false
\end{lstlisting}
&
\begin{lstlisting}[mathescape=true]
let rec eval_o n =
 match n.content with
 | App(Lam(y,c), t) ->
    (* $\betafsym$ *)
    let y' = mk_node t in
    let c' = copy_crumbp y y' c in
    push y' (n @ c') ;
    eval_o y'
 | App
   (Shared{content=
    (Lam _|True|False|Err) as t1},
     t2) ->
    (* $\elsym$ *)
    n.content <- App(t1, t2);
    eval_o n
 | App((True|False|Err), _) ->
    (* $\apesym$ *)
    n.content <- Err ;
    eval_o n
 | IfThenElse(True,c,_)  (* $\iftsym$ *)
 | IfThenElse(False,_,c) (* $\iffsym$ *)
    -> eval_o (n @ c)
 | IfThenElse
    (Shared{content=
     (Lam _|True|False|Err) as t1},
      t2,t3) ->
    (* $\eitesym$ *)
    n.content <-
     IfThenElse(t1,t2,t3) ;
    eval_o n
 | IfThenElse((Lam _|Err),_,_) ->
    (* $\ifesym$ *)
    n.content <- Err ;
    eval_o n
 | Shared{content=
    (Lam _|True|False|Err) as c} ->
    (* $\evsym$ *)
    n.content <- c ;
    pop n
 | Lam _ | Err | True | False ->
   (* $\csym$ *)
   pop n
 | Var _ | App(Var _, _)
 | Shared _ | App(Shared _, _)
 | IfThenElse (Var _,_,_) ->
    (* $\csym$ *)
    pop n
 | _ -> assert false
\end{lstlisting}
\\ \hline
\end{tabular}

\caption{Evaluation: closed (left) vs open (right)}
\label{tab:abnormal-evaluation}
\end{figure*}

The complexity of each case is $O(1)$, but for the $\msym$ rule
which requires a renaming of \crumblep (\verb+copy_crumbp+).
It remains to see how this operation can be implemented with linear complexity.

\paragraph*{Implementation of $\alpha$-renaming.}

We implement $\alpha$-renaming of unevaluated environments by creating a copy of the environment. The representation in memory of the environment is a DAG because terms in the nodes of the environment contain occurrences of \verb|Shared| nodes defined in the same environment. Therefore we need to implement a copy algorithm over DAGs that runs in linear time.

The algorithm consists in using the content field of
nodes to perform the renaming. When a node is being copied, it is  temporarily put in the \verb+copying=true+ status, and its \verb|content| field is changed to point to the corresponding new node.
Then, the rest of the environment is copied recursively. When an occurrence of a node that is being copied is found in the term being copied, it is replaced with the new node stored in the \verb|content| field of the old one. Finally, when the copy is over, the \verb|copying| status of every node is reset to \verb|false| and the previous value of \verb|content| is restored, yielding the original environment.

The auxiliary \verb|copying_node y y' f| function, where \verb|y| is the node to be copied to \verb|y'|, implements the idea above by temporarily putting \verb|y| in \verb|copying=true| status, until \verb|f| is executed.

\begin{lstlisting}
let copying_node y y' f =
  let saved = y.content in
  y.content <- Shared y' ;
  y.copying <- true ;
  let res = f () in
  y.content <- saved ;
  y.copying <- false ;
  res
\end{lstlisting}

The \verb|copy_crumbp v n p| function (\reftab{copy_env}) not only implements the algorithm above by copying \verb|p|, but it also replaces occurrences of \verb|Var v| (the bound variable in a $\tom$ redex) with \verb|Shared n| (the new node pointing to the argument of the redex), \ie{}: let $\cell \defeq \mol_0, \esub{\var_1}{\mol_1}\ldots\esub{\var_k}{\mol_k}$; then 
\verb|copy_crumbp v n p| $ = \mol_0', \esub{\vartwo_1'}{\mol_1'}\ldots\esub{\vartwo_k'}{\mol_k'}$, where $\mol_i' \defeq \mol_i\isub{v}{n}\isub{\var_{i+1}}{\var_{i+1}'}\ldots\isub{\var_k}{\var_k'}$.

The implementation of \verb|copy_crumbp v n p| uses an auxiliary function \verb+copy_env+ that iterates over the environment copying it. The function \verb+copy+ copies a \crumbledt.

\begin{table}[p!]
\begin{lstlisting}
let copy_crumbp v n p =
 let rec copy = function
  | Var v' when v == v' -> Shared n
  | Shared {content; copying} when copying -> content
  | Err | True | False | Var _ | Shared _ as c -> c
  | App(c1,c2) -> App(copy c1,copy c2) 
  | IfThenElse(c,p,q) ->
      IfThenElse(copy c,copy_crumbp p,copy_crumbp q) 
  | Lam(v,e) -> Lam(v,copy_crumbp e)
 and copy_env c e =
  let n' = mk_node (copy e.content) in
  copying_node e n' (fun () ->
  match e.prev with
   | None -> copy c, n', n'
   | Some prev ->
      let c',b',e' = copy_env c prev in
      push n' e' ;
      c',b',n')
 and copy_crumbp (c,e) =
  match e with
   | None -> copy c, None
   | Some (b,e) ->
       let c',b',e' = copy_env c e in
       c', Some (b',e')
 in copy_crumbp p

let (@) n (c,env) =
 n.content <- c ;
 match env with
    None -> n
  | Some (b,e) -> push b n ; e
\end{lstlisting}
\caption{The \texttt{copy\_crumbp} function}
\label{tab:copy_env}
\end{table}

\paragraph*{\Crumbling.}

The code in \reftab{anf} takes an unevaluated environment and returns the corresponding \crumbled unevaluated environment. It generalizes the function $\mytr\cdot$ in the paper that turns \lat{s} into \crumblep{s}.
 It mainly consists of three mutually recursive functions:
\begin{enumerate}
  \item \verb|aux_term c e| translates \verb+c+ to $\celltwo = (\moltwo, \envtwo)$, appends $\envtwo$ to \verb|e| and returns the obtained \crumblep.
  \item \verb|aux_value c e| checks whether \verb+c+ is a value, computing either $\auxtr\cdot$, or $\mytr\cdot$ by calling \verb|aux_term c e|.
  \item \verb|aux_env c e| creates a copy of \verb|(c, e)| in \crumbled form in linear time reusing the same trick of the \verb|copying| flag as in $\alpha$-renaming.
\end{enumerate}

\begin{table}[p!]
\begin{lstlisting}
let dummy = Var (mk_var ())

let iota e = 
 let star = mk_node dummy in
 star @ e

let anf p =
 let rec aux_term c e = match c with
 | Var _ | Err | True | False -> c, e
 | App(v, w) ->
    let v, e = aux_val v e in 
    let w, e = aux_val w e in 
    App(v, w), e
 | IfThenElse(v, p, q) ->
    let c, e = aux_val v e in 
    let p = aux_crumbp p in
    let q = aux_crumbp q in
    IfThenElse(c,p,q), e
 | Lam(x, p) -> Lam(x, aux_crumbp p), e
 | Shared n -> n.content, e
 and aux_val c e = match c with
 | App _ | IfThenElse _ ->
    let n = mk_node dummy in
    let b =
     (match e with
         None -> n
       | Some (b,e) -> push n e ; b) in
    let c, e = aux_term c (Some (b,n)) in
    n.content <- c;
    Shared n, e
 | Var _ | Lam _ | Shared _
 | Err | True | False -> aux_term c e
 and aux_env c e =
   let p = aux_term e.content None in
   match e.prev with
   | None -> aux_term c (snd p)
   | Some prev ->
      let n = mk_node dummy in
      let last = n @ p in
      let (c,env) =
        copying_node e n (fun () -> aux_env c prev) in
      (match env with
          None -> c, Some(n,last)
        | Some (b,e) -> push n e ; c, Some (b,last))
 and aux_crumbp (c,env) =
  match env with
     None -> aux_term c None
   | Some (_,e) -> aux_env c e in
 iota (aux_crumbp p)
\end{lstlisting}
\caption{The \texttt{anf} function}
\label{tab:anf}
\end{table}

\clearpage

\end{document}